\documentclass[a4paper,11pt]{article}
\usepackage[utf8]{inputenc}
\usepackage{placeins}
\usepackage{jheppubMod}
\usepackage{amsmath,amsfonts,amssymb,graphicx,MnSymbol,subfigure}
\usepackage{siunitx}
\usepackage{hyperref} 
\usepackage{tikz}
\usetikzlibrary{shapes.geometric, arrows}
\usepackage[toc,page]{appendix}
\usepackage{adjustbox}
\usepackage{slashed}
\usepackage{multirow}
\usepackage{braket}
\usepackage{physics}
\usepackage{esint}
\usepackage{gensymb}
\usepackage{afterpage}
\usepackage{listings}
\newcommand{\madgraph}{{\tt MadGraph5\_aMC@NLO}}

\title{Probing $\nu$SMEFT at Belle~II with displaced dilepton vertices}

\author[a]{Patrick D. Bolton,}
\author[b]{Fabian Esser,}
\author[b,c]{Luk\'{a}\v{s} Gr\'{a}f,}
\author[d]{Juraj Klari\'{c},}
\author[e]{Suchita Kulkarni,}
\author[f]{Abner Soffer}
\affiliation[a]{Jožef Stefan Institute, Jamova 39, 1000 Ljubljana, Slovenia}
\affiliation[b]{Institute of Particle and Nuclear Physics (IPNP), Faculty of Mathematics and Physics, Charles University Prague, V Hole\v{s}ovi\v{c}k\'{a}ch 2, 180 00 Praha 8, Czech Republic}
\affiliation[c]{Institute of Physics, Silesian University in Opava, Bezru{\v{c}}ovo n{\'a}m{\v{e}}st{\'i} 1150/13, 746 01 Opava, Czech Republic}
\affiliation[d]{Department of Physics, Faculty of Science, University of Zagreb, 10000 Zagreb, Croatia}
\affiliation[e]{Institute of Physics, NAWI Graz, University of Graz, Universit\"atsplatz 5, A-8010 Graz, Austria}
\affiliation[f]{School of Physics and Astronomy, Tel Aviv University, Tel Aviv 69978, Israel}

\emailAdd{patrick.bolton@ijs.si}
\emailAdd{fabian.esser@matfyz.cuni.cz}
\emailAdd{lukas.graf@matfyz.cuni.cz}
\emailAdd{juraj.klaric@phy.hr}
\emailAdd{suchita.kulkarni@uni-graz.at}
\emailAdd{asoffer@tau.ac.il}

\abstract{
The existence of right-handed (RH) neutrinos is strongly motivated by the observation of small neutrino masses and as a means to shed light on the origin of parity violation in the Standard Model (SM). 
Experimental searches for the corresponding mass eigenstates, referred to as heavy neutral leptons (HNLs), have been carried out in a variety of experiments and within different HNL models. 
In the current work we use the SM effective field theory with added RH neutrinos, known as $\nu$SMEFT, to study GeV-scale HNL production in $e^+e^-$ collisions at the Belle~II experiment. 
We focus on leptonic decays of long-lived HNLs that produce a displaced vertex in the Belle~II detector.
Accounting for detection efficiency and backgrounds, we estimate the sensitivity of Belle~II to the new-physics scale of four-fermion, Higgs-current, and dipole operators in $\nu$SMEFT.
We find that for most operators, the search we propose probes large regions of parameter space that are not excluded by other searches.
}
\makeatletter
\gdef\@fpheader{\phantom{a}}
\makeatother

\begin{document}

\maketitle

%%%%%%%%%%%%%%%%%%%%%%%%%%%%%%%%%%%%%%%%%%%%%%%%%%%%%%%%%
\section{Introduction}
\label{sec:intro}
%%%%%%%%%%%%%%%%%%%%%%%%%%%%%%%%%%%%%%%%%%%%%%%%%%%%%%%%%

The existence of right-handed (RH) neutrinos as fundamental fields is a tantalising scenario that remains experimentally unverified, yet is strongly motivated on theoretical grounds.
In the Standard Model (SM), the chiral gauge structure~\cite{Glashow:1961tr,Weinberg:1967tq} predicts the maximal violation of parity ($P$)~\cite{Lee:1956qn,Lee:1957qr,Feynman:1958ty,Sudarshan:1958vf}, so that neutrinos couple to the $W^\pm$ and $Z$ bosons only through their left-handed (LH) chiral component~\cite{Goldhaber:1958nb,GargamelleNeutrino:1973jyy}.
RH neutrinos are consequently absent from the SM weak interactions and are often termed \textit{sterile}; nevertheless, there is good reason to believe that they may still interact with the other fields of the SM.

The most compelling argument for the existence of RH neutrinos, and their relevance phenomenologically, is their connection to neutrino masses.
The observation of neutrino flavour oscillations~\cite{Super-Kamiokande:1998kpq,SNO:2002tuh} has proven unequivocally that at least two neutrinos have tiny, but nonzero masses~\cite{Pontecorvo:1957qd,Maki:1962mu}, which the original formulation of the SM does not account for~\cite{Weinberg:2004kv}.
The neutrino masses could arise after electroweak symmetry breaking from a Yukawa coupling of the Higgs doublet to the lepton doublets and RH neutrinos.
For this to produce light Dirac neutrinos analogous to the other SM fermions would require lepton number $L$, an accidental global symmetry of the SM~\cite{Weinberg:1979sa}, to be conserved by the ultraviolet (UV) theory that supersedes the SM at high energies.

If lepton number is instead violated, the SM gauge group does not forbid a Majorana mass term~\cite{Majorana:1937vz} for the RH neutrinos.
This additional freedom in the model can permit the so-called \textit{seesaw} mechanisms, which explain the small neutrino masses from, e.g., large RH neutrino masses~\cite{Minkowski:1977sc,Yanagida:1979as,Gell-Mann:1979vob,Mohapatra:1979ia,Schechter:1980gr} or approximate $L$ conservation~\cite{Mohapatra:1986aw,Mohapatra:1986bd}.
A prediction of such models is the mixing of LH and RH neutrinos to give light and heavy mass eigenstates, the latter often referred to as heavy neutral leptons (HNLs)~\cite{Shrock:1980ct,Gronau:1984ct,Langacker:1988ur,Buchmuller:1991tu,Atre:2009rg,Bondarenko:2018ptm,Bolton:2019pcu}.
The light and heavy neutrino fields are Majorana fermions and $L$-violating (LNV) processes such as neutrinoless double beta ($0\nu\beta\beta$) decay are allowed~\cite{Goeppert-Mayer:1935uil,Racah:1937qq,Furry:1939qr,Schechter:1981bd,Deppisch:2020ztt} and provide smoking-gun signatures.

However, it could be the case that, even if the HNLs are kinematically accessible, the active-sterile mixing is suppressed or beyond the reach of current and future experiments.
This motivates a more general approach to sensitivity studies which incorporates \textit{effective} interactions of light HNLs and SM fields, which would be generated at low energies by the exchange of heavy degrees of freedom.
This is most commonly performed in the framework of the SM effective field theory (SMEFT)~\cite{Buchmuller:1985jz,Grzadkowski:2010es} extended with RH neutrinos ($\nu$SMEFT, also known as $N_R$SMEFT), which provides a complete basis of operators of dimension $d$ that respect the SM gauge group~\cite{delAguila:2008ir,Aparici:2009fh,Bhattacharya:2015vja,Liao:2016qyd,Li:2021tsq}.

In addition to LNV processes, one of the most promising experimental signatures of HNLs in the GeV mass regime is displaced vertices (DVs)~\cite{Helo:2013esa,Izaguirre:2015pga,Gago:2015vma,Lee:2018pag,Cottin:2018kmq,Abada:2018sfh,Jones-Perez:2019plk,DeVries:2020jbs,Barducci:2020icf,Cottin:2021lzz,Beltran:2021hpq,Bolton:2021pey,Delgado:2022fea,Beltran:2024twr}.
Being feebly-coupled via an active-sterile mixing or an effective operator suppressed by a high scale of new physics (NP), it is natural for HNLs to be long-lived particles (LLPs).
DVs are also advantageous over prompt HNL decay signatures due to the stark suppression of SM backgrounds.
Searches for DV signatures of HNLs have been performed at LEP~\cite{DELPHI:1996qcc,L3:2001zfe}, ATLAS~\cite{ATLAS:2022atq,ATLAS:2025uah}, CMS~\cite{CMS:2022fut,CMS:2023jqi}, and Belle~\cite{Belle:2013ytx,Belle:2024wyk} and have resulted in stringent constraints on the HNL parameter space.
Future sensitivities of FCC-ee, CEPC etc. have also been estimated~\cite{Antusch:2016vyf,Blondel:2014bra,Blondel:2022qqo,Barducci:2022hll,Bernal:2024kqx,Bolton:2025tqw,Ai:2025cpj,Drewes:2025ocf,Fuks:2026hra}.

In the mid-term future, the currently operating Belle~II experiment provides a powerful means to search for feebly-interacting particles, such as  dark photons~\cite{Ferber:2022ewf,Bandyopadhyay:2022klg,Jaeckel:2023huy,Cheung:2024oxh} and HNLs~\cite{Dey:2020juy,Zhou:2021ylt}.
The experiment is sensitive to DV displacements up to $\mathcal{O}(1~\text{m})$, and its $e^+ e^-$-collider environment enables probing different regions of HNL parameter space from those probed at hadron colliders.
Belle~II has already collected almost $1~\text{ab}^{-1}$ of data, and by the end of its operation will have collected $50~\text{ab}^{-1}$.

In this work, we specifically investigate the sensitivity of DV searches at Belle~II to the lowest dimension $\nu$SMEFT operators that lead to the production and decay of Majorana HNLs, assuming one nonzero operator at a time.
For operators involving two HNL fields, we assume the presence of two HNL species with various values of the mass splitting.
All of the considered operators may result in a displaced $e^+e^-$ pair and missing energy in the Belle~II detector.
Some of the operators also lead to displaced $\mu^+\mu^-$, $\tau^+\tau^-$, $e^\pm\mu^\mp$ and $e^\pm\tau^\mp$ signatures.
While hadronic HNL decay modes are also interesting, we do not consider their sensitivity in this work.
For the sensitivity analysis, we pay careful attention to experimental considerations.
This includes the reconstruction efficiency for displaced charged tracks and the expected backgrounds, which come predominantly from photon conversions and misidentified neutral-kaon decays.
We choose appropriate cuts to strongly suppress the background while retaining high signal efficiency to maximise the new-physics reach.
We also devise a constraint-based method to obtain kinematic handles for separating signal from background.

This work is organised as follows.
In Sec.~\ref{sec:pheno}, we introduce the $\nu$SMEFT operators of interest in this study and give expressions relevant for the production of HNLs in $e^+e^-$ collisions and for their decays, from which we also estimate the HNL lifetimes and branching fractions.
In Sec.~\ref{sec:analysis} we conduct our DV sensitivity analysis.
First, in Sec.~\ref{sec:analytics}, we present analytical estimates for the number of signal events and derive the sensitivity of Belle~II to the scale of NP for different HNL masses, mass splittings, and final states.
This estimation is performed with 100\% signal-reconstruction efficiency and in the absence of background.
In Sec.~\ref{sec:DVbackground}, the simulation, suppression, and estimation of the relevant backgrounds are presented, with detailed calculations provided in App.~\ref{app:bkg_estimates}.
Then, in Sec.~\ref{sec:DVsens} we detail the full numerical analysis, which involves signal simulation across the parameter space, realistic simulation of the reconstruction efficiency, and the background yield.
We summarise and conclude in Sec.~\ref{sec:conclusions}.

%%%%%%%%%%%%%%%%%%%%%%%%%%%%%%%%%%%%%%%%%%%%%%%%%%%%%%%%%
\section{$\nu$SMEFT framework}
\label{sec:pheno}
%%%%%%%%%%%%%%%%%%%%%%%%%%%%%%%%%%%%%%%%%%%%%%%%%%%%%%%%%

%%%%%%%%%%%%%%%%%%%%%%%%%%%%%%%%%%%%%%%%%%%%%%%%%%%%%%%%%
\subsection{Operators}
\label{subsec:operators}
%%%%%%%%%%%%%%%%%%%%%%%%%%%%%%%%%%%%%%%%%%%%%%%%%%%%%%%%%

In this work, we consider the SM extended by operators in the $\nu$SMEFT.
A basis of higher dimensional operators is constructed from SM degrees of freedom and an arbitrary number of the RH neutrino fields $N_r$ ($r = 1,2,\ldots$).
The Lagrangian for this EFT can be written as
\begin{align}
\label{eq:vSMEFT}
\mathcal{L} = \mathcal{L}_{\text{SM}} + \bar{N}_r i\slashed{\partial}N_r - \bigg[\bar{l}_r [Y_\nu]_{rs}N_s\tilde{H} + \frac{1}{2}\bar{N}_r^c [M_R]_{rs} N_s + \text{h.c.}\bigg] + \sum_i C_i^{(d)} Q_i^{(d)}\,,
\end{align}
where $C_i^{(d)} \propto 1/\Lambda^{d-4}$ is the Wilson coefficient for the operator $Q_i^{(d)}$, with $\Lambda$ being the scale at which the full UV dynamics enters, $\tilde{H} = i\tau^2 H$, with $\tau^2$ being the second Pauli matrix, and $N_r^c = C\bar{N}_r^T$, with $C$ being the charge-conjugation matrix.

The $\nu$SMEFT Lagrangian provides an effective description valid for processes whose energy and momentum transfer are small compared to $\Lambda$.
In this study, we consider the $d = 5$ and $d = 6$ operators listed in Table~\ref{tab:vSMEFT-operators}, which are relevant phenomenologically at Belle~II.
At the energy scales probed by Belle~II, the operators should be taken in the broken phase of the theory with the fermion and gauge fields rotated to the mass basis and the heavy fields ($W^\pm/Z$) integrated out.

\begin{table}[t!]
\centering
\renewcommand{\arraystretch}{1.25}
\setlength\tabcolsep{2.7pt}
\begin{tabular}{c|c}
\hline 

\multicolumn{2}{c}{$\psi^2 X$} \\ \hline

$Q_{NNB}$ & $g'(\bar{N}^c \sigma^{\mu\nu}N)B_{\mu\nu}$\\
\hline
\end{tabular}
\hspace{1em}
\begin{tabular}{c|c|c|c}
\hline 

\multicolumn{2}{c|}{$\psi^2 H X$} & \multicolumn{2}{c}{$\psi^4$} \\ \hline

$Q_{NB}$ & $g'(\bar{l} \sigma^{\mu\nu} N)\tilde{H} B_{\mu\nu}$ & $Q_{lN}$ & $(\bar{l} \gamma_\mu l)(\bar{N} \gamma^\mu N)$ \\ 

$Q_{NW}$ &  $g(\bar{l} \sigma^{\mu\nu} \tau^I N) \tilde{H} W^{I}_{\mu\nu}$ & $Q_{eN}$ & $(\bar{e} \gamma_\mu e)(\bar{N} \gamma^\mu N)$ \\ \cline{1-2}

\multicolumn{2}{c|}{$\psi^2 H^2 D$} & $Q_{lNle}$ & $(\bar{l}^j N)\epsilon_{jk}(\bar{l}^k e)$ \\ \cline{1-2}

$Q_{HN}$ & $(\bar{N} \gamma_\mu N)(H^{\dagger} i \overleftrightarrow{D}^\mu H)$ &  & \\

$Q_{HNe}$ & $(\bar{N} \gamma_\mu e)(\tilde{H}^{\dagger} i D^\mu H)$ &  &  \\

\hline
\end{tabular}
\caption{Dimension $d = 5$ (left) and $d = 6$ (right) $\nu$SMEFT operators~\cite{delAguila:2008ir} considered in the context of Belle~II in this work. $\psi, X, H$ denote fermion, gauge, Higgs fields, respectively, and $D$ a covariant derivative. Flavour indices are omitted.}
\label{tab:vSMEFT-operators}
\end{table}

In the following, we consider three classes of $d = 5$ and $d = 6$ $\nu$SMEFT operators, first considered in Ref.~\cite{delAguila:2008ir}.
Firstly, dipole operators of the form
\begin{align}
\label{eq:dipole}
\mathcal{L} &\supset \frac{1}{2}g'C_{\underset{pr}{NNB}}
(\bar{N}_{p}^c\sigma^{\mu\nu} N_{r})B_{\mu\nu} \nonumber \\
&\hspace{1.3em} + g'C_{\underset{pr}{NB}}(\bar{l}_p \sigma^{\mu\nu} N_{r})\tilde{H} B_{\mu\nu} + gC_{\underset{pr}{NW}}(\bar{l}_p \sigma^{\mu\nu} \tau^I N_{r})\tilde{H} W^{I}_{\mu\nu} + \text{h.c.} \,,
\end{align}
where $\tau^I$ are the Pauli matrices and $g$ and $g'$ are the $SU(2)_L$ and $U(1)_Y$ gauge couplings, respectively.
For $SU(3)_c$ singlet fields, the convention taken for the SM covariant derivative is $D_\mu = \partial_\mu + i g t^IW^{I\mu} + i g' Y B_\mu$, where $t^I = \tau^I/2$ are the $SU(2)_L$ generators and $Y$ is the hypercharge.
Next, the Higgs current operators,
\begin{align}
\label{eq:Higgs_current}
\mathcal{L} &\supset C_{\underset{pr}{HN}}
(\bar{N}_{p}\gamma_\mu N_{r})(H^{\dagger} i \overleftrightarrow{D}^\mu H) + \Big[C_{\underset{pr}{HNe}}(\bar{N}_{p} \gamma_\mu e_{r})(\tilde{H}^{\dagger} i D^\mu H) + \text{h.c.}\Big] \,,
\end{align}
where we use $(H^{\dagger} i \overleftrightarrow{D}^\mu H) = iH^{\dagger} (D^\mu H) - i(D^\mu H)^\dagger H$.
Lastly, the four-fermion operators
\begin{align}
\mathcal{L} &\supset C_{\underset{prst}{lN}}(\bar{l}_p\gamma_\mu l_r)
(\bar{N}_{s}\gamma^\mu N_{t}) + 
C_{\underset{prst}{eN}}(\bar{e}_{p}\gamma_\mu e_{r})
(\bar{N}_{s}\gamma^\mu N_{t}) \nonumber \\
&\hspace{1.3em} + \Big[C_{\underset{prst}{lNle}}(\bar{l}_p^j N_{r})
\epsilon_{jk}(\bar{l}_s^k e_{t}) + \text{h.c.}\Big] \,.
\label{eq:four_fermion}
\end{align}
In Eqs.~\eqref{eq:dipole}--\eqref{eq:four_fermion}, we have kept the flavour indices $p,r,s,t$ explicit.
All relevant $\nu$SMEFT operators are summarised with omitted flavour indices in Table \ref{tab:vSMEFT-operators}.

Before examining the impact of these SM-invariant operators at low energies, we first note that they induce effective $W^\pm$ and $Z$ interactions involving $N_R$, i.e.,
\begin{align}
\mathcal{L} \supset - g_Z j_Z^\mu Z_\mu - \frac{g}{\sqrt{2}}\Big[j_W^\mu W_\mu^+ + \text{h.c.}\Big]\,,
\label{eq:induced-interactions}
\end{align}
where $g_Z = g/\cos\theta_w$ with $\theta_w$ being the weak mixing angle.
The currents appearing in~Eq.~\eqref{eq:induced-interactions} are
\begin{align}
j_W^\mu &= [W_\nu^L]_{pr}\bar{\nu}_{Lp}\gamma_\mu e_{Lr} + [W_q^L]_{pr}\bar{u}_{Lp}\gamma_\mu d_{Lr} + [W_N^R]_{pr}\bar{N}_{Rp}\gamma_\mu e_{Rr} \,, \nonumber \\
j_Z^\mu &= [Z_f^L]_{pr}\bar{f}_{Lp} \gamma_\mu f_{Lr} + [Z_f^R]_{pr}\bar{f}_{Rp} \gamma_\mu f_{Rr} + [Z_N^R]_{pr} \bar{N}_{Rp} \gamma_{\mu} N_{Rr} \,,
\label{eq:WZ_currents}
\end{align}
for $f = \nu, e, u, d$.
The SM couplings in Eq.~\eqref{eq:WZ_currents} are contained in
\begin{align}
[W_\nu^L]_{pr} = [W_q^L]_{pr} = \delta_{pr}\,,\quad [Z_f^L]_{pr} = (t^3_f - Q_f s_w^2)\delta_{pr}\,,\quad [Z_f^R]_{pr} =  - Q_f s_w^2\delta_{pr} \,,
\label{eq:SM-couplings-induced-interactions}
\end{align}
where $Q_f$ is the electric charge of $f$, $s_w=\sin \theta_w$, and $\delta_{pr}$ is the Kronecker delta.
Similarly, the $\nu$SMEFT operators contribute to the new $W^\pm$ and $Z$ interactions as
\begin{align}
[W_N^R]_{pr} &=  \frac{v^2}{2}C_{\underset{pr}{HNe}} \,, \quad
[Z_N^R]_{pr} = -\frac{v^2}{2}C_{\underset{pr}{HN}} \,,
\end{align}
where $v = 246$~GeV is the Higgs vacuum expectation value.

\begin{table}[t!]
\centering
\renewcommand{\arraystretch}{1.25}
\setlength\tabcolsep{2.7pt}
\begin{tabular}{c|c}
\hline 

\multicolumn{2}{c}{$\psi^2 X$} \\ \hline

$\mathcal{O}_{NN\gamma}$ & $e(\bar{N}_R^c \sigma^{\mu\nu} N_R) F_{\mu\nu}$\\

$\mathcal{O}_{\nu N\gamma}$ & $e(\bar{\nu}_L \sigma^{\mu\nu} N_R) F_{\mu\nu}$\\

\hline
\end{tabular}
\hspace{1em}
\begin{tabular}{c|c}
\hline

\multicolumn{2}{c}{$\psi^4$} \\ \hline

$\mathcal{O}_{Nf}^{V,RL}$ & $(\bar{N}_R \gamma_\mu N_R)(\bar{f}_L \gamma^\mu f_L)$ \\

$\mathcal{O}_{Nf}^{V,RR}$ & $(\bar{N}_R \gamma_\mu N_R)(\bar{f}_R \gamma^\mu f_R)$ \\

$\mathcal{O}_{eNud}^{V,RL}$ & $(\bar{e}_R \gamma_\mu N_R)(\bar{u}_L \gamma^\mu d_L)$ \\

$\mathcal{O}_{\nu Ne}^{S,RL}$ & $(\bar{\nu}_L N_R)(\bar{e}_R e_L)$ \\

$\mathcal{O}_{\nu Ne}^{S,RR}$ & $(\bar{\nu}_L N_R)(\bar{e}_L e_R)$ \\

$\mathcal{O}_{\nu Ne}^{T,RR}$ & $(\bar{\nu}_L \sigma_{\mu\nu} N_R)(\bar{e}_L \sigma^{\mu\nu} e_R)$ \\

\hline
\end{tabular}
\caption{Dimension $d = 5$ dipole (left) and $d = 6$ four-fermion (right) $\nu$LEFT operators induced by the $\nu$SMEFT operators in Table~\ref{tab:vSMEFT-operators}, with $f = \nu, e, u, d$. Flavour indices are again omitted.}
\label{tab:vLEFT-operators}
\end{table}

Integrating $W^\pm$, $Z$ and $H$ out of the theory yields the low energy effective interactions of $N_R$ relevant at Belle~II.
In the low-energy EFT ($\nu$LEFT) we obtain the  $N_R$-containing operators in Table~\ref{tab:vLEFT-operators}.
These are the dipole operators
\begin{align}
\mathcal{L}&\supset \frac{1}{2} d_{\underset{pr}{NN\gamma}} e (\bar{N}_{Rp}^c \sigma^{\mu\nu} N_{Rr}) F_{\mu\nu} + d_{\underset{pr}{\nu N\gamma}} e (\bar{\nu
}_{Lp} \sigma^{\mu\nu} N_{Rr}) F_{\mu\nu} + \text{h.c.} \,,
\label{eq:photon_operators}
\end{align}
and the four-fermion operators of neutral-current type,
\begin{align}
\mathcal{L}&\supset L_{\underset{prst}{Nf}}^{V,RL}(\bar{N}_{Rp} \gamma_\mu N_{Rr})(\bar{f}_{Ls}\gamma^\mu f_{Lt}) + L_{\underset{prst}{Nf}}^{V,RR}(\bar{N}_{Rp} \gamma_\mu N_{Rr})(\bar{f}_{Rs}\gamma^\mu f_{Rt}) \nonumber \\
&\hspace{1.3em}+\Big[L_{\underset{prst}{\nu Ne}}^{S,RL}(\bar{\nu}_{Lp} N_{Rr})(\bar{e}_{Rs}e_{Lt}) + L_{\underset{prst}{\nu Ne}}^{S,RR}(\bar{\nu}_{Lp} N_{Rr})(\bar{e}_{Ls}e_{Rt})\nonumber \\
&\hspace{3em} + L_{\underset{prst}{\nu Ne}}^{T,RR}(\bar{\nu}_{Lp} \sigma_{\mu\nu} N_{Rr})(\bar{e}_{Ls}\sigma^{\mu\nu}e_{Rt}) + \text{h.c.}\Big]\,,
\label{eq:dilepton_operators_nc}
\end{align}
and charged-current type,
\begin{align}
\mathcal{L}&\supset L_{\underset{prst}{eNud}}^{V,RL}(\bar{e}_{Rp} \gamma_\mu N_{Rr})(\bar{u}_{Ls}\gamma^\mu d_{Lt}) + \text{h.c.}\,.
\label{eq:dilepton_operators_cc}
\end{align}
The relevant tree-level matching relations between the $\nu$SMEFT operators in Table~\ref{tab:vSMEFT-operators} and the $\nu$LEFT operators in Table~\ref{tab:vLEFT-operators} are provided in Table~\ref{tab:Matching}.

\begin{table}[t!]
\centering
\renewcommand{\arraystretch}{1.25}
\setlength\tabcolsep{2.7pt}
\begin{tabular}{c|c}
\hline

Operator & Matching \\ \hline

$\mathcal{O}_{NN\gamma}$ & $C_{\underset{pr}{NNB}}$ \\

$\mathcal{O}_{N\nu}^{V,RL}$ & $C_{\underset{stpr}{lN}} - \frac{g_Z^2}{M_Z^2}[Z_N^R]_{pr}[Z_\nu^L]_{st}$ \\

$\mathcal{O}_{Ne}^{V,RL}$ & $C_{\underset{stpr}{lN}} - \frac{g_Z^2}{M_Z^2}[Z_N^R]_{pr}[Z_e^L]_{st}$\\

$\mathcal{O}_{Ne}^{V,RR}$ & $C_{\underset{stpr}{eN}} - \frac{g_Z^2}{M_Z^2}[Z_N^R]_{pr}[Z_e^R]_{st}$ \\

$\mathcal{O}_{Nq}^{V,RL}$ & $- \frac{g_Z^2}{M_Z^2}[Z_N^R]_{pr}[Z_q^{L}]_{st}$ \\

$\mathcal{O}_{Nq}^{V,RR}$ & $- \frac{g_Z^2}{M_Z^2}[Z_N^R]_{pr}[Z_q^{R}]_{st}$ \\ 

\hline
\end{tabular}
\hspace{0em}
\begin{tabular}{c|c}
\hline

Operator & Matching \\ \hline

$\mathcal{O}_{\nu N\gamma}$ & $\frac{v}{\sqrt{2}}\big(C_{\underset{pr}{NB}} + C_{\underset{pr}{NW}}\big)$ \\

$\mathcal{O}_{eNud}^{V,RL}$ & $-\frac{g^2}{2M_W^2}[W_N^R]_{rp}^*[W_q^L]_{st}$ \\

$\mathcal{O}_{\nu Ne}^{S,RL}$ & $\frac{g^2}{M_W^2}[W_N^R]_{rs}^*[W_\nu^L]_{pt}$ \\

$\mathcal{O}_{\nu Ne}^{S,RR}$ & $C_{\underset{prst}{lNle}} + \frac{1}{2}C_{\underset{srpt}{lNle}}$ \\

$\mathcal{O}_{\nu Ne}^{T,RR}$ & $\frac{1}{8}C_{\underset{srpt}{lNle}}$ \\

\hline
\end{tabular}
\caption{Tree-level matching of the $\nu$SMEFT operators in Table~\ref{tab:vSMEFT-operators} to the $\nu$LEFT operators in Table~\ref{tab:vLEFT-operators}, which induce the production and decay of HNLs at Belle~II. Shown are operators containing one (right) and two (left) RH neutrino fields.}
\label{tab:Matching}
\end{table}

In the following, for simplicity, we assume that the Yukawa coupling $Y_\nu$ and therefore the active-sterile mixing between $\nu_L$ and $N_R$ is negligible.
Without loss of generality, the Majorana mass matrix $M_R$ can be taken to be real and diagonal, with the weak eigenbasis states (indices $r,s$) equivalent to the mass eigenbasis states (indices $i,j$), and the Majorana fermions $N_i = N_i^c$.
We consider the case of two Majorana HNLs, $i = 1,2$, with masses $m_{N_1}$, $m_{N_2}$. 
We define the mass-splitting ratio
\begin{align}
\delta = \frac{m_{N_2} - m_{N_1/\nu}}{m_{N_2}} \,,
\end{align}
with the limiting values $\delta = 0$ for $N_1$ and $N_2$ being degenerate, and $\delta = 1$ for a massless $N_1$ or SM neutrino.

\begin{figure}[t!]
    \centering
    \includegraphics[width=0.3\linewidth]{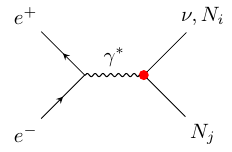}
    \includegraphics[width=0.225\linewidth]{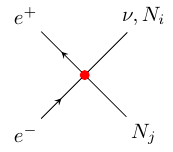}
\caption{Processes leading to the pair and single production of HNLs at Belle~II via the $\nu$SMEFT operators in Table~\ref{tab:vSMEFT-operators}. (Left) Production via an off-shell photon due to the dipole operators $Q_{NNB}$, $Q_{NB}$ and $Q_{NW}$. (Right) Production via low-energy four-fermion interactions resulting from the Higgs current operators $Q_{HN}$ and $Q_{HNe}$ and the four-fermion operators $Q_{lN}$, $Q_{eN}$ and $Q_{lNle}$.}
\label{fig:production}
\end{figure}

\subsection{Focus of this work}
\label{sec:focus-of-this-work}

\begin{figure}[t!]
    \centering
    \includegraphics[width=0.225\linewidth]{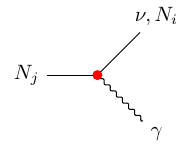}
    \includegraphics[width=0.27\linewidth]{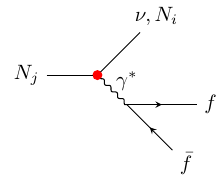}
    \includegraphics[width=0.23\linewidth]{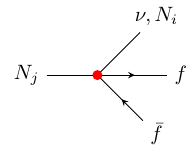}
    \includegraphics[width=0.23\linewidth]{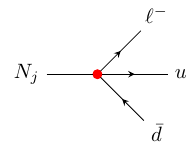}
    \caption{Decays of the HNL $N_j$ via the $\nu$SMEFT operators considered in Table~\ref{tab:vSMEFT-operators}: two-body (left) and three-body Dalitz-type (centre left) decays induced by the dipole operators $Q_{NNB}$, $Q_{NB}$ and $Q_{NW}$ and three-body decays (centre right and right) induced by the Higgs current operators $Q_{HN}$, $Q_{HNe}$ and the four-fermion operators $Q_{lN}$, $Q_{eN}$ and $Q_{lNle}$. The hadronisation of quarks leads to (multi-)hadronic final states.}
\label{fig:decays}
\end{figure}

The $\nu$LEFT operators in Eqs.~\eqref{eq:photon_operators}, \eqref{eq:dilepton_operators_nc} and \eqref{eq:dilepton_operators_cc} can induce the production of one or two HNLs via $e^+e^-(\to \gamma^*)\to N_1 N_2$ or $\nu N_2$, shown diagrammatically in Fig.~\ref{fig:production}.
These processes can then be followed by the two-body decays $N_2\to N_1\gamma/\nu\gamma$, three-body leptonic decays $N_2(\to N_1 \gamma^*/\nu \gamma^*)\to N_1 \ell^+\ell^-/\nu \ell^+\ell^-$ and hadronic decays from the partonic process $N_2(\to N_1 \gamma^*/\nu \gamma^*)\to N_1 q\bar{q}/\nu q\bar{q}$, all depicted in Fig.~\ref{fig:decays}.

The focus of this work is the signature of a leptonic $\ell^+\ell^-$ DV.
Nonetheless, all decay modes are required for an accurate determination of the signal decay branching fraction and the total HNL width, which governs its decay length in the Belle~II detector. 
In Sec.~\ref{subsec:production} we provide analytical expressions for the HNL production cross sections induced by these operators.
The analytical expressions for the HNL decay rates appear in Sec.~\ref{subsec:decays}.

Furthermore, we consider only the case of one nonzero $\nu$SMEFT operator at a time.
The requirement that this operator induce both the production and decay of HNLs at Belle~II limits the possible flavour structures of the operators in the considered two-HNL scenario.
In particular, the requirement that the HNLs couple to the incoming $e^+e^-$ is the most constraining factor.
For $Q_{HNe}$, $Q_{lN}$ and $Q_{eN}$, all lepton fields must be of electron flavour.
Meanwhile, for $Q_{lNle}$, $e_t$ and only one of the lepton doublets $l_p^j$ or $l_s^k$ needs to be electron flavoured, while the other doublet, leading to an undetected final-state neutrino, may have any flavour.

In total, the viable flavour structures of the Wilson coefficients are
\begin{gather}
C_{NNB} = C_{\underset{12}{NNB}}\,,~C_{HN} = C_{\underset{12}{HN}}\,,~C_{lN} = C_{\underset{ee12}{lN}}\,,~C_{eN} = C_{\underset{ee12}{eN}} \,, \nonumber \\
C_{NB}^\rho = C_{\underset{\rho 2}{NB}}\,,~C_{NW}^\rho = C_{\underset{\rho 2}{NW}}\,,~C_{HNe} = C_{\underset{2e}{HNe}}\,,~C_{lNle}^{e\rho} = C_{\underset{e2\rho e}{lNle}}\,,~C_{lNle}^{\rho e} = C_{\underset{\rho 2ee}{lNle}} \,,
\label{eq:vSMEFT_WCs}
\end{gather}
for $\rho = e, \mu, \tau$.
In Table~\ref{tab:final_states_leptonic}, we summarise the leptonic final state signatures resulting from the operators in Eq.~\eqref{eq:vSMEFT_WCs}.
The table also indicates whether each final state is pursued in our realistic sensitivity estimate in Sec.~\ref{sec:DVsens}.
Generally, final states with a $\tau$ lepton are not pursued, as the analytical sensitivity in Sec.~\ref{sec:analytics} shows their inferiority even before accounting for the impact of $\tau$ reconstruction.

\begin{table}[t!]
\centering
\renewcommand{\arraystretch}{1.25}
\setlength\tabcolsep{2.7pt}
\begin{tabular}{c|c|c|c}
\hline 
Signature & Final state & $\nu$SMEFT coefficients & Pursued  \\ 
\hline

\multirow{2}{*}{$e^+ e^-$ + $E_{\rm miss}$} & $N_1 N_1 e^+ e^-$ & $C_{NNB}$, $C_{HN}$, $C_{lN}$, $C_{eN}$ & \multirow{2}{*}{\textcolor{blue}{$\checkmark$}} \\
& $\nu \nu e^+ e^-$ & $C_{NB}^\rho$, $C_{NW}^\rho$, $C_{HNe}$, $C_{lNle}^{e\rho}$, $C_{lNle}^{\rho e}$ & \\
\hline
\multirow{2}{*}{$\mu^+ \mu^-$ + $E_{\rm miss}$} & $N_1 N_1 \mu^+ \mu^-$ & $C_{NNB}$, $C_{HN}$ & \multirow{2}{*}{\textcolor{blue}{$\checkmark$}} \\
& $\nu \nu \mu^+ \mu^-$ & $C_{NB}^\rho$, $C_{NW}^\rho$  & \\
\hline
\multirow{2}{*}{$\tau^+ \tau^-$ + $E_{\rm miss}$} & $N_1 N_1 \tau^+ \tau^-$ & $C_{NNB}$, $C_{HN}$ & \multirow{2}{*}{\textcolor{red}{$\times$}} \\
 & $\nu \nu \tau^+ \tau^-$ & $C_{NB}^\rho$, $C_{NW}^\rho$  & \\
\hline
$e^{\pm} \mu^{\mp}$ + $E_{\rm miss}$ & $\nu \nu e^{\pm} \mu^{\mp}$ & $C_{HNe}$, $C_{lNle}^{ e \mu}$, $C_{lNle}^{\mu e}$ & \textcolor{blue}{$\checkmark$} \\
\hline
$e^{\pm} \tau^{\mp}$ + $E_{\rm miss}$ & $\nu \nu e^{\pm} \tau^{\mp}$ & $C_{HNe}$, $C_{lNle}^{e \tau}$, $C_{lNle}^{\tau e}$ & \textcolor{red}{$\times$} \\
\hline
\end{tabular}
\caption{The experimental signatures of a dilepton-DV plus missing energy, their corresponding full final state, and the contributing Wilson coefficients.
In the last column we indicate whether the final state is pursued in our realistic sensitivity estimate.}
\label{tab:final_states_leptonic}
\end{table}
%

%%%%%%%%%%%%%%%%%%%%%%%%%%%%%%%%%%%%%%%%%%%%%%%%%%%%%%%%%
\section{$\nu$SMEFT Sensitivity 
of the proposed search}
\label{sec:analysis}
%%%%%%%%%%%%%%%%%%%%%%%%%%%%%%%%%%%%%%%%%%%%%%%%%%%%%%%%%

In the following, we take the one-operator-at-a-time approach and derive the sensitivities to individual Wilson coefficients $C_i$ from DV searches at Belle~II, setting all other Wilson coefficients to zero.
We note in passing that a realistic UV completion will potentially produce more than one operator at a time.
In this case, one should calculate the matching of the UV theory onto the $\nu$SMEFT and derive the bounds directly in terms of the masses and couplings of the renormalisable UV theory.
The limits derived in this work should thus be seen as sensitivity estimates of the new physics scales which can be probed by Belle~II.
Using the scaling relation $C_i = \tilde{C}_i/\Lambda^{d-4}$, we can set $\tilde{C}_i = 1$ and translate the bounds on the Wilson coefficient into bounds on the NP scale $\Lambda$.
Note that for any choice of $\tilde{C}_i \neq 1$, the bounds on $\Lambda$ can be easily obtained by rescaling the bounds for $\tilde{C}_i =1$.

\begin{figure}
    \centering
    \includegraphics[width=0.3\linewidth]{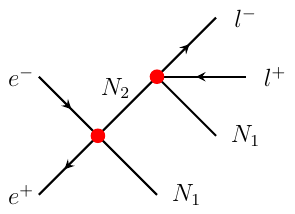}
    \includegraphics[width=0.3\linewidth]{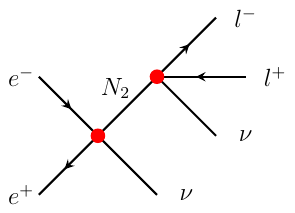}
\caption{Processes leading to the production and leptonic decays of HNLs at Belle~II via the $\nu$SMEFT operators in Table~\ref{tab:vSMEFT-operators}. (Left) Production of $N_1 N_2$ and subsequent decay $N_2 \rightarrow N_1 \ell^+ \ell^-$, mediated by the operators $Q_{HN}$, $Q_{lN}$, $Q_{eN}$ and $Q_{NNB}$. (Right) Production of $N_2$ in association with a SM neutrino $\nu$ and subsequent decay $N_2 \rightarrow \nu \ell^+ \ell^-$, mediated by the operators $Q_{HNe}$, $Q_{lNle}$, $Q_{NB}$ and $Q_{NW}$.}
\label{fig:diagrams_production_leptonic_decay}
\end{figure}
%

%%%%%%%%%%%%%%%%%%%%%%%%%%%%%%%%%%%%%%%%%%%%%%%%%%%%%%%%%
\subsection{Analytic sensitivity estimates for no background and 100\% efficiency}
\label{sec:analytics}
%%%%%%%%%%%%%%%%%%%%%%%%%%%%%%%%%%%%%%%%%%%%%%%%%%%%%%%%%

As a first step, we analytically estimate the expected number of signal events for each operator, accounting for the fiducial-volume acceptance but assuming 100\% reconstruction efficiency and no background. Simulation-based sensitivity estimates that do account for background and efficiency appear in Sec.~\ref{sec:DVsens}. The analytical estimate sets the scale on the maximal sensitivity achievable under ideal conditions, which will be required to infer the scan range for the simulation-based analysis in Sec.~\ref{sec:DVsens}. 

\begin{figure}[t!]
    \centering
    \includegraphics[width=0.49\linewidth]{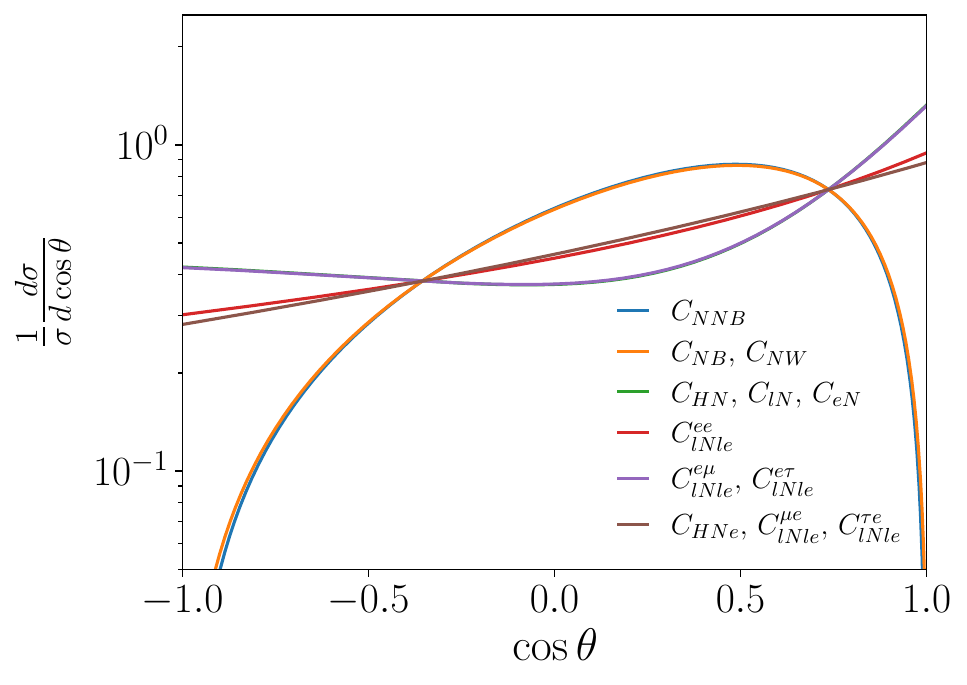}
    \includegraphics[width=0.49\linewidth]{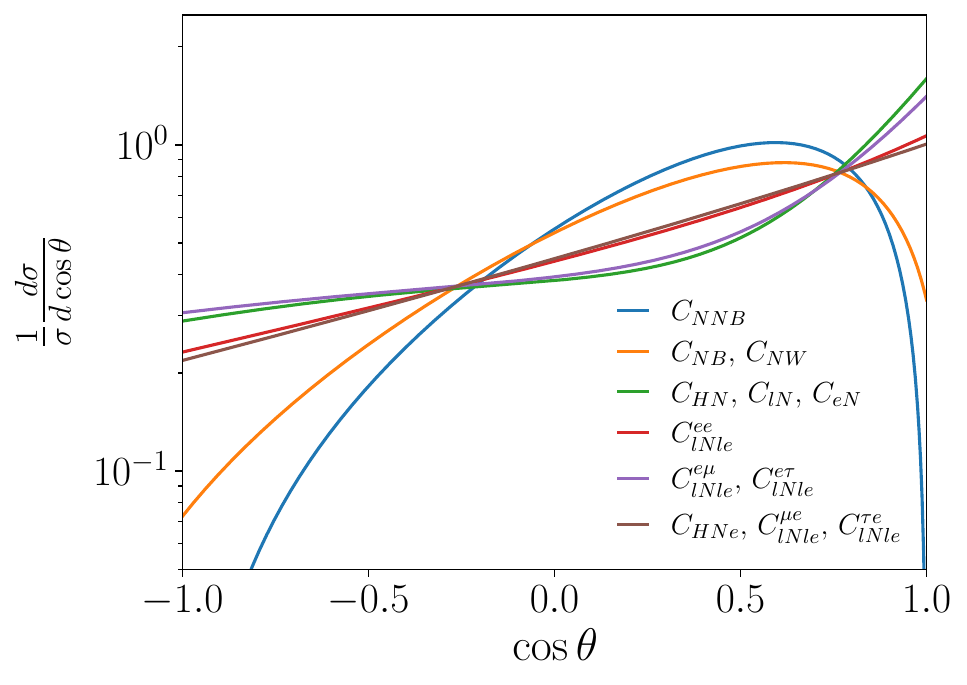}
    \caption{Differential cross sections for the process $e^+e^- \to N_1 N_2/\nu N_2$ for the $\nu$SMEFT operators considered in this work, where $\theta$ is the angle between the outgoing state $N_2$ and the incoming $e^-$ direction in the laboratory frame, for a heavier HNL mass $m_{N_2} = 1$~GeV (left) and $m_{N_2} = 4$~GeV (right). A mass splitting ratio of $\delta = 0.1$ is used for the case of two-HNL production, and effectively $\delta=1$ for HNL-neutrino production.}
\label{fig:dsigma}
\end{figure}

The differential cross section in the polar CM angle $\theta^*$ for the production of $N_1 N_2$ and $\nu N_2$ are given in Eqs.~\eqref{eq:diff_xsec_NN} and \eqref{eq:diff_xsec_vN}. 
In Fig.~\ref{fig:dsigma}, we plot the normalised differential cross sections $(d\sigma/d\cos\theta)/\sigma$ induced by nonzero values of the $\nu$SMEFT coefficients in Eq.~\eqref{eq:vSMEFT_WCs}, after matching to the $\nu$LEFT operators in Table~\ref{tab:vLEFT-operators} and transforming Eqs.~\eqref{eq:diff_xsec_NN} and \eqref{eq:diff_xsec_vN} to the laboratory frame.

The partial decay widths for the leptonic final states $N_2 \to N_1 \ell^+\ell^-$ and $N_2 \to \nu \ell^+\ell^-$ are given in Eqs.~\eqref{eq:decay_rate_Ni} and \eqref{eq:decay_rate_nu}, respectively. Together with the total decay width $\Gamma_{N_2}$ in Eq.~\eqref{eq:total_decay_width}, this allows us to determine the branching fractions for each $\nu$SMEFT operator and channel, shown in Figs.~\ref{fig:branching_1} and~\ref{fig:branching_2}.

To estimate the number of signal events, we multiply the relevant differential cross section with the probability that the HNL decays within the radial range $1.2 < R < 60$~cm, which is instrumented with  the Belle~II Pixel Detector (PXD), Silicon Vertex Detector (SVD) and Central Drift Chamber (CDC).
The lower boundary of this range suppresses the background from prompt tracks, defined as those produced at the collider interaction point (IP), as well as those from particle-material interactions in the beampipe.
The upper value ensures that the DV tracks are well reconstructed, as outgoing tracks originating at $R \leq 60$~cm  can pass through 28 CDC layers, well above the usual track-quality requirement of 20 CDC hits.

We calculate the number of signal events by numerically integrating
\begin{align}
    &S_{\ell^+\ell^-} \nonumber\\
    &\hspace{-0.75em} = \mathcal{L} \int_{\cos\theta_{\rm max}}^{\cos\theta_{\rm min}} d \cos\theta \frac{d \sigma}{d \cos \theta^*} \text{B}_{\ell^+\ell^-} \left( \exp \left[ \frac{- \Gamma_{N_2} L_{\rm min}}{(\beta\gamma)_{N_2} \sin \theta} \right] - \exp \left[\frac{- \Gamma_{N_2} L_{\rm max}}{(\beta\gamma)_{N_2} \sin \theta} \right] \right) \frac{d \cos \theta^*}{d \cos \theta}\,,
\label{eq:NDV_analytical}
\end{align}
where $\mathcal{L} = 50~\text{ab}^{-1}$ is the  integrated luminosity corresponding to the full Belle~II data set, and the angular range $\theta_{\rm min} = 43.6 \degree$ and $\theta_{\rm max} = 101.5 \degree$ corresponds to the cuts chosen in Sec.~\ref{sec:DVbackground} to reduce background.
In Eq.~\eqref{eq:NDV_analytical}, we take into account the boost from the CM frame to the laboratory frame by expressing $\cos\theta^*$ in terms of $\theta$ using Eq.~\eqref{eq:CMtolab} and multiplying by the Jacobian factor in Eq.~\eqref{eq:Jacobian}.
The branching fraction for the leptonic final state $\ell^+\ell^-$ is given by
\begin{align}
    \text{B}_{\ell^+\ell^-} = \frac{\Gamma_{\ell^+\ell^-}}{\Gamma_{N_2}}\,,
\end{align}
with the total decay width $\Gamma_{N_2}$ in Eq.~\eqref{eq:total_decay_width} and the individual decay widths $\Gamma_{\ell^+\ell^-}$ into leptonic final states $\ell^+\ell^-$ in Eqs.~\eqref{eq:decay_rate_Ni} and \eqref{eq:decay_rate_nu}, respectively.
The difference between the two exponential factors gives the probability that $N_2$ covers a transverse distance between $L_{\rm min} =1.2$~cm and $L_{\rm max}=60$~cm before decaying.
Here, the factor $(\beta\gamma)_{N_2} \sin \theta/\Gamma_{N_2}$ describes the transverse distance that $N_2$ travels before decaying, where $\sin \theta$ picks the transverse component and $(\beta\gamma)_{N_2} = |\vec{p}_{N_2}|/m_{N_2}$ accounts for the Lorentz boost of $N_2$.
The three-momentum of $N_2$ in the laboratory frame $|\vec{p}_{N_2}|$ is given as a function of the angle $\theta$ in the laboratory frame in Eq.~\eqref{eq:pNj_costheta}.

\begin{figure}[t!]
\centering
\includegraphics[width=0.4\linewidth]{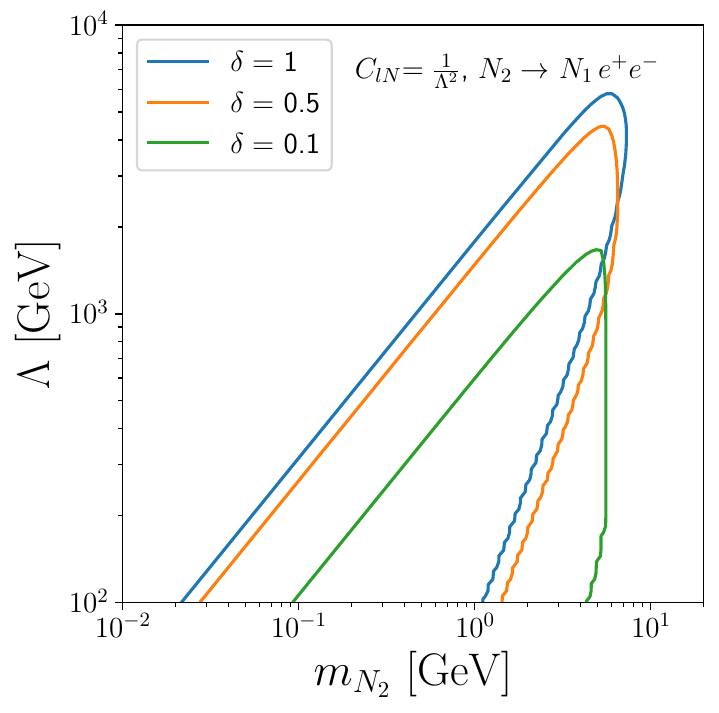}
\includegraphics[width=0.4\linewidth]{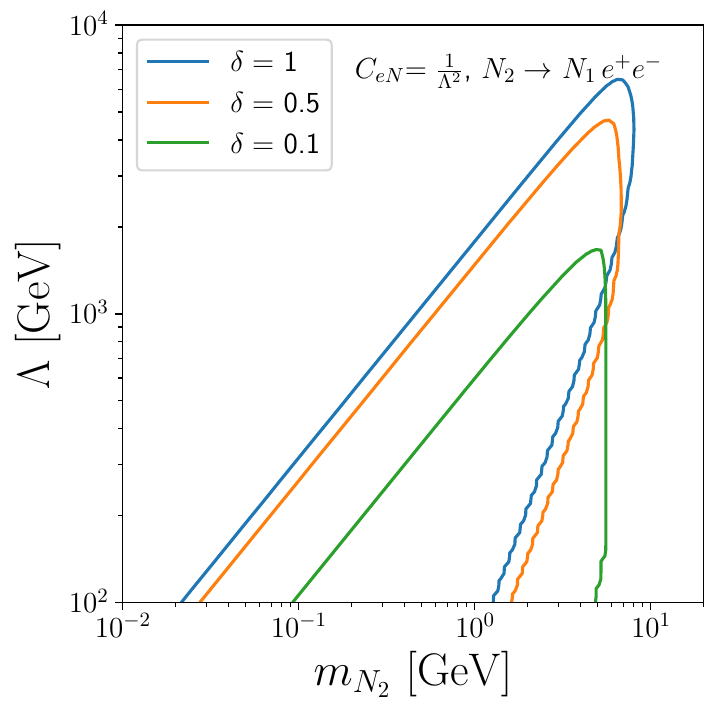}
\includegraphics[width=0.4\linewidth]{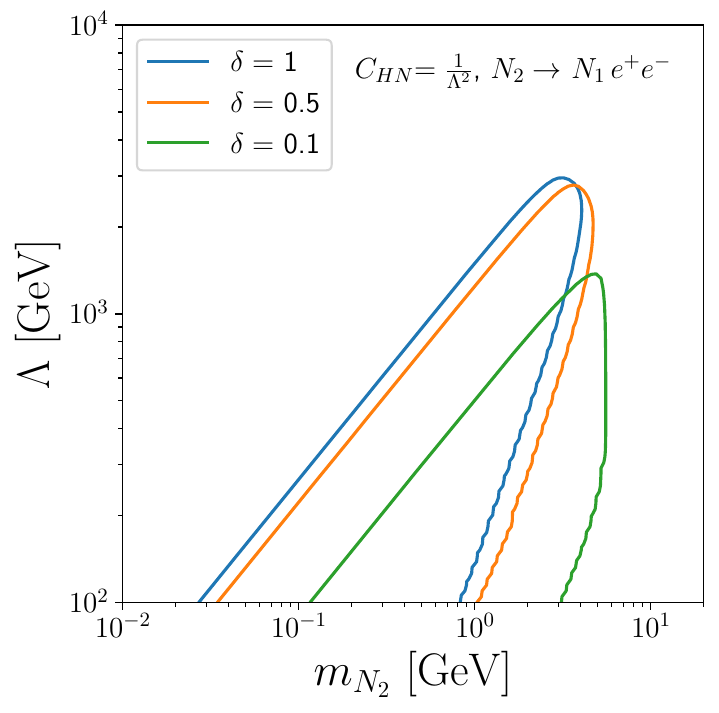}
\includegraphics[width=0.4\linewidth]{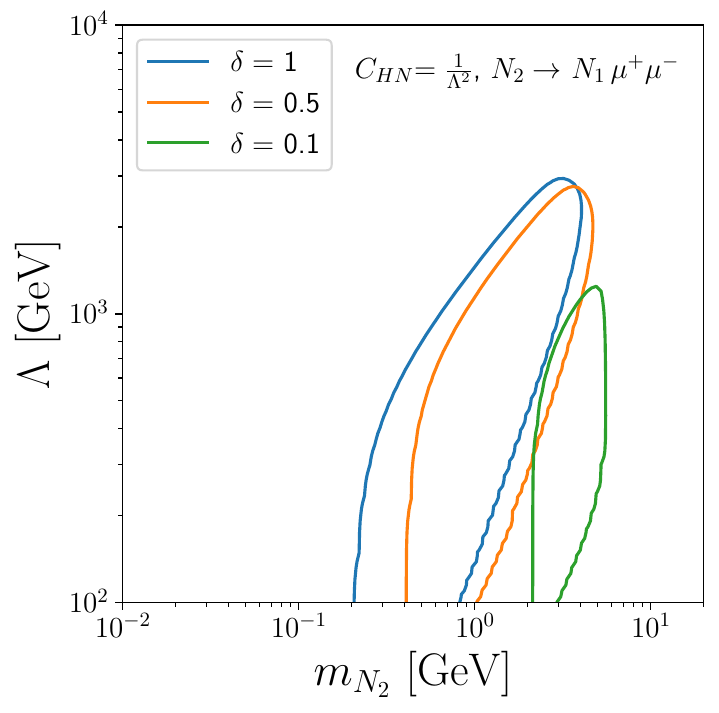}
\caption{Analytic sensitivity contours for zero background and 100\% efficiency in the $\Lambda$ vs. $m_{N_2}$ plane for different mass splittings $\delta$ and for operators that mediate the decays $N_2\to N_1 \ell^+ \ell^-$.
The operator $Q_{HN}$ also allows the decay $N_2 \to N_1  \tau^+\tau^-$, but sensitivity in this channel is lost due to the high mass threshold  required for production. }
\label{fig:DV_SensPlot_N1N2}
\end{figure}
\begin{figure}[t!]
\centering
\includegraphics[width=0.4\linewidth]{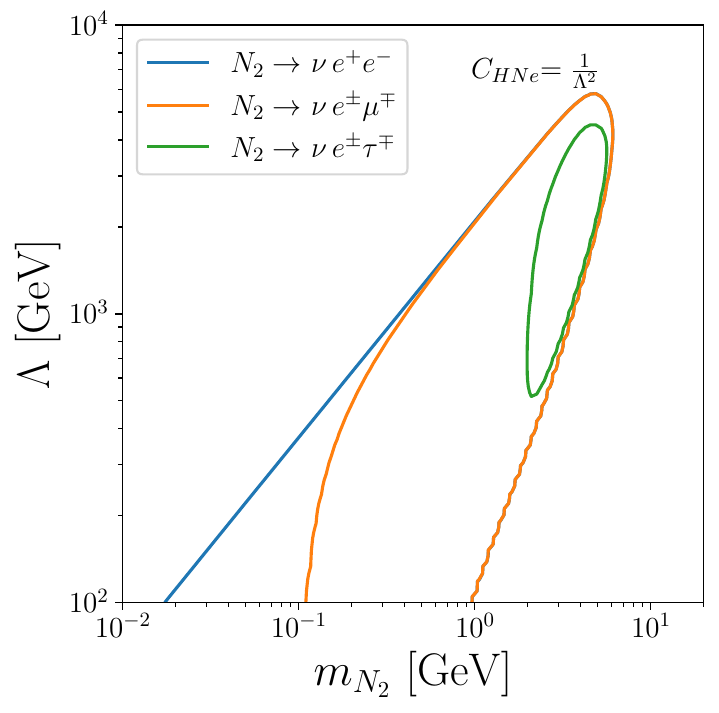}  
\includegraphics[width=0.4\linewidth]{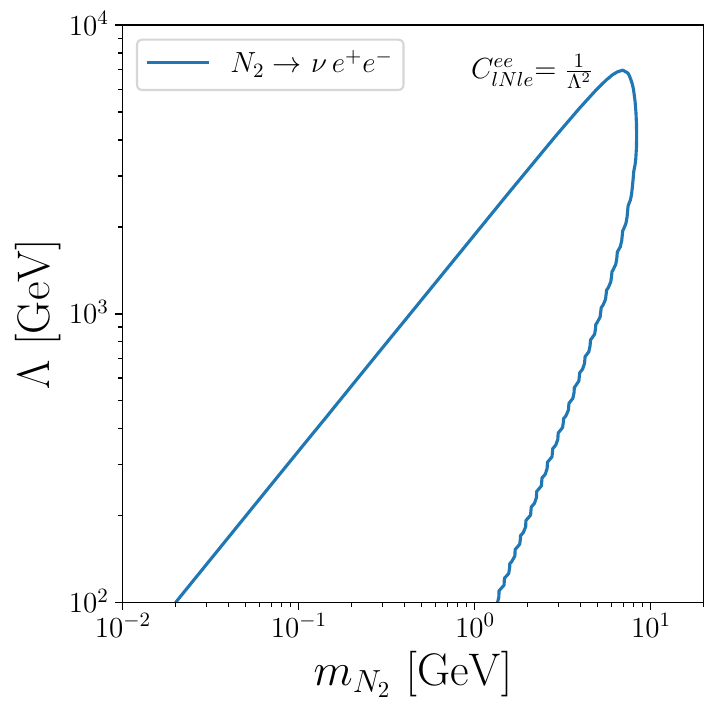}
\includegraphics[width=0.4\linewidth]{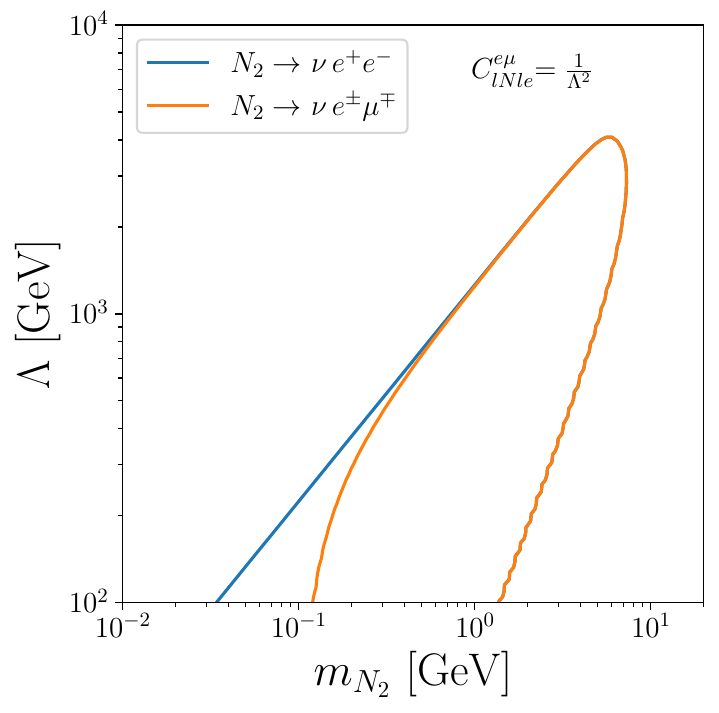}
\includegraphics[width=0.4\linewidth]{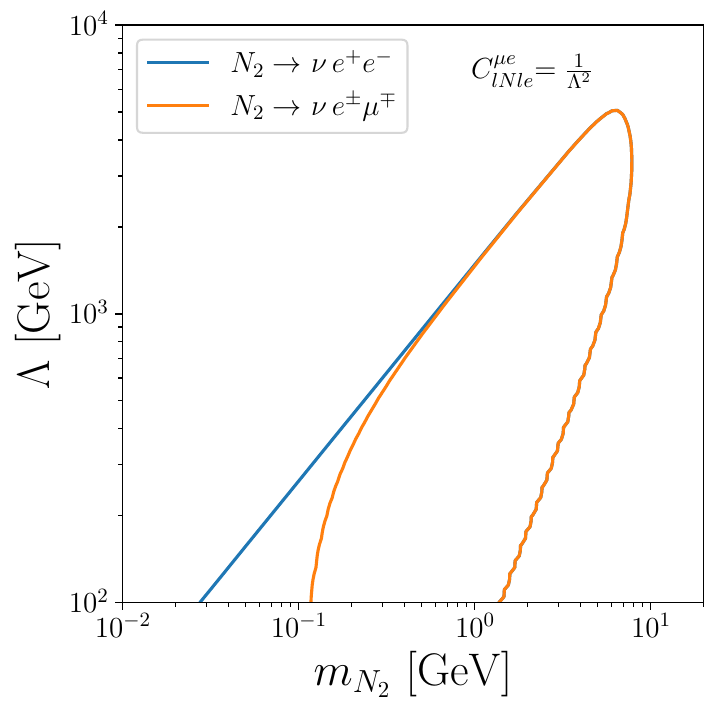}
\includegraphics[width=0.4\linewidth]{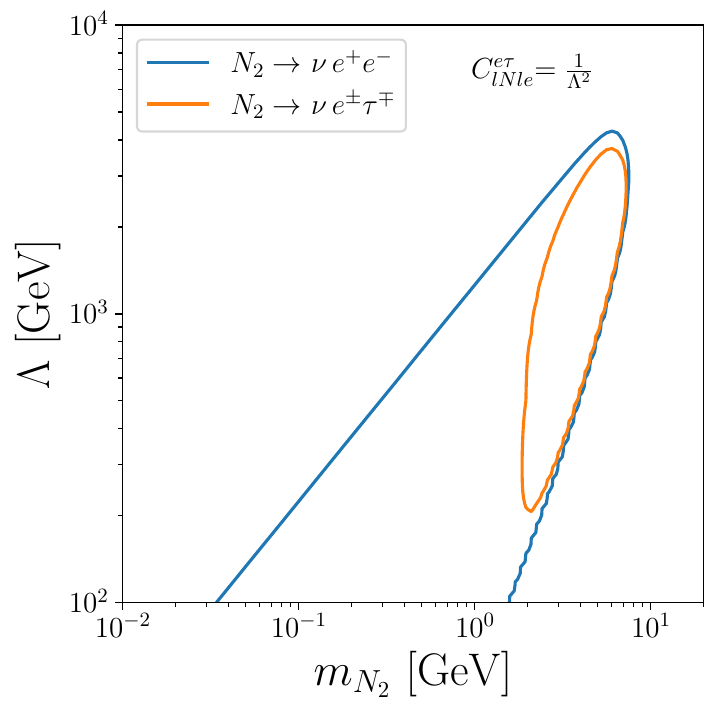} 
\includegraphics[width=0.4\linewidth]{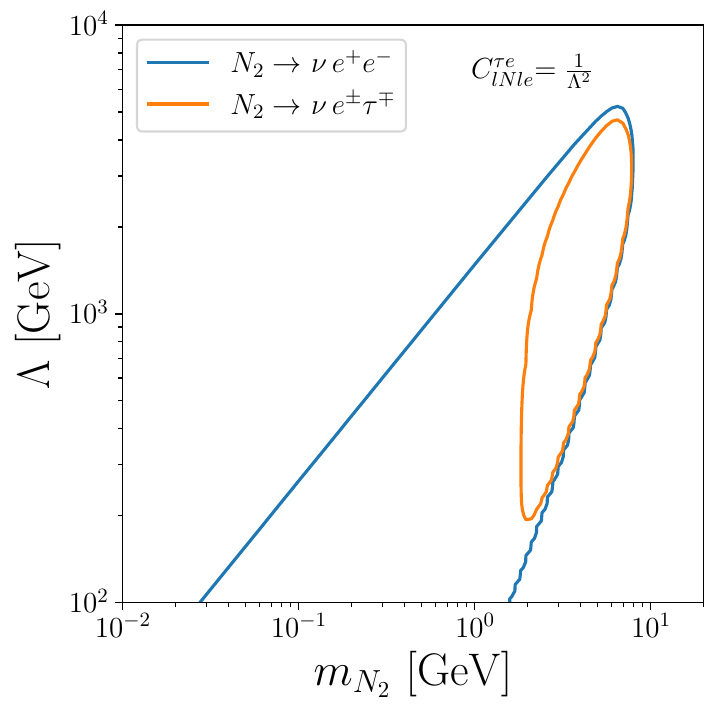}
\caption{Analytic sensitivity contours for zero background and 100\% efficiency in the $\Lambda$ vs. $m_{N_2}$ plane for operators that mediate the decay $N_2\to \nu \ell^+ \ell^-$.  }
\label{fig:DV_SensPlot_nuN2}
\end{figure}
\begin{figure}[t!]
\centering
\includegraphics[width=0.4\linewidth]{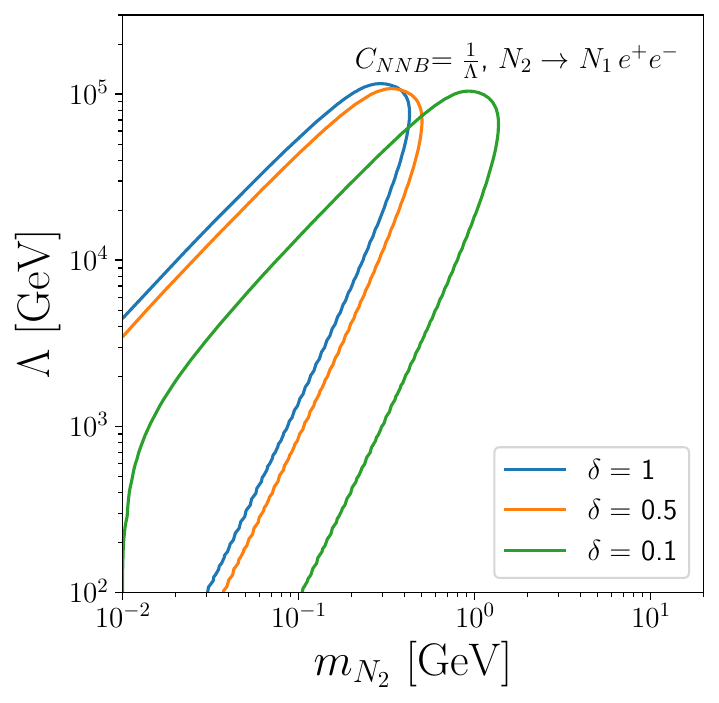}
\includegraphics[width=0.4\linewidth]{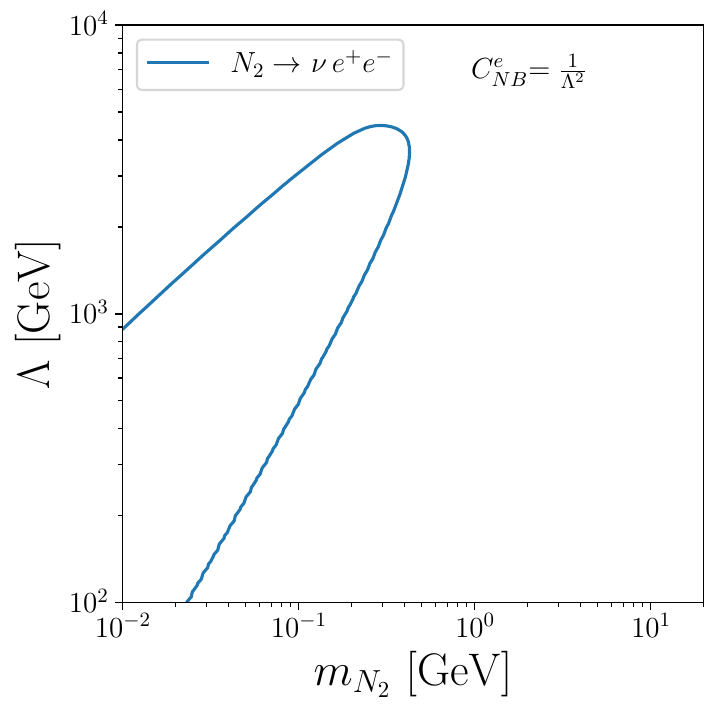}
\caption{Analytic sensitivity contours for zero background and 100\% efficiency in the $\Lambda$ vs. $m_{N_2}$ plane for the dipole operators $Q_{NNB}$ and $Q_{NB}^e$. }
\label{fig:DV_SensPlot_dipole}
\end{figure}

The total number $n$ of observed signal and background events follows a Poisson distribution
\begin{align}
    P(n|\mu) = \frac{\mu^n e^{-\mu}}{n!}\,,
\end{align}
where $\mu = S+B$, with $S$ and $B$ being the mean expected yields for signal and background, respectively.
For this estimate, we assume zero observed events, i.e. $n = 0$, and can calculate the $95 \%$ confidence level exclusion limit by demanding that the probability to observe zero events be $5\%$, i.e.
\begin{align}
    P(0|S+B) = e^{-(S+B)} = 0.05\,.
\end{align}
This implies
\begin{align}
    S+B = -\ln(0.05) \approx 2.996 \quad \Rightarrow S \approx 3 - B \,.
\end{align}
If we expect $B=0$ background events, this simplifies to $S=3$ events. The region in $(m_{N_2}, \delta, \Lambda)$ space enclosed by the $S=3$ sensitivity contour corresponds to the excludable parameter space.

The resulting sensitivity contours are shown in Fig.~\ref{fig:DV_SensPlot_N1N2} for the operators that couple to $N_1$ and $N_2$, and in Fig.~\ref{fig:DV_SensPlot_nuN2} for the operators that couple to $N_2$ and a SM neutrino (or antineutrino, due to the Majorana nature of the HNL).
In Fig.~\ref{fig:DV_SensPlot_nuN2} it is evident that for the operator $Q_{H N e}$, the sensitivity of the $e^\mp \tau^\pm $ final state is smaller than that of $e^+ e^-$ and $e^\pm \mu^\mp$, due to the limited phase space.
The phase space is even more limited for the $\tau^+\tau^-$ final state, which can be mediated by $Q_{HN}$, so it is not shown here.

Finally, in Fig.\ \ref{fig:DV_SensPlot_dipole} we show the sensitivity estimates for the two dipole operators $Q_{NNB}$ and $Q_{NB}^e$, which can mediate same-flavour $\ell^+ \ell^-$ final states via Dalitz-type decays, cf.\ the second diagram in Fig.~\ref{fig:decays}. The plots show only $e^+ e^-$ final states since for $\mu^+ \mu^-$ and $\tau^+ \tau^-$ the mass threshold lies above the range of masses $m_{N_2}$ that can be probed for these operators. Note that the curve for $C_{NW}^e$ is identical to the one for $C_{NB}^e$, since the operators have the same matching relation to the photon dipole coupling. Similarly, different flavours of these operators, $C_{NB}^{\rho}$ and $C_{NW}^{\rho}$ for $\rho = \mu, \tau$, result in exactly the same sensitivities, as they differ only by the flavour of the SM neutrinos. While the scale $\Lambda$ that can be probed for the $d = 5$ operator $Q_{NNB}$ is much higher than for the other operators, its sensitivity is limited to low masses, where the signal yield will be drastically reduced after background-suppression cuts detailed in Secs.~\ref{sec:DVbackground}. 

%%%%%%%%%%%%%%%%%%%%%%%%%%%%%%%%%%%%%%%%%%%%%%%%%%%%%%%%%
\subsection{Background prediction and mitigation strategies}
\label{sec:DVbackground}
%%%%%%%%%%%%%%%%%%%%%%%%%%%%%%%%%%%%%%%%%%%%%%%%%%%%%%%%%

To obtain realistic experimental sensitivities of Belle~II to the Wilson coefficients $C_i$ as a function of the $N_2$ mass and the mass difference $\delta$, we estimate the dominant background event yields for the relevant DV signatures and devise event-selection cuts to suppress them.
In Sec.~\ref{sec:DVbackground-cutbased} we describe the background sources and propose cuts that strongly suppress the background relative to the signal.
The resulting background yields and signal efficiencies enter the revised sensitivity estimates in Sec.~\ref{sec:DVsens}.

Beyond the cut-based method, we describe in Sec.~\ref{sec:constraint} the application of kinematic constraints that provide further background suppression and enable measurement of $m_{N_2}$ and $\delta$ if signal is detected. 
The full application of this method is left for experimental studies and not used in Sec.~\ref{sec:DVsens}.

%%%%%%%%%%%%%%%%%%%%%%%%%%%%%%%%%%%%%%%%%%%%%%%%%%%%%%%
\subsubsection{Cut-based background suppression}
\label{sec:DVbackground-cutbased}
%%%%%%%%%%%%%%%%%%%%%%%%%%%%%%%%%%%%%%%%%%%%%%%%%%%%%%%

The signal signature is a displaced vertex (DV) formed from two oppositely-charged leptons ($e^+e^-$, $\mu^+\mu^-$ or $e^\pm \mu^\mp$) in events with little or no additional detector activity. 
We require that the DV radial position, $R$, be in the range $1.2~\text{cm} < R < 60~\text{cm}$, as motivated in Sec.~\ref{sec:analytics}.
DVs located in dense material regions, such as silicon detector layers, may be vetoed to suppress material-interaction background.
However, we take a simplified and conservative approach by including no such veto and considering all detector material in our estimates.
For consistency with the signal signature, it is further required that the event contain no high-quality tracks beyond those that emanate from the DV, and only limited neutral-particle activity in the electromagnetic calorimeter (ECL) and in the $K^0_L$-and-muon detector (KLM).
As an example, in Ref.~\cite{Belle:2024wyk}, events were rejected if they contained additional tracks, a high-quality $\pi^0\to \gamma\gamma$ candidate, or if the total laboratory frame energy of photons, each with energy greater than 0.2~GeV, exceeded 1~GeV.
The last requirement reflects the fact that some neutral activity must be accepted, due to the presence of beam background and the long integration time of the ECL.
In addition, the cut
\begin{equation}
|\vec p_\mu|> 0.7~\text{GeV}\,
\label{eq:pMuCut}
\end{equation}
on the momentum of muon candidates ensures that they can be identified with high confidence in the KLM~\cite{Abashian:2002bd}.
Additional cuts are motivated below in the following discussion on background sources.

\begin{figure}
    \centering
    \raisebox{2mm}{\includegraphics[width=0.2\linewidth]{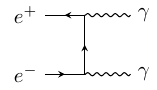}}
    \includegraphics[width=0.2\linewidth]{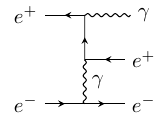} 
    \raisebox{2.2mm}{\includegraphics[width=0.23\linewidth]{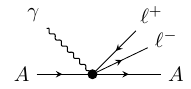}}\\\vspace{0.3em}
    \includegraphics[width=0.25\linewidth]{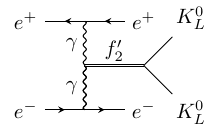}
    \raisebox{-2.7mm}{\includegraphics[width=0.25\linewidth]{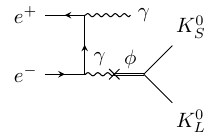}}
    \includegraphics[width=0.23\linewidth]{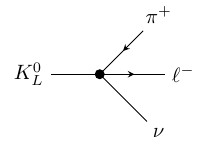}
\caption{(Above) Representative diagrams for the dominant sources of photons at the IP ($e^+e^-\to \gamma\gamma$ and $e^+e^- \to e^+e^-\gamma$) which give rise to background DVs when they undergo conversions to charged lepton pairs $\ell^+\ell^-$ on nuclei $A$ in the Belle~II detector (top right). (Below) Representative diagrams for the dominant sources of $K_L^0$ mesons which undergo displaced decays via $K_L^0\to \pi^\pm \ell^\mp \nu$. The cross in the centre diagram denotes the production of a $\phi$ meson via an off-shell photon.}
\label{fig:background_diagrams}
\end{figure}

We first consider the background from photon conversions on nuclei $A$ in the Belle~II detector, $\gamma A\to \ell^+\ell^- A$, shown in Fig.~\ref{fig:background_diagrams}, which constitute background for the $e^+e^-$ and $\mu^+\mu^-$ final states.
The expected number of photon conversion events are estimated in detail in App.~\ref{app:conversion} for the transverse displacement range $1.2~\text{cm}< R < 60~\text{cm}$, which includes materials in the Belle~II pixel vertex detector (PXD), silicon vertex detector (SVD), CDC inner wall and CDC bulk.
We find that the dominant sources of conversion photons from the IP are $e^+e^-\to \gamma\gamma$ and $e^+e^-\to e^+e^-\gamma$, shown in Fig.~\ref{fig:background_diagrams}.
For $e^+e^-\to \gamma\gamma$, we require that the second photon is not observed. 
To estimate the probability that the photon escapes detection, we estimate from Fig.~3 of Ref.~\cite{Belle-II-photon-eff} that for photons with energy $E_\gamma > 0.5~\text{GeV}$, the detection efficiency in the angular range $50^\circ < \theta_\gamma < 110^\circ$ of the barrel ECL  is at least 99\%.
The second photon is guaranteed to be in this region if the polar angle $\theta_{\rm DV}$ of the DV (with respect to the IP) induced by the conversion process satisfies $43.6^\circ < \theta_{\text{DV}} < 101.5^\circ$.
We thus apply this cut, and assume a 1\% probability of the other photon going undetected, leading to 100-fold suppression of this background.
While the endcap ECL regions can also be used, we exclude it due to lack of published data about its photon detection efficiency.
For $e^+e^-\to e^+e^-\gamma$, we require the outgoing $e^+e^-$ to not pass through the ECL, i.e. $\theta_{e^\pm} < 17^\circ$ and $\theta_{e^\pm} > 150^\circ$.
In App.~\ref{app:conversion}, we further show that the photon conversion backgrounds can be suppressed by applying the kinematic cuts
\begin{align}
&|\vec{p}_{\ell\ell}| > 0.7~\text{GeV}  \,, \nonumber \\
&\alpha_{\ell\ell} > \begin{cases}
12.5^\circ~(10^\circ) &~\ell\ell = ee \\
3^\circ~(2^\circ) &~\ell\ell = \mu\mu,e\mu
\end{cases} \,,
\label{eq:conversion-cuts}
\end{align}
where $|\vec{p}_{\ell\ell}|$ is the laboratory frame pair momentum of the DV leptons and $\alpha_{\ell\ell}$ is the pointing angle between the line from the IP to the DV and the momentum of the DV tracks, i.e.
\begin{align}
\label{eq:pointing_angle}
\cos\alpha_{\ell\ell} \equiv \hat{r}_{\text{DV}}\cdot\hat{p}_{\ell\ell}  =  \frac{\vec{r}_{\text{DV}}\cdot\vec{p}_{\ell\ell}}{|\vec{r}_{\text{DV}}||\vec{p}_{\ell\ell}|} \,.
\end{align}
For each $\alpha_{\ell\ell}$ cut in Eq.~\eqref{eq:conversion-cuts}, the number outside (inside) the parentheses corresponds to the cut for conversions in the range $1.2~\text{cm} < R < 17~\text{cm}$ ($17~\text{cm} < R < 60~\text{cm}$), corresponding to the region before (inside) the CDC bulk. With these cuts, the number of photon-conversion events is of the same order of magnitude in the two $R$ regions.

While the tight $\alpha_{\ell \ell}$ cuts in Eq.~\eqref{eq:conversion-cuts} reduce the total number of background events to $\mathcal{O}(100)$, as we show below, they also tightly affect the signal, as discussed in Sec.~\ref{sec:DVsens}. Therefore, we consider also the loose $\alpha_{\ell\ell}$ cuts
\begin{align} 
&\alpha_{\ell\ell} > \begin{cases}
10^\circ~(8^\circ) &~\ell\ell = ee \\
0.5^\circ~(0^\circ) &~\ell\ell = \mu\mu,e\mu
\end{cases} \,.
\label{eq:loose-cuts}
\end{align}
A priori, it is not obvious if the tight or the loose cuts will be able to probe a larger parameter space, and we compare the final sensitivity estimates for both sets of cuts in Sec.~\ref{sec:DVsens}. 
The tight and loose sets of cuts are summarised in Table~\ref{tab:DV_cuts}.
\begin{table}[t!]
\centering
\begin{tabular}{c|c}
\hline 
Tight cuts & Loose cuts \\
\hline
\multicolumn{2}{c}{$ 1.2~\text{cm} < R < 60~\text{cm}$} \\
\multicolumn{2}{c}{$43.6^\circ < \theta_{\text{DV}} < 101.5^\circ $} \\ 
\multicolumn{2}{c}{$|\vec{p}_{\ell\ell}| > 0.7~\text{GeV} $} \\
\multicolumn{2}{c}{$|\vec p_\mu| >0.7~\textrm{GeV}$} \\
\hline
 $\alpha_{\ell\ell} > \begin{cases}
12.5 ^\circ~(10 ^\circ) & ~\ell\ell = ee \\ 
3^\circ~(2^\circ) & ~\ell\ell = \mu \mu, e \mu  \\
\end{cases}$ 
& $\alpha_{\ell\ell} > \begin{cases}
10 ^\circ~(8 ^\circ) & ~\ell\ell = ee \\ 
0.5^\circ~(0^\circ) & ~\ell\ell = \mu \mu, e \mu  \\
\end{cases}$ \\
\hline
\end{tabular}
\caption{Summary of the cuts on the radial position $R$ of the DV, its polar angle $\theta_{\textrm{DV}}$, the dilepton momentum and muon momentum. 
The cuts on the pointing angle $\alpha_{\ell\ell}$ come in two sets--tight and loose--with the values outside (inside) the parentheses referring to the region $1.2~\text{cm} < R < 17~\text{cm}$ ($17~\text{cm} < R < 60~\text{cm}$).}
\label{tab:DV_cuts}
\end{table}

In addition to photon conversions, a significant source of background involves decays of long-lived $K_L^0$ mesons which decay via $K_L^0 \to \pi^\pm \ell^\mp \nu$.
If the charged pion is misidentified as a charged lepton, the decay contributes as background for the $e^+e^-$, $\mu^+\mu^-$ or $e^\pm\mu^\mp$ final state.
For the search signature, the dominant production mechanisms of $K_L^0$ mesons are shown in the lower row of Fig.~\ref{fig:background_diagrams}.
The first mechanism is the $\gamma\gamma$-fusion process $e^+e^- \to e^+e^- K_L^0 K_L^0$, with the final-state $e^+e^-$ scattered at low angles, out of the instrumented region of the detector.
The second is the process $e^+e^- \to  \gamma K_S^0 K_L^0$, which we expect to be dominated by $\phi \to K_S^0 K_L^0$.
The estimation of these background yields is described in detail in App.~\ref{app:KL-background}.

In Table~\ref{tab:background_reductions} we show the resulting expected background yields, separated by source and signal-search channel, for the full Belle~II integrated luminosity of $50~\textrm{ab}^{-1}$.
The results are shown in three column groups as cuts are progressively applied.
The left column group shows the yields for only the cuts on $R$ and $\theta_{\text{DV}}$ (see Table~\ref{tab:DV_cuts}).
In the middle column group, the cuts on $|\vec p_{\ell\ell}|$ and $|\vec p_{\mu}|$ are also applied, leading to modest background reduction. 
However, as shown in the rightmost column group, the background yields are strongly suppressed after adding the cuts on the pointing angle $\alpha_{\ell\ell}$, for which we consider two options: tight and loose.

\begin{table}
    \centering
    \setlength{\tabcolsep}{4pt}
    \begin{tabular}{c||c|c|c||c|c|c||c|c|c}
    \hline
    \multirow{3}{*}{Background source} & \multicolumn{3}{c||}{$1.2~\text{cm} < R < 60~\text{cm}$} & \multicolumn{3}{c||}{$|\vec{p}_{\ell\ell}| > 0.7~\text{GeV}$} & \multicolumn{3}{c}{Tight $\alpha_{\ell\ell}$ cuts}\\

    & \multicolumn{3}{c||}{$43.6^\circ < \theta_{\text{DV}} < 101.5^\circ$} & \multicolumn{3}{c||}{$|\vec{p}_{\mu}| > 0.7~\text{GeV}$} & \multicolumn{3}{c}{[Loose $\alpha_{\ell\ell}$ cuts]}\\
    \cline{2-10}

    & $e^+e^-$ & $\mu^+\mu^-$ & $e^\pm\mu^\mp$ & $e^+e^-$ & $\mu^+\mu^-$ & $e^\pm\mu^\mp$ & $e^+e^-$ & $\mu^+\mu^-$ & $e^\pm\mu^\mp$ \\ \hline\hline

    $e^+e^- \to \gamma\gamma$ & \multirow{2}{*}{$3.5 \times 10^{7}$} &  \multirow{2}{*}{$130$} & \multirow{2}{*}{--} & \multirow{2}{*}{$3.5 \times 10^{7}$} & \multirow{2}{*}{$130$} & \multirow{2}{*}{--} & $< 1$ & $< 1$ & \multirow{2}{*}{--}\\
    $\gamma A \to \ell^+\ell^- A$ & & & & & & & $[< 1]$ & $[63]$ & \\ \hline

    $e^+e^- \to e^+e^-\gamma$ & \multirow{2}{*}{$4.5 \times 10^{9}$} & \multirow{2}{*}{$930$} & \multirow{2}{*}{--} & \multirow{2}{*}{$9.6\times10^{8}$} & \multirow{2}{*}{$470$} & \multirow{2}{*}{--} & $110$ & $95$ & \multirow{2}{*}{--}\\
    $\gamma A \to \ell^+\ell^- A$ & & & & & & & $[480]$ & $[450]$ &\\ \hline

    $e^+e^- \to e^+e^- K_L^0 K_L^0$ & \multirow{2}{*}{$45$} & \multirow{2}{*}{$420$} & \multirow{2}{*}{$660$} & \multirow{2}{*}{$8.2$} & \multirow{2}{*}{$1.1$} & \multirow{2}{*}{$47$} & $< 1$ & $< 1$ & $38$\\
    $K_L^0\to \pi^\pm \ell^\mp \nu$ & & & & & & & $[1.6]$ & $[1.1]$ & $[47]$ \\ \hline

    \multirow{2}{*}{Total} & \multirow{2}{*}{$4.5 \times 10^{9}$} & \multirow{2}{*}{$1500$} & \multirow{2}{*}{$660$} & \multirow{2}{*}{$1.0\times10^{9}$} & \multirow{2}{*}{$600$} & \multirow{2}{*}{$47$} & $110$ & $95$ & $38$ \\ 
    
     &  &  &  &  &  &  & $[480]$ & $[510]$ & $[47]$ \\\hline    
    \end{tabular}
\caption{The total number of events from each background source and final state ($e^+ e^-$, $\mu^+ \mu^-$, $e^\pm \mu^\mp$) for a luminosity of $50~\text{ab}^{-1}$ after successive kinematic cuts, from left to right. The tight and loose $\alpha_{\ell\ell}$ cuts referred to in the rightmost column are given in Table~\ref{tab:DV_cuts}.}
\label{tab:background_reductions}
\end{table}

In Table~\ref{tab:backgrounds},  the final background yields shown in  Table~\ref{tab:background_reductions} are separated into two dilepton-mass ranges, below and above $m_{\ell\ell}=0.5$~GeV. 
These are defined as our low-mass and high-mass search regions.
As can be seen from the dependence of the photon conversion background on $m_{\ell\ell}$ in Fig.~\ref{fig:conversion_bkg}, these events have low invariant masses after the cuts. We find that for $50~\textrm{ab}^{-1}$, the $e^+e^-$ background yield in the region $m_{\ell\ell} > 0.5$~GeV is less than 1 event. 
For the $\mu^+\mu^-$ background, which has looser cuts on $\alpha_{\ell\ell}$, 5\% of the events are in this region.
The $K_L^0$ background events all satisfy $m_{\ell\ell} < 0.5$~GeV due to the $K_L^0$ mass, and mostly appear at smaller masses due to the energy taken up by the neutrino and the difference particularly between the electron and pion masses.

We now briefly discuss other possible backgrounds, which can be lowered to negligible levels with appropriate cuts, and therefore are not shown in Table~\ref{tab:background_reductions}:
\begin{itemize}
\item Background DVs from the decay $K_S^0\to \pi^+\pi^-$ are strongly suppressed by the $\alpha_{\ell\ell}$ cuts and by requiring that the invariant mass $m_{\ell\ell}$ be sufficiently far from the known $K_S^0$ mass.
As an example, in Ref.~\cite{Belle:2024wyk}, $K_S^0$ suppression was achieved without an $\alpha_{\ell\ell}$ cut and only a broad cut on $m_{\ell\ell}$.

Further suppression arises from  the probability that both pions fake leptons, whose largest value is $10^{-3}$, in the case of the $\mu^+\mu^-$ channel.
The $\alpha_{\ell\ell}$ cut and muon identification may be satisfied if the pions decay in flight.
In this case, the decay must occur quickly in order for the muons to form a vertex. 

\item Initial-state-radiation and $\gamma\gamma$-fusion events can also give rise to neutron-antineutron final states, although with smaller cross sections than those of the corresponding $K_L^0$ production processes discussed above.
Soft antinucleons have a very large cross section for annihilation with detector nucleons~\cite{Cline:1971bh}.
The annihilation may produce two tracks and several neutrons, creating a DV with a large pointing angle.
However, dilepton production in this process is very rare.
Therefore, we expect this background to be no larger than the $K_L^0$ background.

\item An additional source of background is a prompt track that undergoes hard scattering due to material interactions, appearing as two tracks with a large opening angle.
Pion and kaon decays in flight produce a similar signature, which may be mistaken for a dilepton DV if the hadron track is misidentified as a lepton.
Both backgrounds can be identified by a kink-finding algorithm. 
Based on Ref.~\cite{Belle:2024wyk}, we expect this to be a very minor background.

\item A background DV can be caused by a cosmic-ray muon reconstructed as two back-to-back tracks.
This background can be very effectively removed by rejecting track pairs that are back-to-back in the laboratory frame and have similar momenta, given the good angular and momentum resolutions discussed in Appendix~\ref{app:alpha-resolution}.
\end{itemize}
\begin{table}
    \centering
    \begin{tabular}{c|c|c}
    \hline
    \multirow{2}{*}{Final state} & \multicolumn{2}{c}{Total background after cuts} \\\cline{2-3}
     & $m_{\ell\ell} < 0.5$~GeV & $m_{\ell\ell} > 0.5$~GeV \\
    \hline
    $e^+ e^-$ & $110$ $[480]$ & $< 1$ $[< 1]$ \\
    $\mu^+ \mu^-$ & $90$ $[490]$ & $5$ $[20]$ \\
    $e^{\pm} \mu^{\mp}$ & $38$ $[47]$ & $< 1$ $[< 1]$  \\
    \hline
    \end{tabular}
\caption{Expected numbers of background events in $50~\textrm{ab}^{-1}$ of Belle~II data after cuts for the three considered final states. Estimates for \textit{tight} and \textit{loose} $\alpha_{\ell\ell}$ cuts are shown without and with square parentheses, respectively.}
\label{tab:backgrounds}
\end{table}
%

%%%%%%%%%%%%%%%%%%%%%%%%%%%%%%%%%%%%%%%%%%%%%%%%
\subsubsection{Additional background suppression and measurement of $m_{N_2}$ and $\delta$}
\label{sec:constraint}
%%%%%%%%%%%%%%%%%%%%%%%%%%%%%%%%%%%%%%%%%%%%%%%%

Beyond the cut-based background reduction discussed above, we show here how to use kinematics to further distinguish signal from background and also obtain the values of $m_{N_2}$ and $\delta$ if signal is observed.

As detailed in App.~\ref{app:kinematics}, for the HNL production $e^+e^-\to N_1 N_2/\nu N_2$ and decay $N_2 \to N_1 \ell^+\ell^-/\nu \ell^+\ell^-$, there are enough (specifically, 12) constraints from momentum conservation, the reconstructed $N_2$ decay vertex position, the known beam energies, and $N_2$ being on-shell to find two solutions for $m_{N_2}^\pm$, purely from the known initial-state beam energies $E_{e^+}$ and $E_{e^-}$ and the final-state kinematic quantities $\theta_{\text{DV}}$, $E_{\ell\ell}$, $|\vec{p}_{\ell\ell}|$ and $\alpha_{\ell\ell}$.
However, there are not enough constraints to resolve an ambiguity in $\delta$, which remains unknown.
Thus, one must calculate $m_{N_2}^+$ and $m_{N_2}^-$ with an assumed value of $\delta$.

To study the signal and background distributions in the $(m_{N_2}^+, m_{N_2}^-)$ plane, we simulate signal and background samples as described above.
In the signal simulation, we simulate the process $e^+e^- \to N_1 N_2$ followed by $N_2 \to N_1 e^+e^-$ in \madgraph~\cite{Alwall:2014hca}, using a \texttt{UFO}~\cite{Degrande:2011ua} model file generated in \texttt{FeynRules}~\cite{Alloul:2013bka}, as described in App.~\ref{app:feynrules}. 
A mass $m_{N_2}=2$~GeV and the mass splitting $\delta = 1, \, 0.5, \, 0.1$ were chosen for this study. 
To account for the impact of the beam-particle energy spread on the calculation, we sample the beam energies from Gaussian distributions with the central values $E_{e^+} = 4.000$~GeV and $E_{e^-} = 7.007$~GeV and spread $\sigma_{E_{e^\pm}}/E_{e^\pm} = 8 \times 10^{-4}$~\cite{Belle-II:2010dht}.
We then retain only the events that satisfy the loose cuts.
To ensure the existence of real and positive solutions, we require $S\ge 0$ and $T\ge 0$, where $S$ and $T$ are defined in Eqs.~(\ref{eq:S}) and~(\ref{eq:T}), respectively.
We then calculate $m_{N_2}^+$, $m_{N_2}^-$ as in App.~\ref{app:kinematics} with assumed $\delta$ values of 1, 0.5, and 0.1.

\begin{figure}[t]
\centering
\includegraphics[width=0.328\linewidth]{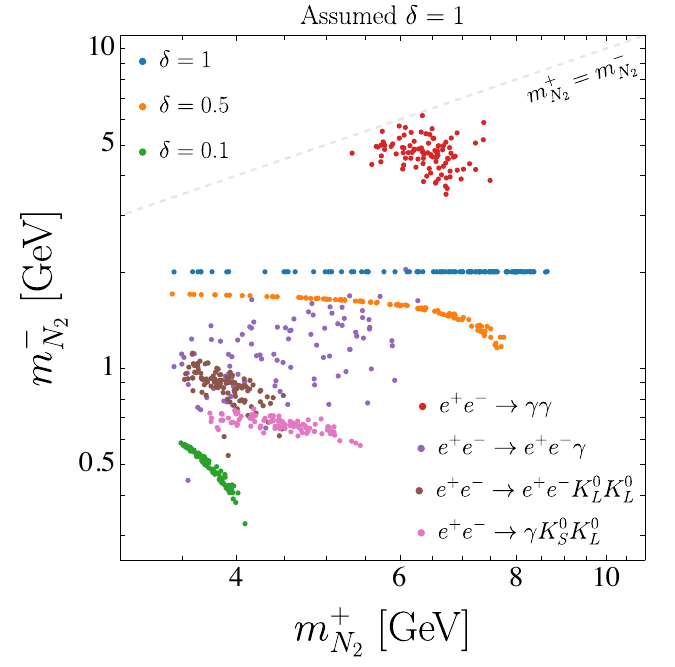}
\includegraphics[width=0.328\linewidth]{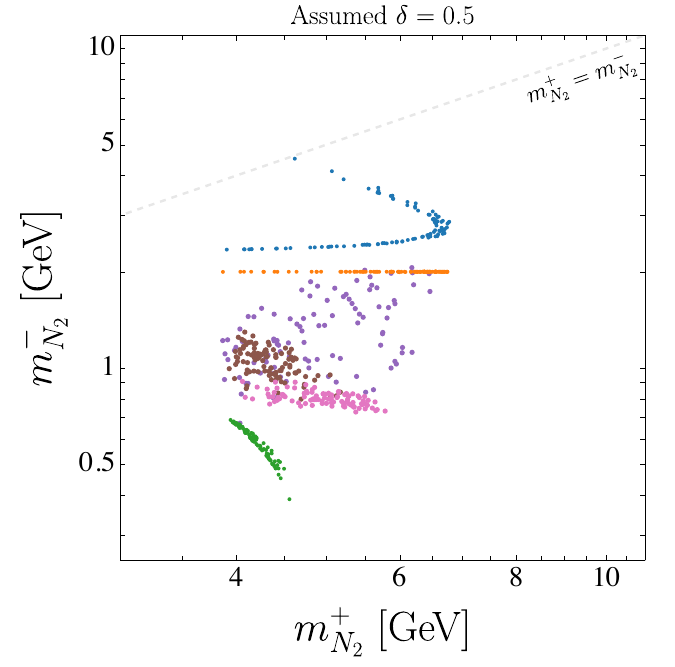}
\includegraphics[width=0.328\linewidth]{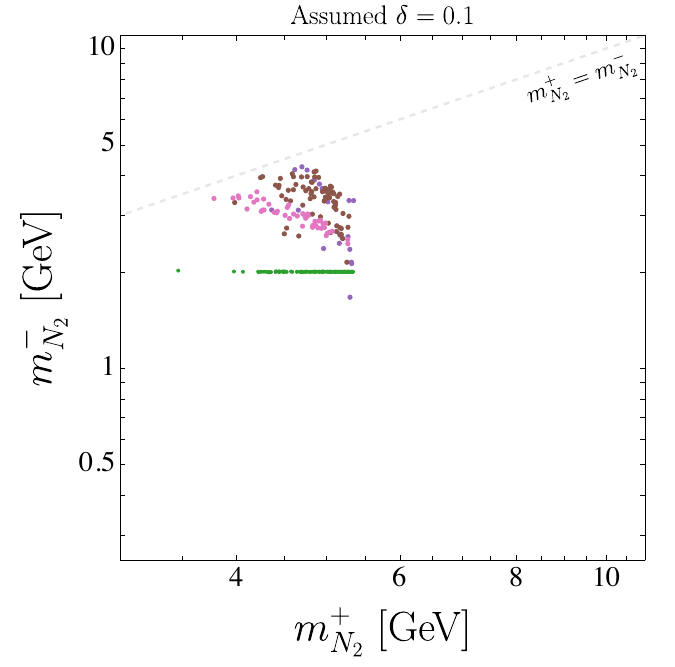}
\includegraphics[width=0.328\linewidth]{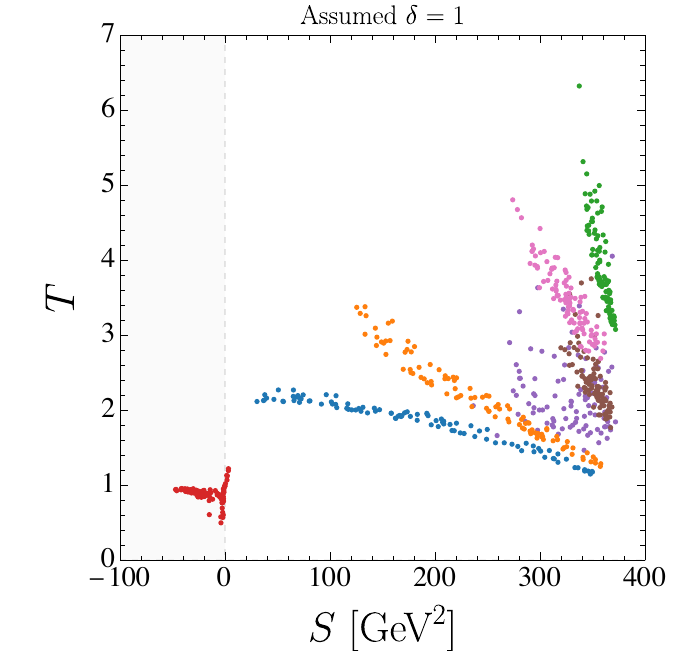}
\includegraphics[width=0.328\linewidth]{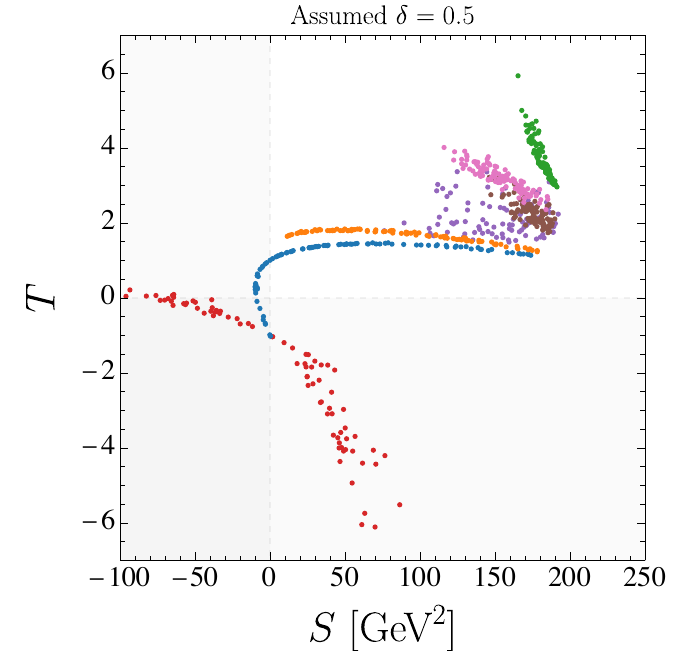}
\includegraphics[width=0.328\linewidth]{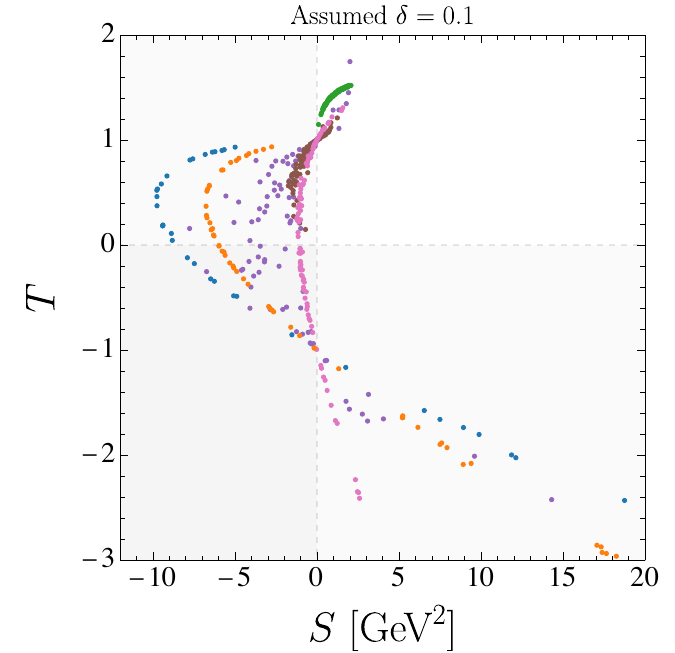}
\caption{(Above) Distributions of the kinematic-constraint solutions $m^-_{N_2}$ and $m^+_{N_2}$ for the $N_2$ mass, calculated as in App.~\ref{app:kinematics}.
Results are shown for simulated signal events with true $N_2$ mass $m_{N_2} = 2$~GeV and mass splittings $\delta = 1$ (blue), $0.5$ (orange), and $0.1$ (green), as well as for the four simulated background samples: photon conversions from $e^+e^-\to \gamma\gamma$ (red) and $e^+e^- \to e^+e^-\gamma$ (purple), and $K_L^0\to\pi^\pm\ell^\mp\nu$ decays from $e^+e^-\to e^+e^-K_L^0K_L^0$ (brown) and $e^+e^-\to\gamma K_S^0 K_L^0$ (pink). 
The solutions are found with the assumed value of $\delta$ being 1 (left), 0.5 (centre), and 0.1 (right). 
All events satisfy $S > 0$ and $T > 0$ (see App.~\ref{app:kinematics}), which is necessary to yield real and positive $m_{N_2}$ solutions. (Below) the corresponding $T$ vs. $S$ distributions for all  simulated signal and background events.}
\label{fig:kinematics}
\end{figure}

The resulting signal and background distributions in the $(m_{N_2}^+, m_{N_2}^-)$ plane are shown in the upper part of Fig.~\ref{fig:kinematics}, where the left, middle, and right plots show the results for assumed $\delta$ values of 1, 0.5, and 0.1, respectively. 
First considering the signal distributions, one observes from these plots that, when the assumed value of $\delta$ equals the true one, the $m^-_{N_2}$ distribution clusters around the true value $m_{N_2}$, while the  $m^+_{N_2}$ distribution is much broader.
When the assumed value of $\delta$ is incorrect, the distribution in the $(m_{N_2}^+, m_{N_2}^-)$ plane becomes curved.
Thus, one can calculate $m_{N_2}^+$ and $m_{N_2}^-$ for different values of assumed $\delta$ and identify the correct one as that for which the distribution in the $(m_{N_2}^+, m_{N_2}^-)$ is a straight line.
For that value of $\delta$, one obtains the $N_2$ mass $m_{N_2}$ from the value of $m_{N_2}^-$ at which the events cluster. 

Directing our attention to the background distributions, we see that they are generally different from those of the signal and do not sharply cluster at a given $m_{N_2}^2$ value.
The overlap between the signal and background depends on the true values of $(m_{N_2}$ and $\delta)$.
Nonetheless, we conclude that at least for large part of the $(m_{N_2}, \, \delta)$ parameter space, the background can be further reduced relative to that in Table~\ref{tab:backgrounds}.
Due to the complexity of carrying out this procedure for the full parameter space and the need to perform it with full detector simulation, we still take Table~\ref{tab:backgrounds} as the benchmark for our sensitivity estimate in Sec.~\ref{sec:DVsens}. 

To complete this study, we show in the lower part of Fig.~\ref{fig:kinematics} the distributions of the signal and background samples in the $(S, \, T)$ plane, with the three plots again corresponding to assumed $\delta=1$, 0.5, 0.1. 
Furthermore, for each signal or background sample and each value of assumed $\delta$, Table~\ref{tab:solution_fractions} gives the fraction of events retained by the $S\geq0$, $T\geq0$.

\begin{table}
    \centering
    \begin{tabular}{c|c|c|c|c|c|c|c}
    \hline
    \multirow{4}{*}{Assumed $\delta$} & \multicolumn{7}{c}{Fraction of events yielding physical $m_{N_2}$ solutions} \\\cline{2-8}
     & \multicolumn{3}{c|}{Benchmark signals} & \multicolumn{4}{c}{Backgrounds} \\
      & \multicolumn{3}{c|}{$m_{N_2} = 2$~GeV} & \multicolumn{4}{c}{$e^+e^-\to X$} \\ \cline{2-8}
     & $\delta = 1$ & $\delta = 0.5$ & $\delta = 0.1$ & $\gamma\gamma$ & $e^+e^-\gamma$ & $e^+e^-K_L^0K_L^0$ & $\gamma K_S^0K_L^0$ \\
    \hline
    $1$ & $1$ & $1$ & $1$ & $0.05$ & $1$ & $1$ & $1$ \\
    $0.5$ & $0.72$ & $1$ & $1$ & $0$ & $1$ & $1$ & $1$ \\
    $0.1$ & $0$ & $0$ & $1$ & $0$ & $0.11$ & $0.24$ & $0.08$ \\
    \hline
    \end{tabular}
\caption{For the three benchmark signal scenarios and four considered backgrounds, the fraction of events which yield physical solutions for the $N_2$ mass (therefore satisfying $S \geq 0$ and $T \geq 0$), for three different assumed values of $\delta$.}
\label{tab:solution_fractions}
\end{table}
%

%%%%%%%%%%%%%%%%%%%%%%%%%%%%%%%%%%%%%%%%%%%%%%%%%%%%%%%%%
\subsection{Realistic sensitivity estimates}
\label{sec:DVsens}
%%%%%%%%%%%%%%%%%%%%%%%%%%%%%%%%%%%%%%%%%%%%%%%%%%%%%%%%%

This subsection presents our main result: the sensitivity of our proposed search taking into account the expected background and the signal-reconstruction efficiency.
Determining the efficiency requires correctly accounting for its dependence on $m_{N_2}$ and $\delta$, but also on the $N_2$ lifetime, which in turn depends on the scale $\Lambda$.

To generate signal events, we have implemented the effective interactions and the matching of the $\nu$SMEFT to the $\nu$LEFT operators into \texttt{FeynRules} and created a \texttt{UFO} file, as detailed in App.~\ref{app:feynrules}. 
Utilising \madgraph, we generate signal events while switching on one operator at a time. 
Events are generated on a grid of $m_{N_2}$ and $\delta$ values, using the beam energies $E_{e^+} = 4.000$~GeV and ${E_{e^-} = 7.007}$~GeV and with the $e^+e^-$ collision taking place at the origin.
The NP scale is set to $\Lambda_0 = 1$ TeV in the event generation. Afterwards, we also scan over $\Lambda$ in a range motivated by the analytic estimates in Sec.~\ref{sec:analytics} and multiply the resulting cross section for $\Lambda_0$ by $(\Lambda_0/\Lambda)^4$. 

At each grid point we generate $10^4$ signal events of either $e^+ e^- \rightarrow N_1 N_2, \, N_2 \rightarrow N_1 \ell^+ \ell^-$ for the operators $Q_{lN}$, $Q_{eN}$ and $Q_{HN}$, or ${e^+ e^- \rightarrow \nu N_2, \, N_2 \rightarrow \nu \ell^+ \ell^-}$ (where $\nu$ can be an antineutrino) for $Q_{HNe}$ and $Q_{lNle}^{ee}$, $Q_{lNle}^{e \mu}$, $Q_{lNle}^{\mu e}$, $Q_{lNle}^{e\tau}$ and $Q_{lNle}^{\tau e}$. 
The generated final states for each operator are those given in Table~\ref{tab:final_states_leptonic}, except that final states with $\tau$ leptons are not pursued.
This is because, as shown in Fig.~\ref{fig:DV_SensPlot_nuN2}, the analytically determined parameter-space region that is probed by final states with a $\tau$ does not extend beyond that probed by $e^+e^-$.
In addition, note that Fig.~\ref{fig:DV_SensPlot_nuN2} does not account for the leptonic branching fraction for the $\tau$, which would reduce signal statistics by a factor of three per $\tau$ lepton. 
One can make up for much of the lost branching fraction with the decays $\tau^-\to \pi^- (\pi^0) \nu$.
However, this introduces hadronic background, requiring a background estimate that is not justified given the arguments herein.

The generated events that pass the cuts summarised in Table~\ref{tab:DV_cuts} are categorised as \textit{low mass} if they have $m_{\ell\ell} <0.5$~GeV and \textit{high-mass} if $m_{\ell\ell} >0.5$~GeV.
Each event is passed through the enhanced version v.2.0.0~\cite{bertholet_soffer_2024} of the \texttt{TrackEff} package~\cite{Bertholet:2025lcr} to determine whether the two DV leptons would be successfully reconstructed in the real experiment.
We use the default setting of \texttt{TrackEff}, in which a track is considered reconstructed if it passes at least 20 layers in the CDC, given its production point and momentum vector. 
The fractions of events with both tracks reconstructed in the low- and high-mass regions are the corresponding efficiencies $\epsilon^{\rm low}$ and $\epsilon^{\rm low}$.
The total number of signal events in final state $\ell^+\ell^-$ and mass region $r$ is then calculated from
\begin{equation}
    S_{\ell^+\ell^-}^{r} = \mathcal{L} \cdot \sigma \cdot \text{B}_{\ell^+\ell^-} \cdot \epsilon_r \,,
\end{equation}
where $\mathcal{L}=50~\textrm{ab}^{-1}$ is the integrated luminosity, $\sigma$ denotes the total cross section, and $\text{B}_{\ell^+\ell^-}$ is the branching fraction of the $N_2$ decay to $\ell^+\ell^-$. Note that, compared to the analytic estimate in Eq.~\eqref{eq:NDV_analytical}, the integration over the polar angle is substituted here by selecting the simulated events in the correct angular range, and that the acceptance formerly expressed with exponential factors is now handled by the proper detector geometry through the efficiencies $\epsilon_r$. 

In each region with zero background events, i.e. in the high-mass region for $e^+e^-$ and $e^{\pm}\mu^{\mp}$  (see Table~\ref{tab:backgrounds}), we calculate the sensitivity contours as in Sec.~\ref{sec:analytics}: we assume the Poison distribution and take the 95\% confidence-level parameter-space region to be that for which three signal events would be observed.
In all other cases, i.e. all low-mass regions and the high-mass region for $\mu^+ \mu^-$, we expect a finite number of background events, such that observing 0 events is quite unlikely. Therefore, we determine the exclusion limits on the signal yield $S$ (dropping the indices $\ell^+\ell^-$ and $r$ for brevity) from a hypothesis test of the background-only hypothesis $H_1$ against the background plus signal hypothesis $H_0$. We calculate the likelihood ratio
\begin{equation}
    \lambda(n) = \frac{p(n|S+B)}{p(n|B)} = e^{-S} \left(\frac{S+B}{B}\right)^n
\end{equation}
where $B$ is the background yield, and $n$ is the number of observed events.
We then calculate the test statistic
\begin{equation}
    q(n) = - 2 \ln (\lambda) = 2 \Big[ n \ln \left(\frac{B}{S+B}\right) +S \Big]\,,
\end{equation}
from which the signal-observation significance is calculated as $Z= \sqrt{q}$.
In the median expected dataset under the background hypothesis, the so-called Asimov dataset, the number of observed events is $n_A = B$~\cite{Cowan:2010js}, so the  Asimov significance is
\begin{equation}
    Z_A = \sqrt{2 \Big[ S - B \ln \left(1+\frac{S}{B}\right)\Big]}\,.
\label{eq:Assimov_sig}
\end{equation}
For every value of $m_{N_2}$ and $\delta$ on the simulation grid we determine the range of $\Lambda$ values for which $Z_A > 1.64$, corresponding to the one-sided $95\%$ CL exclusion limit. 
The $(m_{N_2},\, \delta,\, \Lambda)$ region satisfying this condition is the excludable parameter space for this final state and $m_{\ell\ell}$ region. 

\begin{figure}
\centering
\includegraphics[width=0.55\linewidth]{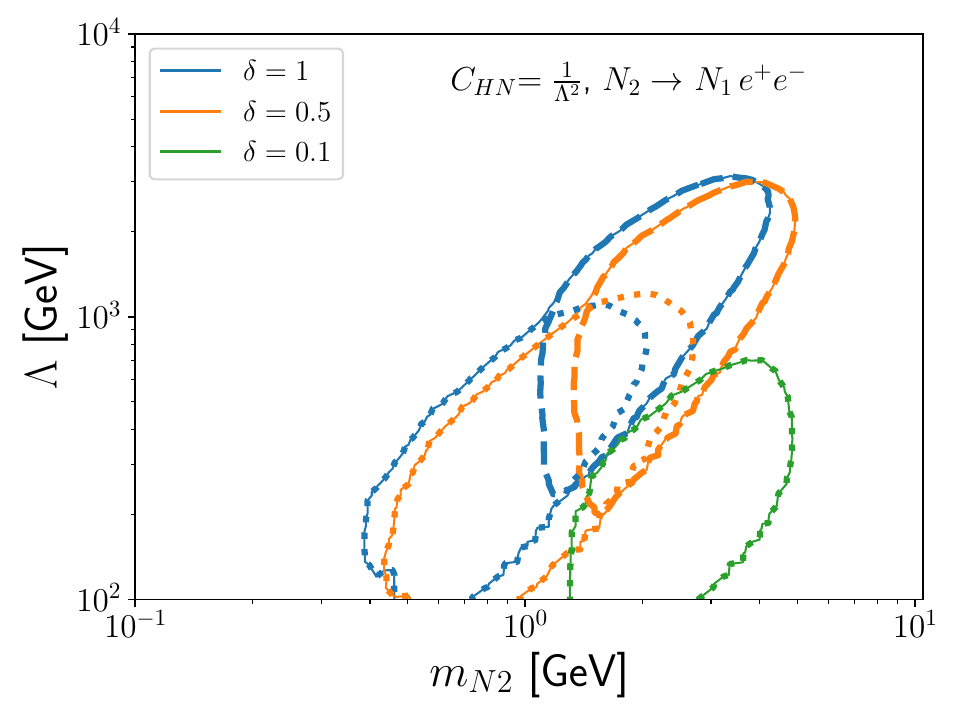}
\caption{Realistic sensitivity contours in  the $\Lambda$ vs. $m_{N_2}$ plane for several values of $\delta$ in the case of the $Q_{HN}$ operator and the $e^+e^-$ final state. 
Shown are the contours for the high-mass (dashed) and low-mass (dotted) regions and for their combination (solid).}
\label{fig:DV_MG_CHN_mll_combined}
\end{figure}

For a given final state, we also calculate the combined significance from the low- and high-mass regions as
\begin{align}
Z_A^{\text{comb}} = \sqrt{Z_{\text{low}}^2 + Z_{\text{high}}^2} \,.
\label{eq:mll_combined_sig}
\end{align}
For the two high-mass regions with a background expectation of less than 1 event, we take the background to be $0$, setting the significance $Z_{\text{high}} = \sqrt{S_{\text{high}}}$ in Eq.~(\ref{eq:mll_combined_sig}).
As an example of the impact of combining the low- and high-mass regions, we show in Fig.~\ref{fig:DV_MG_CHN_mll_combined} the sensitivity contours for the two regions individually and for their combination for one of the operators.
One observes that combining the regions increases the excludable parameter space only marginally relative to the union of the individual excludable parameter spaces.
Therefore, to avoid clutter, we do not show the combined sensitivity contour in the subsequent sensitivity plots.  

\begin{figure}[t]
\centering
\includegraphics[width=0.45\linewidth]{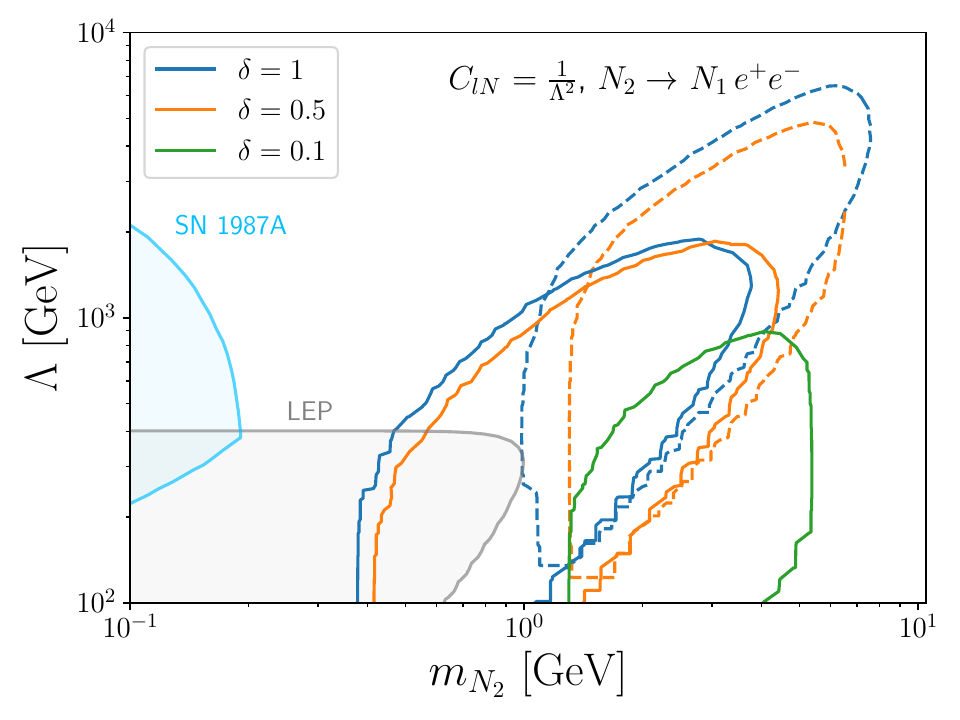}
\includegraphics[width=0.45\linewidth]{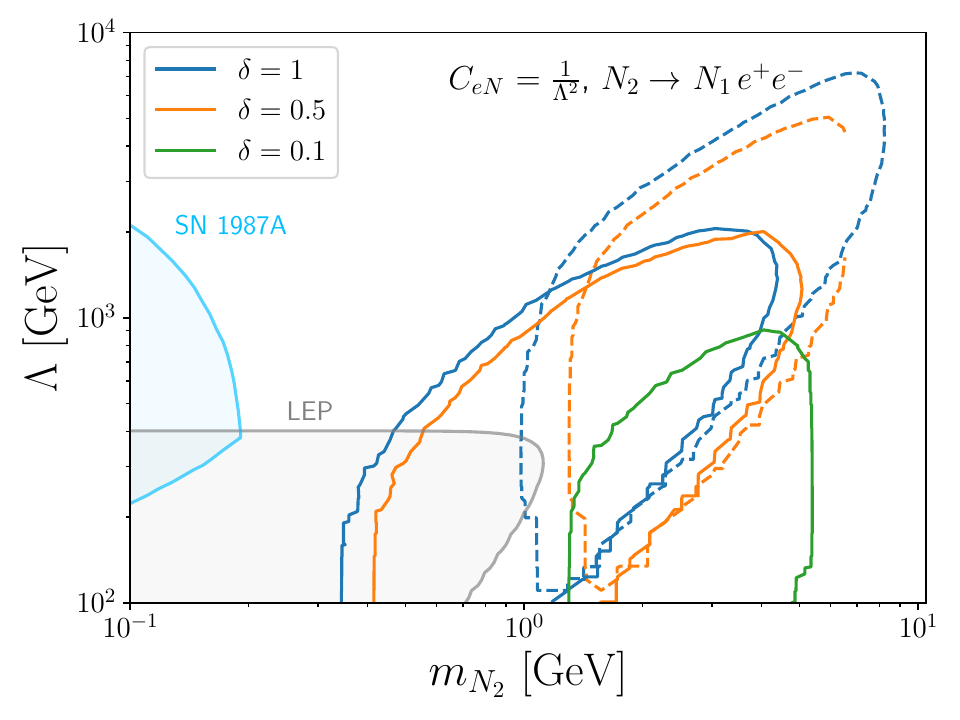}
\includegraphics[width=0.45\linewidth]{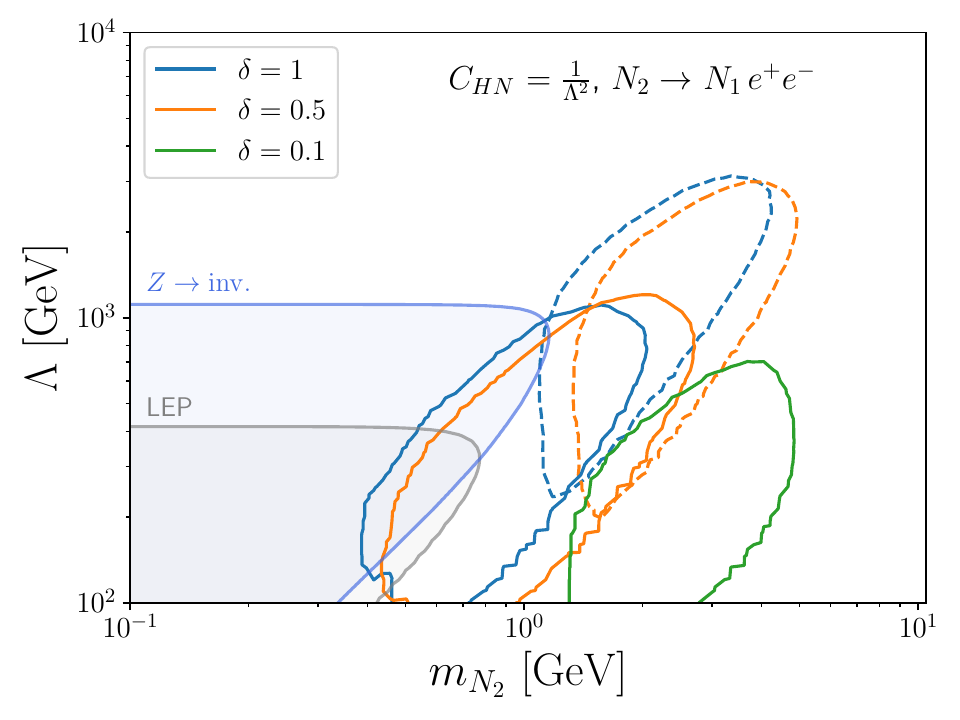}
\includegraphics[width=0.45\linewidth]{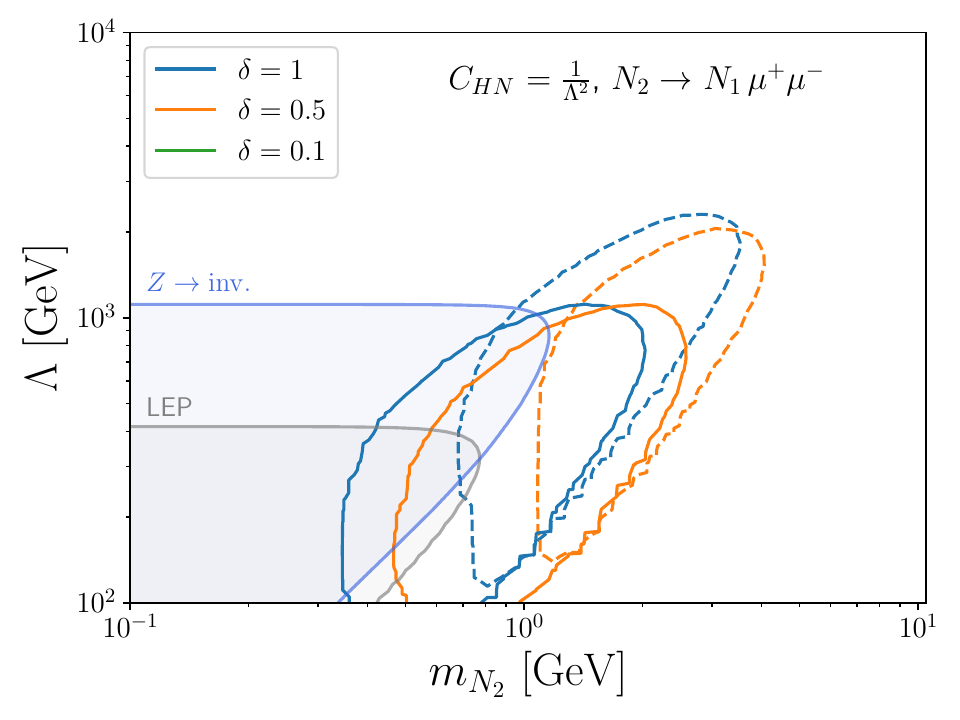}
\caption{Realistic sensitivity contours in the $\Lambda$ vs. $m_{N_2}$ plane from $e^+ e^- \to N_1 N_2$ processes for different mass splittings $\delta$.  
The operator that is switched on and the final state are indicated in each plot. 
The tight cuts (Table~\ref{tab:DV_cuts}) are used in the calculation.
The solid and dashed curves are for the $m_{\ell\ell}<0.5$~GeV and $m_{\ell\ell}>0.5$~GeV regions, respectively. 
Also shown are existing limits from monophoton searches at LEP (grey)~\cite{DELPHI:2003dlq}, supernovae (cyan)~\cite{DeRocco:2019jti}, and invisible decays of the $Z$ boson (blue)~\cite{Janot:2019oyi}.}
\label{fig:DV_MG_N1N2}
\end{figure}

The sensitivity contours for the tight DV cuts are shown in Fig.~\ref{fig:DV_MG_N1N2} for the three operators that produce two HNLs. 
These operators mediate decays into an electron pair and, in the case of $Q_{HN}$, also a muon pair. 
The figures also show the existing constraints, which arise from monophoton searches at LEP (grey)~\cite{DELPHI:2003dlq}, supernovae (cyan)~\cite{DeRocco:2019jti}, and invisible decays of the $Z$ boson (blue)~\cite{Janot:2019oyi}. 
In all these cases, the sensitivity of the search proposed here far exceeds that of the existing constraints.

\begin{figure}[t]
\centering
\includegraphics[width=0.45\linewidth]{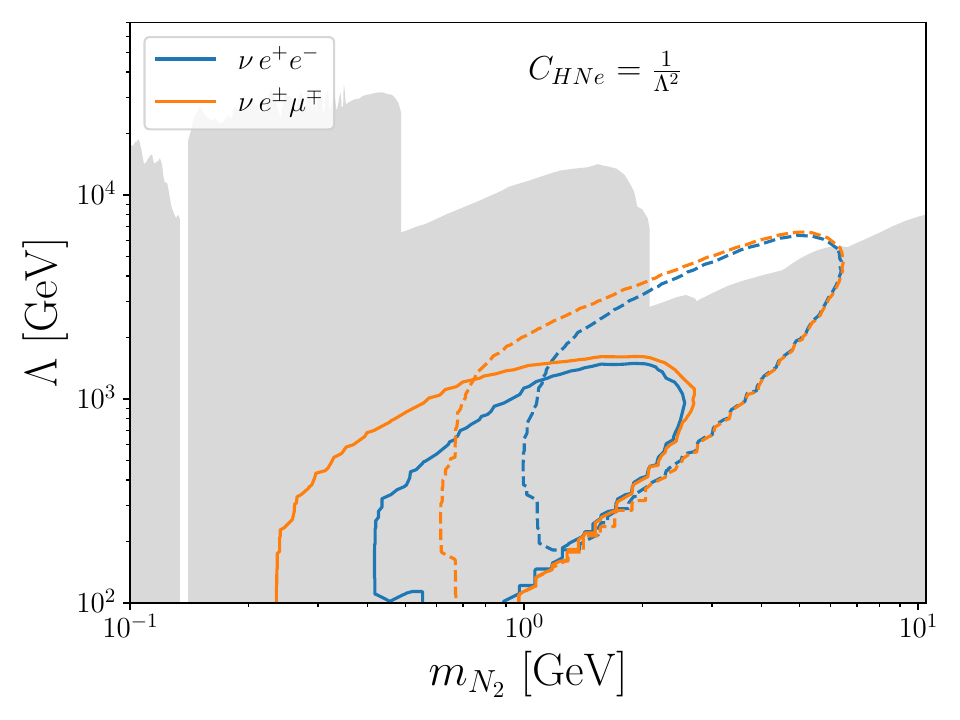}
\includegraphics[width=0.45\linewidth]{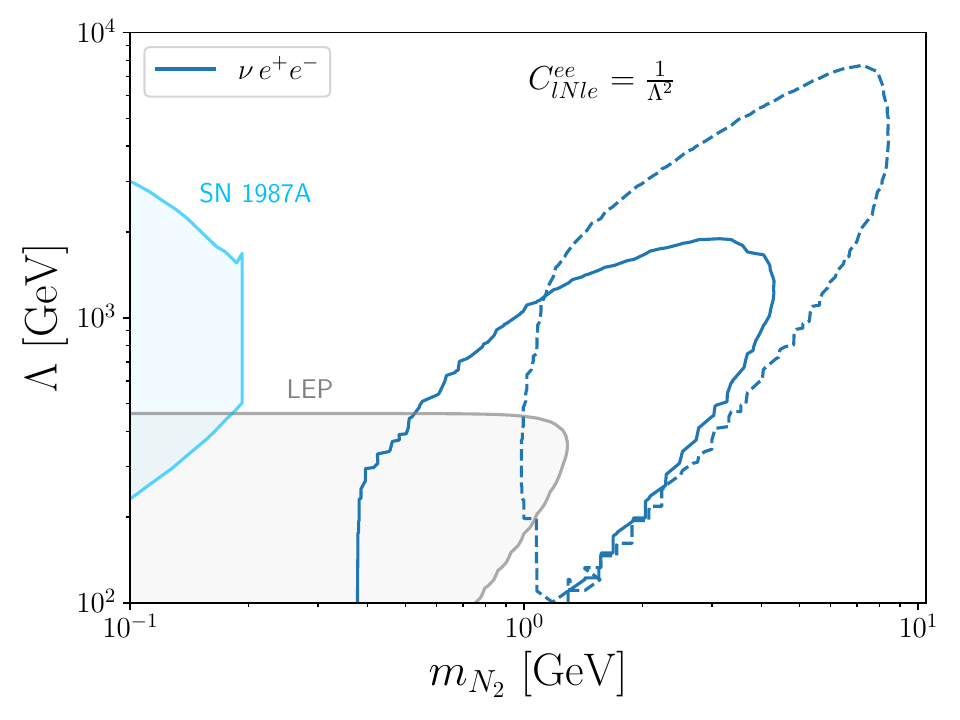}
\includegraphics[width=0.45\linewidth]{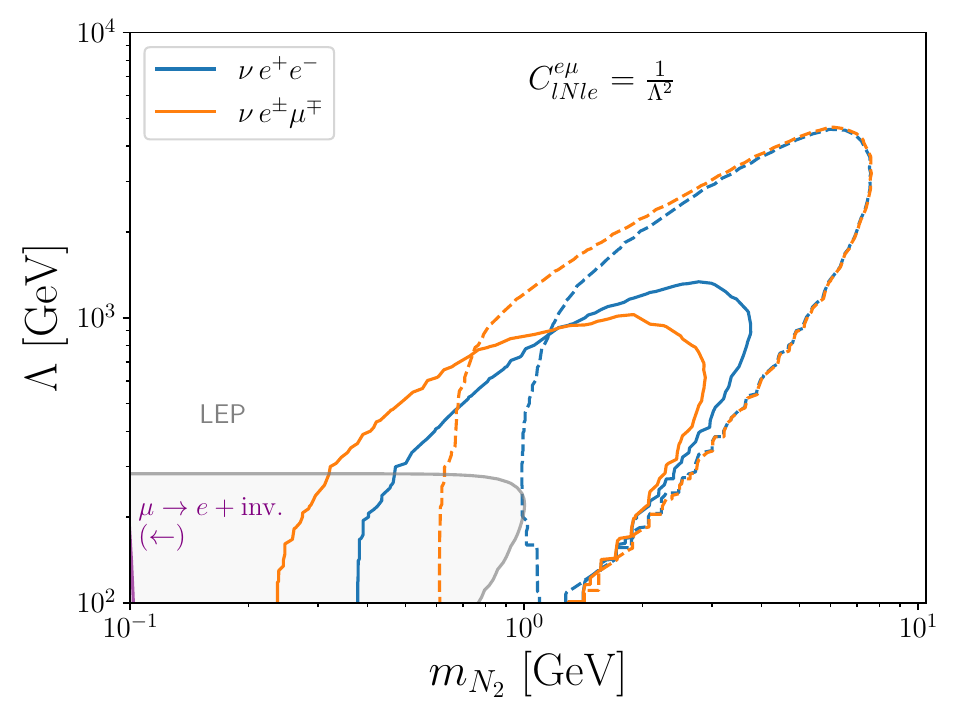}
\includegraphics[width=0.45\linewidth]{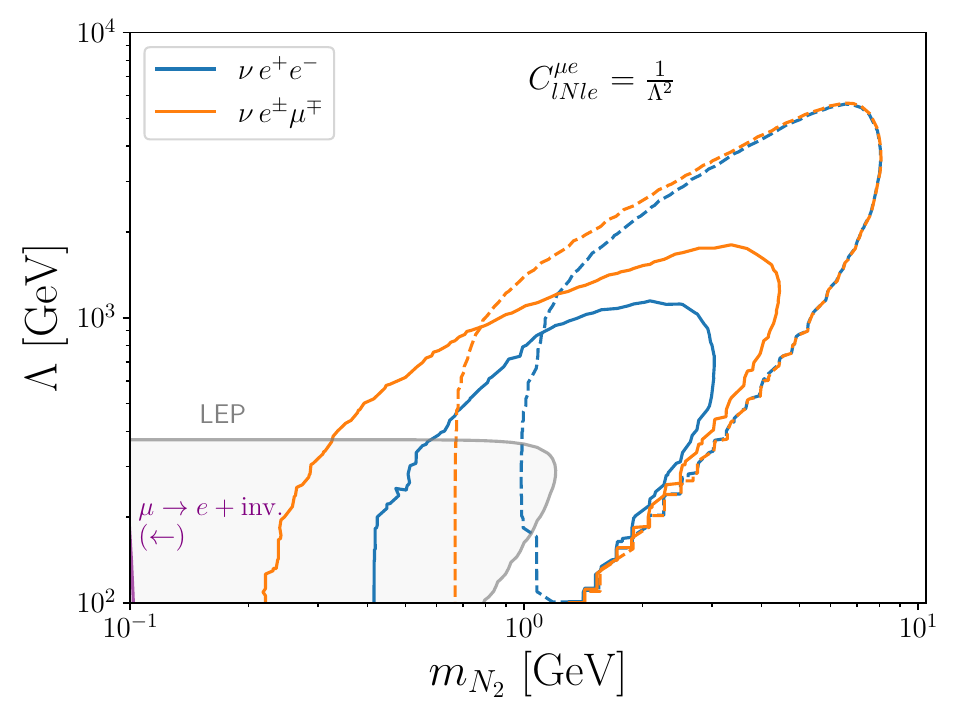}
\includegraphics[width=0.45\linewidth]{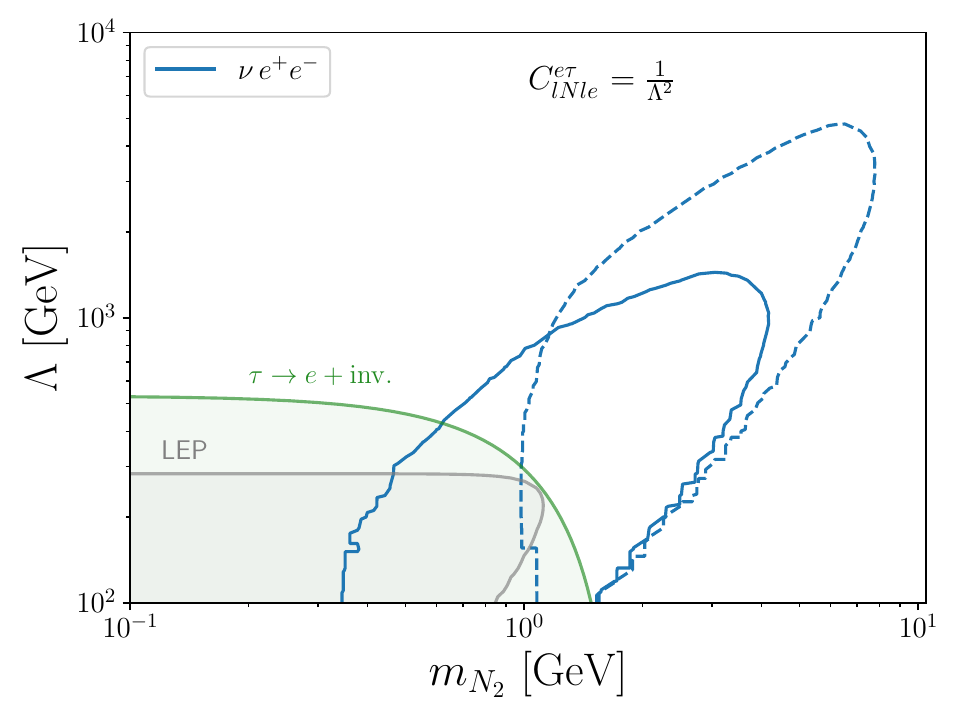}
\includegraphics[width=0.45\linewidth]{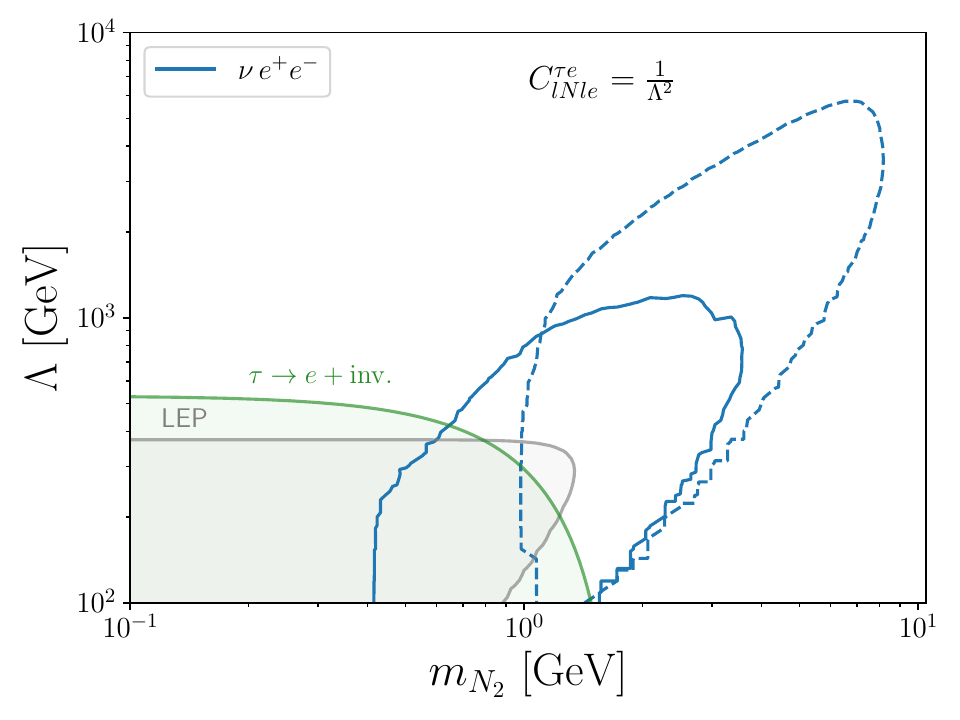}
\caption{Realistic sensitivity contours in the $\Lambda$ vs. $m_{N_2}$ plane from $e^+ e^- \to \nu N_2$ processes.
The operator that is switched on and the final state are indicated in each plot. 
The tight cuts in Table~\ref{tab:DV_cuts} are used in the calculation.
The solid and dashed curves are for the $m_{\ell\ell}<0.5$~GeV and $m_{\ell\ell}>0.5$~GeV regions, respectively. 
Also shown are existing constraints from LEP (grey), supernovae (cyan), and the decays $\mu \to e+\text{invisible}$ (purple) and $\tau \to e+\text{invisible}$ (green)~\cite{Fernandez-Martinez:2023phj}.
For the $C_{HNe}$ operator, the union of several existing constraints is shown in grey. The individual-experiment constraints appear in Fig.~\ref{fig:CHNe_constraints}.}
\label{fig:DV_MG_nuN2}
\end{figure}

Fig.~\ref{fig:DV_MG_nuN2} shows the sensitivity contours for the six operators that produce one HNL and a SM neutrino.
Also shown are existing constraints from LEP (grey), supernovae (cyan), and the decays $\mu \to e+\text{invisible}$ (purple) and $\tau \to e+\text{invisible}$ (green)~\cite{Fernandez-Martinez:2023phj}. 
In this case as well, the proposed search goes much beyond the existing constraints, except in the case of the W-current operator $C_{HNe}$, for which existing constraints arise from many complementary experiments. 
The top left panel of Fig.~\ref{fig:DV_MG_nuN2} shows in grey the combined parameter space excluded by the difference experiments in grey. 
The constraints from the individual experiments are shown in Fig.~\ref{fig:CHNe_constraints} in App.~\ref{app:CHNe}. 

It is striking that, in contrast to the analytic estimate in Sec.~\ref{sec:analytics}, $e^{\pm} \mu^{\mp}$ and $\mu^+ \mu^-$ final states are more sensitive to small masses $m_{N_2}$ than the $e^+ e^-$ final state. This can be explained by the looser cuts needed for final states with muons to reduce the photon conversion background to the same level.

One can see from the $C_{lNle}^{\mu e}$ plot in Fig.~\ref{fig:DV_MG_nuN2} that in the low-mass region the maximal new physics scale that can be probed for the $e^\pm \mu^\mp$ final state is larger than that for $e^+e^-$, while the opposite is true for 
$C_{lNle}^{e \mu}$. 
This difference originates from the different differential cross sections for these operators, as seen in Fig.~\ref{fig:dsigma}, and their impact on the $\theta_{\textrm{DV}}$ efficiency.

\begin{figure}
\centering
\includegraphics[width=0.45\linewidth]{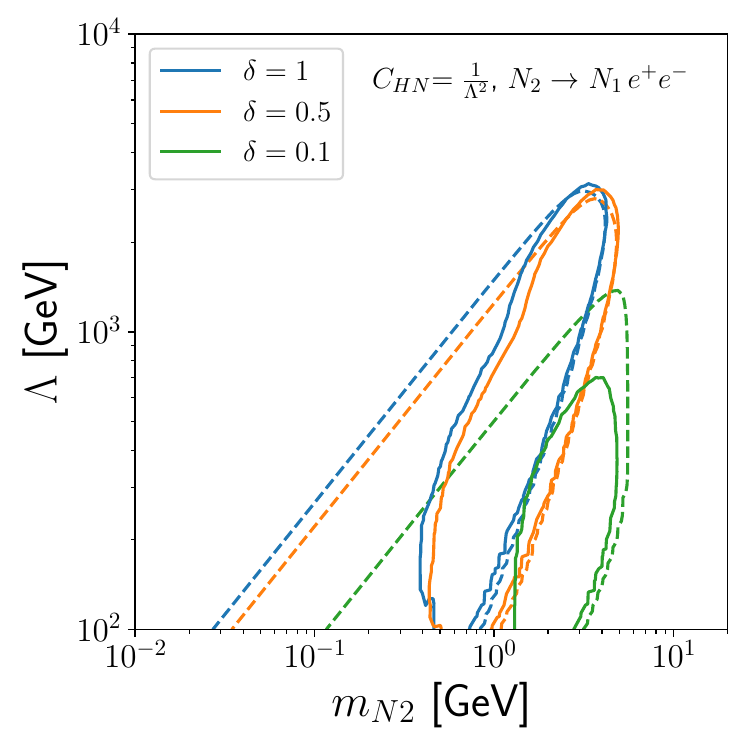}
\includegraphics[width=0.45\linewidth]{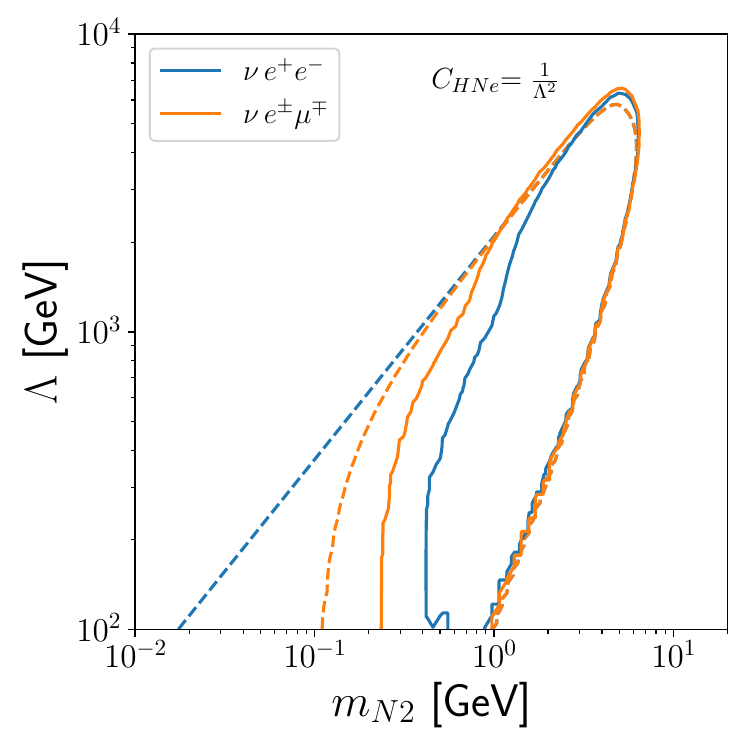}
\caption{Comparison of the full numerical estimation of the DV signal sensitivity (solid) with the analytical estimates (dashed) for two example operators: $Q_{HN}$ with different mass splittings (left) and $C_{HNe}$ with different final states (right).}
\label{fig:DV_num_vs_ana}
\end{figure}

As an example of the difference between this realistic sensitivity estimate and the analytic and background-free estimate in Sec.~\ref{sec:analytics}, we compare the two types of sensitivity contours in Fig.~\ref{fig:DV_num_vs_ana} for two selected operators. 
One observes that the realistic estimate loses sensitivity primarily at low $m_{N_2}$, corresponding to low $m_{\ell\ell}$, both due to the cuts applied on the signal and the higher background in this region. 

Reclaiming some of the lost sensitivity at low $m_{N_2}$ is what motivates exploring the loose cuts listed in Table~\ref{tab:DV_cuts}.
We compare the sensitivity contours with the loose and tight cuts in Figs.~\ref{fig:DV_MG_N1N2_tight+loose} and~\ref{fig:DV_MG_nuN2_tight+loose} in Appendix~\ref{app:loose_cuts}.
These figures show that the excludable parameter space can be increased towards smaller $m_{N_2}$ by using the loose cuts, with the largest increase for the operators $C_{lNle}^{e \mu}$ and $C_{lNle}^{\mu e}$, but generally no dramatic increase outside the already-excludable parameter-space regions.
However, at the same time the highest scale $\Lambda$ that can be probed becomes smaller for some operators and final states using the loose cuts, thus motivating sticking to the use of the tight cuts.

\begin{figure}
\centering
\includegraphics[width=0.6\linewidth]{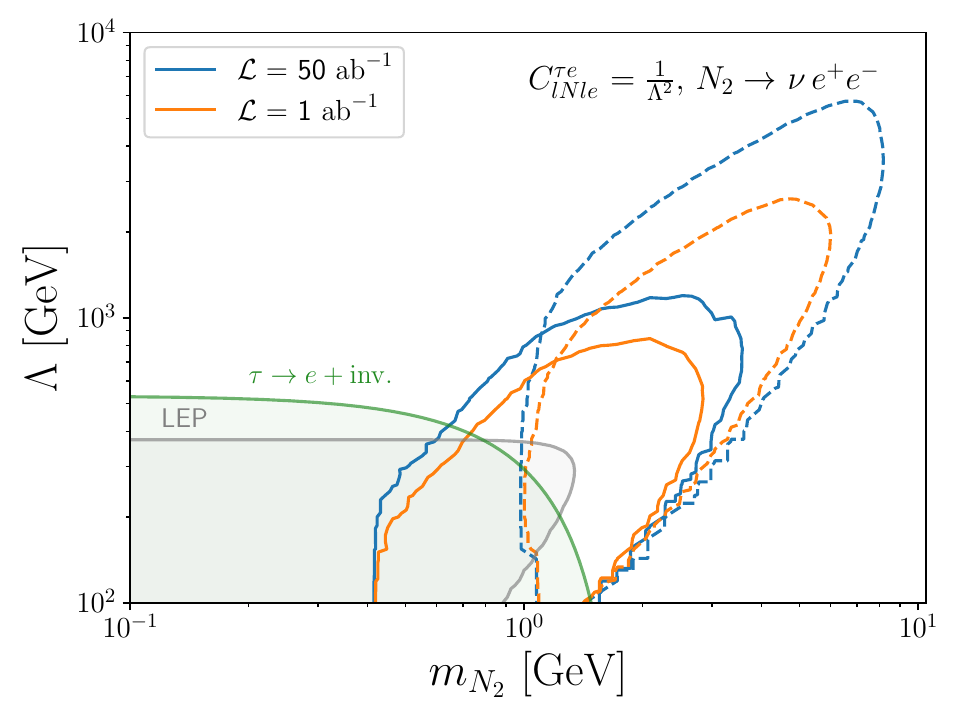}
\caption{Comparison of sensitivity contours for the full integrated luminosity $\mathcal{L} = 50~\textrm{ab}^{-1}$ (blue) and $\mathcal{L} = 1~\textrm{ab}^{-1}$ (yellow) for the operator $Q_{lNle}^{\tau e}$.}
\label{fig:DV_lumi}
\end{figure}

Finally, we consider the sensitivity of a search performed with the current integrated luminosity of $\mathcal{L}\approx 1~\mathrm{ab}^{-1}$.
As an example, we compare in Fig.~\ref{fig:DV_lumi} the sensitivity contours for $Q_{lNle}^{\tau e}$ for $\mathcal{L}=50~\mathrm{ab}^{-1}$, which are identical to those in Fig.~\ref{fig:DV_MG_nuN2}, with contours for $\mathcal{L}=1~\mathrm{ab}^{-1}$, where signal and background yields are lower by a factor of 50. 
As expected, the maximum sensitivity to $\Lambda$ is reduced by a factor of a few. 
However, the excludable parameter space is still much larger than that of the current constraints.
This conclusion holds for all operators and motivates conducting the search with the already available data.

%

%%%%%%%%%%%%%%%%%%%%%%%%%%%%%%%%%%%%%%%%%%%%%%%%%%%%%%%%%
\section{Conclusions}
\label{sec:conclusions}
%%%%%%%%%%%%%%%%%%%%%%%%%%%%%%%%%%%%%%%%%%%%%%%%%%%%%%%%%

This article presents a sensitivity study of the couplings of heavy neutral leptons (HNLs) to the electroweak and lepton sector of the Standard Model at Belle~II.
The effective interactions are parametrised model-independently by an effective field theory, dubbed $\nu$SMEFT or $N_R$SMEFT, which extends the SM by fermion fields that transform as a singlet under the SM gauge groups. 
We focus on Majorana fermions $N_i$ with multiple generations, as required to explain the observed neutrino masses and mixing angles. 
Considering four-fermion, Higgs-current and dipole operators, we study the production of $N_i$ in $e^+e^-$ collisions at Belle~II, and subsequent leptonic decays.
Considering the case of a long lived $N_2$, we propose a search for the experimental signature of a dilepton displaced vertex and missing energy. 
Displaced vertices constitute a promising signature, since HNLs coupled through effective operators suppressed by a high new-physics scale $\Lambda$ can be naturally long-lived, and since the SM background is significantly smaller than for prompt-decay signatures.

We provide analytic formulae for the production cross section and decay rates of the HNLs in $\nu$SMEFT and derive maximal sensitivity estimates for the new-physics scale $\Lambda$ that can be probed in the proposed search.
Next, we thoroughly estimate the leading background sources and rates in each channel. 
These are found to be dominated by conversion to $\ell^+\ell^-$ of photons produced via $e^+e^-\to\gamma\gamma$ and $e^+e^-\to e^+e^-\gamma$, as well as  $K_L\to \pi^\pm \ell^\mp \nu$ decays from  $e^+e^- \to e^+e^- K_L^0 K_L^0$ and $e^+e^- \to \gamma K_S^0 K_L^0$. 
We find that kinematic cuts can effectively mitigate these backgrounds, reducing their yields by several orders of magnitude while significantly limiting the signal sensitivity only at low HNL masses. 
Furthermore, we demonstrate how the $N_2$ HNL mass and momentum vector can be reconstructed from the observed final state, up to an ambiguity in the HNL mass splitting $\delta$.
This provides additional discrimination of signal from background and enables measurement of the HNL masses if significant signal is observed.

Then, we simulate signal events and derive bounds on $\Lambda$ that can be realistically obtained, taking into account the cuts, the nonzero backgrounds, and the track-reconstruction efficiency based on a geometrical detector simulation. 
We find that the proposed search can probe new-physics scales of order $3$ TeV for neutral-Higgs-current operators and up to $7$~TeV for scalar four-fermion operators. Detailed plots show the sensitivities for different values of $m_{N_2}$ and $\delta$. 
Comparing with existing limits, we identify a large open parameter space that can be probed by the proposed search at Belle~II.
The fact that $\Lambda$ is much larger than the energy scale of the experiment confirms the validity of the EFT approach. 

\acknowledgments
P.~D.~B. is supported by the Slovenian Research Agency under the research core funding No.~P1-0035 and in part by the research grants N1-0253 and J1-4389.
F.~E. is supported by Charles University through the project PRIMUS/24/SCI/013.
L.~G. acknowledges support from the Dutch Research Council (NWO) under project number VI.Veni.222.318 and from Charles University through the project number PRIMUS/24/SCI/013.
J.~K. is supported by the HRZZ grant UIP-2025-02-3909 and the MZOM project number 910-06/25-01/00041.
S.~K. is supported by the FWF project number P 36947-N.
A.~S. is supported by the Israel Science Foundation with grant 206/23, the United States-Israel Binational Science Fund grant 2020044, and the Horizon 2020 Marie Sklodowska-Curie RISE project JENNIFER2 grant agreement No. 822070. 
P.~D.~B acknowledges support from the University of Graz and Charles University for visits to both during the completion of this work.
F.~E. is grateful to the University of Graz and the Jožef Stefan Institute in Ljubljana for their hospitality, where part of this work was carried out.
J.~K. is grateful to the Charles University and the Jožef Stefan Institute in Ljubljana for the hospitality during the completion of this work.

\appendix

%%%%%%%%%%%%%%%%%%%%%%%%%%%%%%%%%%%%%%%%%%%%%%%%%%%%%%%%%
\section{Analytical formulae}
\label{app:formulae}
%%%%%%%%%%%%%%%%%%%%%%%%%%%%%%%%%%%%%%%%%%%%%%%%%%%%%%%%%

%%%%%%%%%%%%%%%%%%%%%%%%%%%%%%%%%%%%%%%%%%%%%%%%%%%%%%%%%
\subsection{HNL production}
\label{subsec:production}
%%%%%%%%%%%%%%%%%%%%%%%%%%%%%%%%%%%%%%%%%%%%%%%%%%%%%%%%%

From the $\nu$LEFT interactions in Table~\ref{tab:vLEFT-operators}, the differential cross section for the production of two Majorana HNLs via $e^+e^-(\to \gamma^*)\to N_i N_j$ is given by
\begin{align}
\label{eq:diff_xsec_NN}
\frac{d\sigma}{d\cos\theta^*} &= \frac{s \sqrt{\lambda_{ij}}}{128\pi (1+\delta_{ij})} \nonumber\\
&\hspace{1.3em} \times \bigg\{\frac{8}{s}\Big(1 - \Delta_{ij}^2 - \lambda_{ij}\cos^2\theta^*\Big)\big|e^2 d_{\underset{ij}{NN\gamma}}\big|^2 \nonumber \\
&\hspace{3.3em} + \Big(1 - \Delta_{ij}^2 + \lambda_{ij}\cos^2\theta^*\Big)\Big[\big|L_{\underset{ijee}{Ne}}^{V,RL}\big|^2 + \big|L_{\underset{ijee}{Ne}}^{V,RR}\big|^2\Big] \nonumber \\
&\hspace{3.3em} - 4\,\text{Re}\bigg[\frac{m_{N_i} m_{N_j}}{s}\bigg(\frac{8}{s}\big(e^2 d_{\underset{ij}{NN\gamma}}\big)^2 + \big(L_{\underset{ijee}{Ne}}^{V,RL}\big)^2 + \big(L_{\underset{ijee}{Ne}}^{V,RR}\big)^2\bigg) \nonumber \\
&\hspace{6.8em} + \frac{2m_{N_i}}{s}\big(1 - \Delta_{ij}\big) \big(e^2d_{\underset{ij}{NN\gamma}}\big) \big(L_{\underset{ijee}{Ne}}^{V,RL} + L_{\underset{ijee}{Ne}}^{V,RR}\big)^*\bigg] + (i\leftrightarrow j)\bigg\} \,,
\end{align}
where $\Delta_{ij} \equiv (m_{N_i}^2 - m_{N_j}^2)/s$, $\lambda_{ij} \equiv \lambda(1,m_{N_i}^2/s,m_{N_j}^2/s)$ with the Källén function $\lambda(x,y,z) = x^2 + y^2 + z^2 - 2(xy + yz + zx)$, and  $\theta^*$ is the polar angle between the momenta of the outgoing HNL $N_j$ and that of the incoming beam electron in the centre-of-mass (CM) frame.
Integrating over $\cos\theta^*$, the total cross section is
\begin{align}
\sigma&= \frac{s \sqrt{\lambda_{ij}}}{48\pi(1+\delta_{ij})} \nonumber \\
&\hspace{1.3em}\times\bigg\{\frac{4}{s}\Big(1 + \Sigma_{ij} - 2\Delta_{ij}^2\Big)\big|e^2 d_{\underset{ij}{NN\gamma}}\big|^2 \nonumber \\
&\hspace{3.3em} + \bigg(1 - \frac{\Sigma_{ij}}{2} - \frac{\Delta_{ij}^2}{2}\bigg)\Big[\big|L_{\underset{ijee}{Ne}}^{V,RL}\big|^2 + \big|L_{\underset{ijee}{Ne}}^{V,RR}\big|^2\Big] \nonumber \\
&\hspace{3.3em} - 3\,\text{Re}\bigg[\frac{m_{N_i} m_{N_j}}{s}\bigg(\frac{8}{s}\big(e^2 d_{\underset{ij}{NN\gamma}}\big)^2 + \big(L_{\underset{ijee}{Ne}}^{V,RL}\big)^2 + \big(L_{\underset{ijee}{Ne}}^{V,RR}\big)^2\bigg) \nonumber \\
&\hspace{6.8em} + \frac{2m_{N_i}}{s}\big(1 - \Delta_{ij}\big) \big(e^2d_{\underset{ij}{NN\gamma}}\big) \big(L_{\underset{ijee}{Ne}}^{V,RL} + L_{\underset{ijee}{Ne}}^{V,RR}\big)^*\bigg] + (i\leftrightarrow j)\bigg\} \,,
\end{align}
where $\Sigma_{ij} \equiv(m_{N_i}^2 + m_{N_j}^2)/s$.
The differential cross section for the production of a single HNL via $e^+e^-(\to
\gamma^*)\to \nu N_j$ is instead given by
\begin{align}
\frac{d\sigma}{d\cos\theta^*} &= \frac{s}{64\pi}\bigg(1 - \frac{m_{N_j}^2}{s}\bigg)^2\bigg\{\frac{8}{s}\bigg(1 + \frac{m_{N_j}^2}{s} - \bigg(1 - \frac{m_{N_j}^2}{s}\bigg)\cos^2\theta^*\bigg)\big|e^2 d_{\underset{\rho j}{\nu N\gamma}}\big|^2 \nonumber\\
&\hspace{8.5em}  + \big|L_{\underset{\rho jee}{\nu Ne}}^{S,RL}\big|^2 + \big|L_{\underset{\rho jee}{\nu Ne}}^{S,RR}\big|^2\nonumber \\
&\hspace{8.5em}  + 16\bigg(\frac{m_{N_j}^2}{s} + \bigg(1 - \frac{m_{N_j}^2}{s}\bigg)\cos^2\theta^*\bigg)\big|L_{\underset{\rho jee}{\nu Ne}}^{T,RR}\big|^2\bigg\} \,,
\label{eq:diff_xsec_vN}
\end{align}
which yields the total cross section
\begin{align}
\sigma &= \frac{s}{6\pi}\bigg(1 - \frac{m_{N_j}^2}{s}\bigg)^2\bigg\{\bigg(1 + \frac{2m_{N_j}^2}{s}\bigg)\bigg[\frac{1}{s}\big|e^2 d_{\underset{\rho j}{\nu N\gamma}}\big|^2 + \big|L_{\underset{\rho jee}{\nu Ne}}^{T,RR}\big|^2\bigg] \nonumber\\
&\hspace{8.5em}+ \frac{3}{16}\Big[\big|L_{\underset{\rho jee}{\nu Ne}}^{S,RL}\big|^2 + \big|L_{\underset{\rho jee}{\nu Ne}}^{S,RR}\big|^2\Big] \bigg\} \,.
\end{align}

To account for the unequal beam energies at Belle~II, we also determine the differential cross sections with respect to the laboratory-frame angle $\theta$ between the $N_j$ momentum $p_{N_j}$ and the beam-electron momentum.
To perform the boost, it is convenient to express $p_{N_j}$ as a function of $\theta$,
\begin{align}
\label{eq:pNj_costheta}
|\vec{p}_{N_j}| = \frac{\beta E_{N_j}^* \cos \theta + \sqrt{|\vec{p}_{N_j}^*|^2 - \left(\gamma \beta m_{N_j} \sin \theta\right)^2}}{\gamma \left(1 - \beta^2 \cos^2 \theta \right)} \,,
\end{align}
where the energy and momentum of $N_j$ in the CM frame are given by
\begin{align}
E_{N_j}^* = \frac{s - m_{N_i/\nu}^2 + m_{N_j}^2}{2 \sqrt{s}} \,, \quad
|\vec{p}_{N_j}^*| = \frac{\lambda^{1/2}(s, m_{N_i/\nu}^2, m_{N_j}^2)}{2 \sqrt{s}}\,.
\end{align}
The parameters $\gamma$ and $\beta$ of the Lorentz boost are determined from the energies of the positron beam, $E_{e^+} = 4.000$~GeV, and the electron beam, $E_{e^-} = 7.007$~GeV, as
\begin{align}
\gamma = \frac{E_{e^-} + E_{e^+}}{\sqrt{s}} \,, \quad \beta = \frac{E_{e^-} - E_{e^+}}{E_{e^-} + E_{e^+}} \,.
\end{align}
The differential production cross sections $d\sigma/d\cos\theta$ is then  determined by expressing $\cos\theta^*$ in terms of $\theta$ using the relation
\begin{align}
\label{eq:CMtolab}
\cos\theta^* = \frac{|\vec{p}_{N_j}| \cos\theta - \gamma \beta E_{N_j}^*}{\gamma |\vec{p}_{N_j}^*|}\,
\end{align}
and multiplying by the functional determinant
\begin{align}
\label{eq:Jacobian}
    \frac{d \cos \theta^*}{d \cos \theta} = \frac{(E_{N_j}^*)^2 - \gamma^2 m_{N_j}^2 + 2 \gamma \beta E_{N_j}^* |\vec{p}_{N_j}| \cos\theta}{\gamma^2 |\vec{p}_{N_j}^*| \left(1 - \beta^2 \cos^2 \theta\right) \sqrt{|\vec{p}_{N_j}^*|^2 - \left(\gamma \beta m_{N_j} \sin \theta\right)^2}} \,,
\end{align}
where $|\vec{p}_{N_j}|$ is taken from Eq.~\eqref{eq:pNj_costheta}.

%%%%%%%%%%%%%%%%%%%%%%%%%%%%%%%%%%%%%%%%%%%%%%%%%%%%%%%%%
\subsection{HNL decays}
\label{subsec:decays}
%%%%%%%%%%%%%%%%%%%%%%%%%%%%%%%%%%%%%%%%%%%%%%%%%%%%%%%%%

We now give expressions for the HNL decay rates.
Firstly, the two-body decays $N_j \to N_i\gamma/\nu_\rho\gamma$ (shown in Fig.~\ref{fig:decays}, left) have the partial decay rates
\begin{align}
\Gamma_{N_i \gamma} = \frac{m_{N_j}^3}{2\pi} (1 - y_i^2)^3 \big|e d_{\underset{ij}{NN\gamma}}\big|^2  \,, \quad \Gamma_{\nu_\rho \gamma} &= \frac{m_{N_j}^3}{2\pi} \big|e d_{\underset{\rho j}{\nu N\gamma}}\big|^2 \,,
\end{align}
where we use the shortand $\Gamma_{X} \equiv \Gamma(N_j \to X)$ and define $y_i = m_{N_i}/m_{N_j}$.

The three-body decays of type $N_j \to N_i f_\alpha \bar{f}_\alpha$ are induced by Dalitz-type decays in the presence of the dipole operator $Q_{NNB}$ ($f = e, u, d$), an off-shell $Z$ from the operator $Q_{HN}$ ($f = \nu, e, u, d$), and the four-fermion operators $Q_{lN}$ and $Q_{eN}$ ($f = e$).
The decay rate is given by
\begin{align}
\Gamma_{N_i f_\alpha \bar{f}_\alpha} &= \frac{N_f m_{N_j}^5}{768\pi^3}\bigg\{\frac{8 I_{d,1}^{\alpha\alpha i}}{m_{N_j}^2}\big|e^2 Q_f d_{\underset{ij}{NN\gamma}}\big|^2 + I_1^{i\alpha\alpha}\Big[\big|L_{\underset{ij\alpha\alpha}{Nf}}^{V,RL}\big|^2 + \big|L_{\underset{ij\alpha\alpha}{Nf}}^{V,RR}\big|^2\Big] \nonumber \\
&\hspace{5em} + \text{Re}\bigg[\frac{8 I_{d,2}^{\alpha\alpha i}}{m_{N_j}^2}\big(e^2 Q_f d_{\underset{ij}{NN\gamma}}\big)^2 + I_2^{\alpha\alpha i}L_{\underset{ij\alpha\alpha}{Nf}}^{V,RL}L_{\underset{ij\alpha\alpha}{Nf}}^{V,RR*} \nonumber \\
&\hspace{7.5em} + \frac{1}{2}I_3^{\alpha\alpha i} \Big(\big(L_{\underset{ij\alpha\alpha}{Nf}}^{V,RL}\big)^2 + \big(L_{\underset{ij\alpha\alpha}{Nf}}^{V,RR}\big)^2\Big) + 2I_4^{\alpha\alpha i}L_{\underset{ij\alpha\alpha}{Nf}}^{V,RL}L_{\underset{ij\alpha\alpha}{Nf}}^{V,RR}  \nonumber \\
&\hspace{7.5em} - \frac{8I_{d,3}^{\alpha \alpha i}}{m_{N_j}}\big(e^2 Q_f d_{\underset{ij}{NN\gamma}}\big)\big(L_{\underset{ij\alpha\alpha}{Nf}}^{V,RL} + L_{\underset{ij\alpha\alpha}{Nf}}^{V,RR}\big) \nonumber \\
&\hspace{7.5em} - \frac{8I_{d,4}^{\alpha \alpha i}}{m_{N_j}}\big(e^2 Q_f d_{\underset{ij}{NN\gamma}}\big)\big(L_{\underset{ij\alpha\alpha}{Nf}}^{V,RL} + L_{\underset{ij\alpha\alpha}{Nf}}^{V,RR}\big)^*\bigg]\bigg\} \,,
\label{eq:decay_rate_Ni}
\end{align}
where $N_f$ is the appropriate colour factor for the final state fermion and we introduce the shorthand notation $I_X^{abc} = I_X(y_a, y_b, y_c)$, where the functions $I_X$ are given by
\begin{align}
I_1(x,y,z) &= 12\int_{(x+y)^2}^{(1-z)^2} ds (1 + z^2 - s) (s - x^2 - y^2) F(s,x,y,z)\,, \nonumber \\
I_2(x,y,z) &= 24xy\int_{(x+y)^2}^{(1-z)^2} ds (1 + z^2 - s) F(s,x,y,z)\,, \nonumber \\
I_3(x,y,z) &= 24z\int_{(x+y)^2}^{(1-z)^2} ds (s - x^2 - y^2) F(s,x,y,z)\,, \nonumber \\
I_4(x,y,z) &= 48xyz\int_{(x+y)^2}^{(1-z)^2} dsF(s,x,y,z)\,,
\end{align}
with $F(s,x,y,z) = \lambda^{1/2}(1, s, z^2)\lambda^{1/2}(s, x^2, y^2)/s$, and
\begin{align}
I_{d,1}(x,y,z) &= \int_{(x+y)^2}^{(1-z)^2} \frac{ds}{s^3} \left(2(1-z^2)^2 - s(1+z^2) - s^2\right) \left(s - (x - y)^2\right) \left(2s + (x + y)^2\right) F(s,x,y,z)\,, \nonumber \\
I_{d,2}(x,y,z) &= 6z\int_{(x+y)^2}^{(1-z)^2} \frac{ds}{s^23} \left(s - (x - y)^2\right) \left(2s + (x + y)^2\right) F(s,x,y,z)\,, \nonumber \\
I_{d,3}(x,y,z) &= \frac{I_{d,4}(x,y,z)}{z}=\frac{3}{2}\int_{(x+y)^2}^{(1-z)^2} \frac{ds}{s^2}  (1 + z^2 - s) \left(s - (x - y)^2\right) \left(2s + (x + y)^2\right) F(s,x,y,z)\,, \nonumber \\
I_{d,5}(x,y,z) &= (x+y) z \int_{(x+y)^2}^{(1-z)^2} \frac{ds}{s} \left(s - (x - y)^2\right) F(s,x,y,z)\,.
\end{align}
Next, the three-body decay $N_j \to \nu_\rho \ell_\alpha^- \ell_\beta^+$ induced by the Dalitz-type process with the dipole operators $Q_{NB}$ and $Q_{NW}$ ($\alpha = \beta$), an off-shell $W^\pm$ from the operator $Q_{HNe}$, and the four-fermion operator $Q_{lNle}$.
The decay rate is
\begin{align}
\Gamma_{\nu_\rho \ell_\alpha^- \ell_\beta^+} &= \frac{m_{N_j}^5}{1536\pi^3}\bigg\{\frac{8 I_{d,1}^{\alpha\beta 0}}{m_{N_j}^2}\big|e^2 d_{\underset{\rho j}{\nu N\gamma}}\big|^2 \delta_{\alpha\beta} + \frac{1}{4}I_1^{0\alpha\beta}\Big[\big|L_{\underset{\rho j\alpha\beta}{\nu Ne}}^{S,RL}\big|^2 + \big|L_{\underset{\rho j\alpha\beta}{\nu Ne}}^{S,RR}\big|^2 + 48\big|L_{\underset{\rho j\alpha\beta}{\nu Ne}}^{T,RR}\big|^2\Big]\nonumber \\
&\hspace{2.5em} - \,\text{Re}\bigg[\frac{1}{2} I_2^{\alpha\beta 0}L_{\underset{\rho j\alpha\beta}{\nu Ne}}^{S,RL}L_{\underset{\rho j\alpha\beta}{\nu Ne}}^{S,RR*} - \frac{48I_{d,5}^{\alpha\beta 0}}{m_{N_j}}\big(e^2 d_{\underset{ij}{NN\gamma}}\big)L_{\underset{\rho j\alpha\beta}{\nu Ne}}^{T,RR*}\delta_{\alpha\beta}\bigg] + (\alpha\leftrightarrow \beta)\bigg\}\,.
\label{eq:decay_rate_nu}
\end{align}
Dalitz-type decays to quark final states, $N_j \to \nu_\rho q_\alpha\bar{q}_\alpha$, induced by the dipole operators $Q_{NB}$ and $Q_{NW}$, have the decay rate
\begin{align}
\Gamma_{\nu_\rho q_\alpha \bar{q}_\alpha} &= \frac{m_{N_j}^3}{32\pi^3}I_{d,1}^{\alpha\alpha 0}\big|e^2 Q_q d_{\underset{\rho j}{\nu N\gamma}}\big|^2 \,.
\end{align}
Finally, via an off-shell $W^\pm$, the operator $Q_{HNe}$ induces the three body decays $N_j \to \ell_\rho^-u_\alpha\bar{d}_\beta$ and $N_j \to \ell_\rho^+ \bar{u}_\alpha d_\beta$ (shown in Fig.~\ref{fig:decays}, right), with the decay rate
\begin{align}
\Gamma_{\ell_\rho^- u_\alpha \bar{d}_\beta} &= \Gamma_{\ell_\rho^+ \bar{u}_\alpha d_\beta} = \frac{m_{N_j}^5}{512\pi^3}I_1^{\rho\alpha\beta}\big|V_{\alpha\beta} L_{\underset{\rho j\alpha\beta}{eNud}}^{V,RL}\big|^2 \,.
\end{align}
For quark final states, we assume that when the minimum momentum transfer $q_{\text{min}}^2$ to the quark pair is larger than the scale $Q = 1$~GeV, quark-hadron duality allows to estimate the decay rates to inclusive multi-hadron final states as
\begin{align}
\Gamma_{N_i +\text{hadr.}} &= (1 + \Delta_{\text{QCD}}) \sum_{\alpha,\beta} R_{i\alpha\alpha} \Gamma_{N_i q_\alpha\bar{q}_\alpha} \,, \nonumber \\
\Gamma_{\nu_\rho + \text{hadr.}} &= (1 + \Delta_{\text{QCD}}) \sum_{\alpha,\beta} R_{\rho\alpha\alpha}\Gamma_{\nu_\rho q_\alpha\bar{q}_\alpha} \,, \nonumber \\
\Gamma_{\ell_{\rho}^\mp + \text{hadr.}} & = (1 + \Delta_{\text{QCD}})  \sum_{\alpha,\beta} R_{\rho\alpha\beta} \Gamma_{\ell_{\rho}^- u_\alpha\bar{d}_\beta} \,,
\label{eq:decays_hadronic}
\end{align}
with the perturbative QCD corrections taken from the three-loop factor applied to hadronic $\tau$ decays~\cite{Gorishnii:1990vf},
\begin{align}
1 + \Delta_{\text{QCD}} = \frac{\Gamma(\tau \to \nu_\tau+\text{hadr.})}{\sum_q\Gamma(\tau \to \nu_\tau + u \bar{q})|_{\text{tree}}} = 1 + \frac{\alpha_s}{\pi} + 5.2\frac{\alpha_s^2}{\pi^2} + 26.4\frac{\alpha_s^3}{\pi^3} \,,
\end{align}
where $\alpha_s$ is evaluated at $(q_{\text{min}}^2)^{1/2}$.
In Eq.~\eqref{eq:decays_hadronic}, we estimate the suppression of the inclusive hadronic decay rates close to the lightest hadronic thresholds by applying the factors $R_{abc} = \sqrt{1 - (y_a + y_b + y_c)^2}$, where $m_b$ and $m_c$ are the masses of the lightest hadronic states associated with quark flavours $b$ and $c$.

\begin{figure}[t!]
    \centering
    \includegraphics[width=0.325\linewidth]{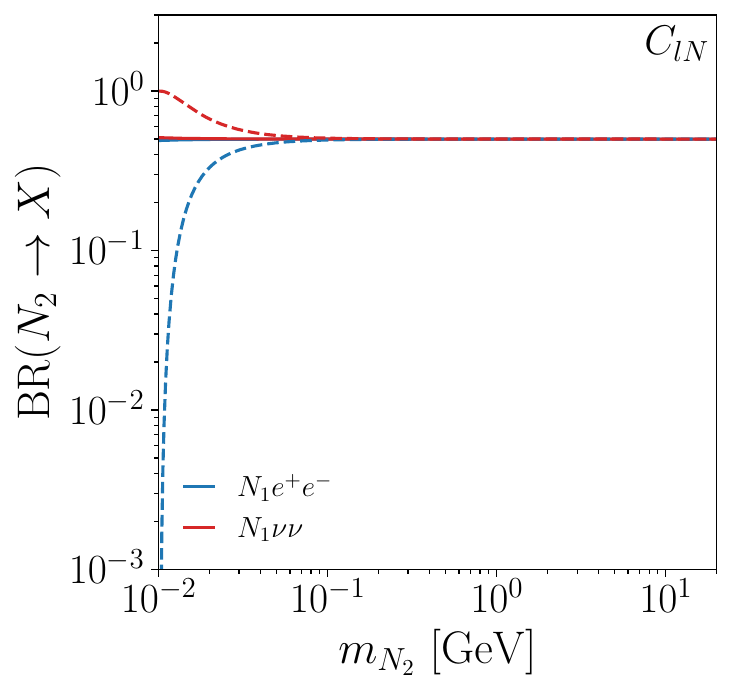}
    \includegraphics[width=0.325\linewidth]{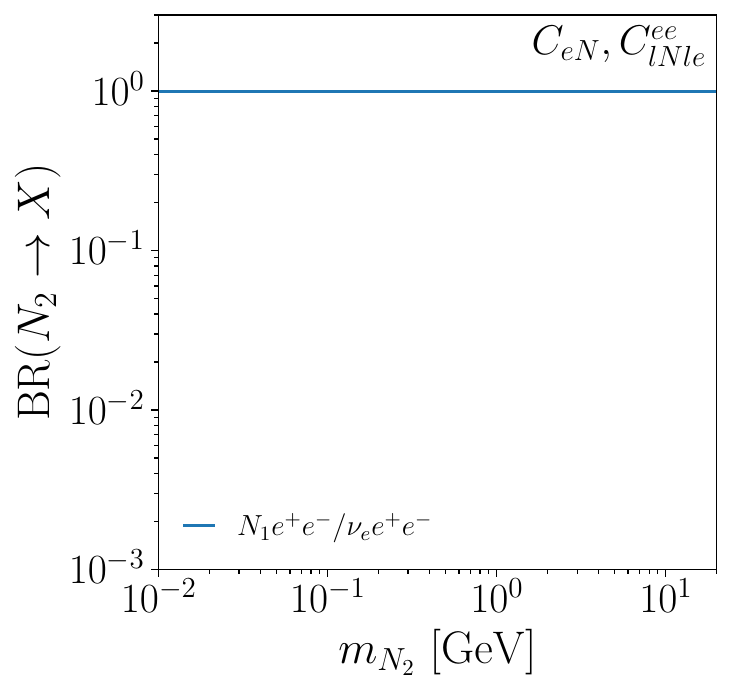}\\
    \includegraphics[width=0.325\linewidth]{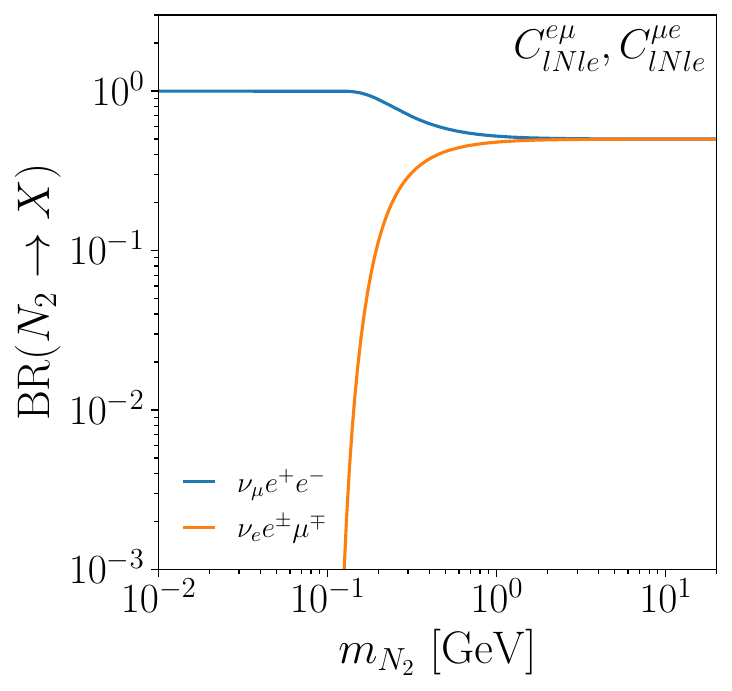}
    \includegraphics[width=0.325\linewidth]{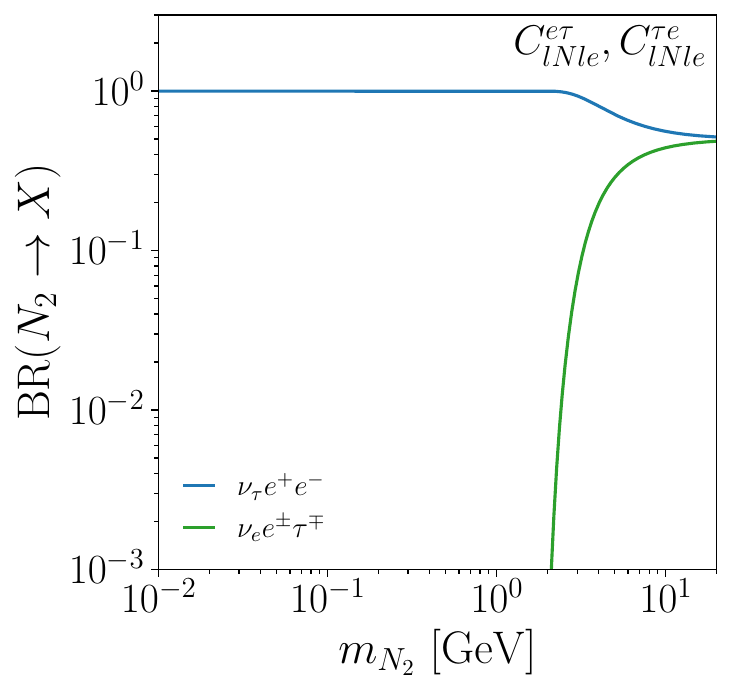}
    \caption{Branching fractions of the heavier HNL $N_2$ as a function of $m_{N_2}$ for the nonzero Wilson coefficients $C_{lN}$ (upper left), $C_{eN}$ and $C_{lNle}^{ee}$ (upper right), $C_{lNle}^{e\mu}$ and $C_{lNle}^{\mu e}$ (lower left) and $C_{lNle}^{e\tau}$ and $C_{lNle}^{\tau e}$ (lower right). For decays involving $N_1$ in the final state, we show the branching fractions for the mass splitting ratios $\delta = 1$ (solid) and $\delta = 0.1$ (dashed).}
\label{fig:branching_1}
\end{figure}
\begin{figure}[t!]
    \centering
    \includegraphics[width=0.325\linewidth]{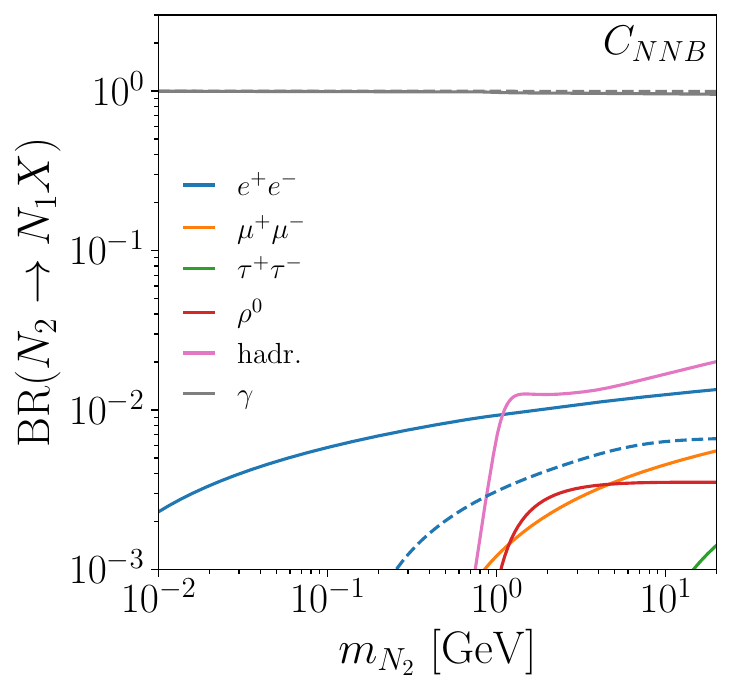}
    \includegraphics[width=0.325\linewidth]{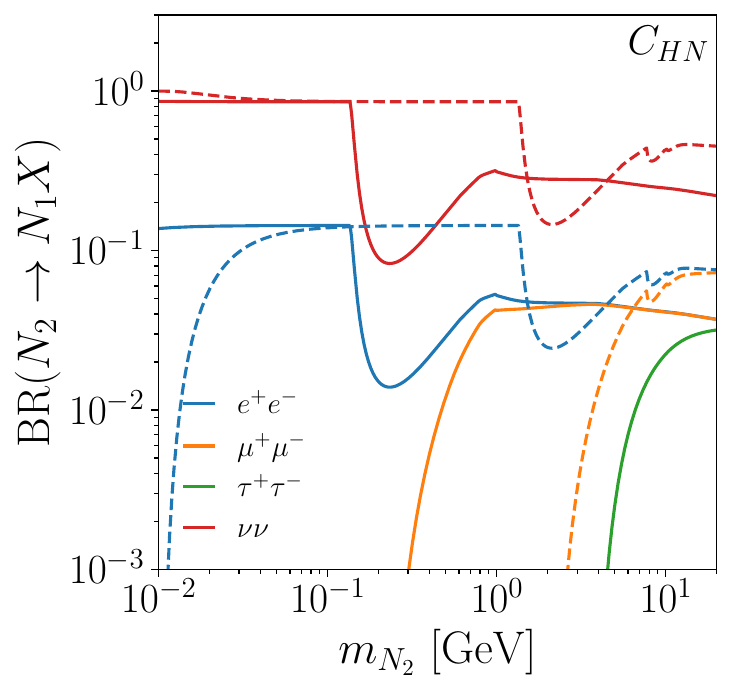}
    \includegraphics[width=0.325\linewidth]{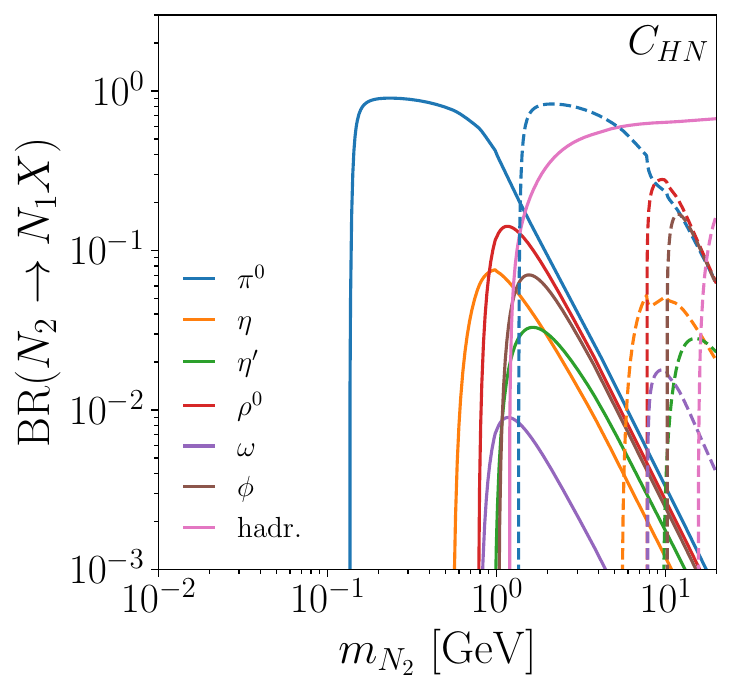}\\
    \includegraphics[width=0.325\linewidth]{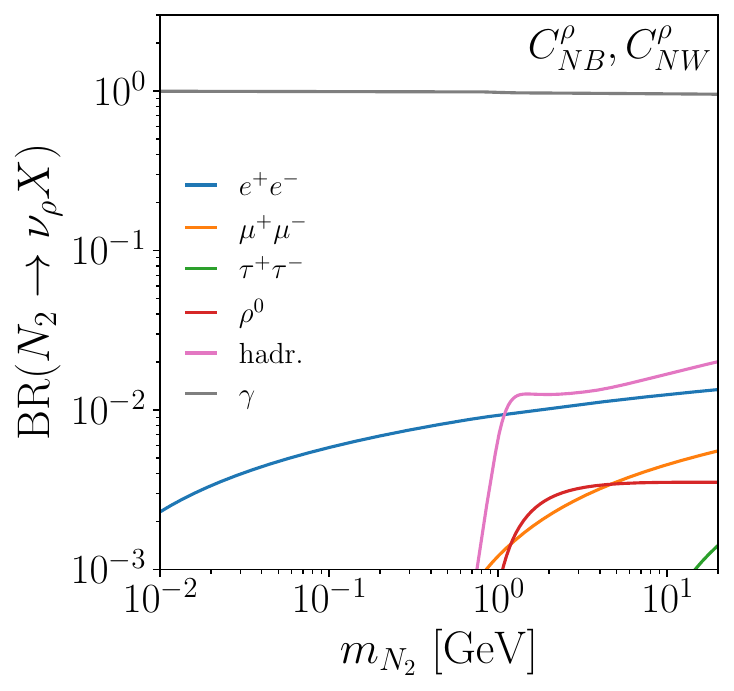}
    \includegraphics[width=0.325\linewidth]{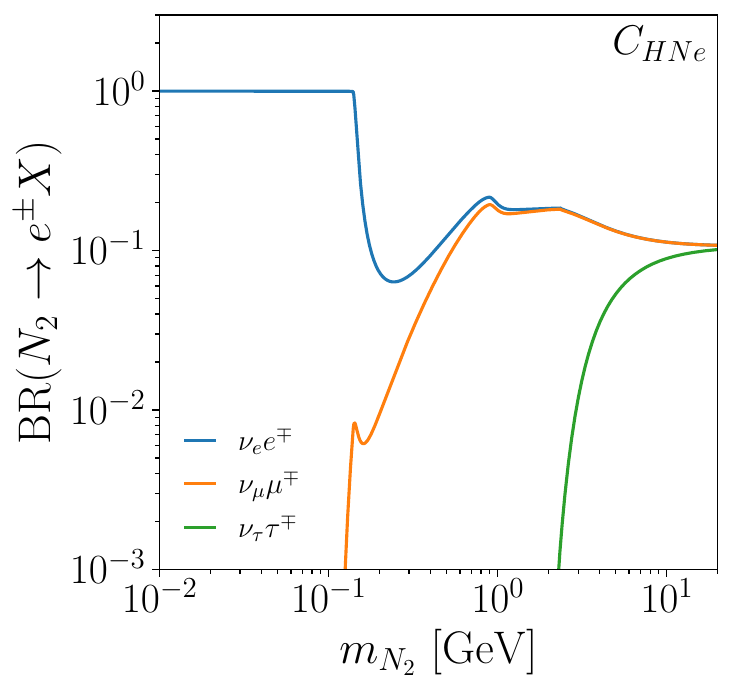}
    \includegraphics[width=0.325\linewidth]{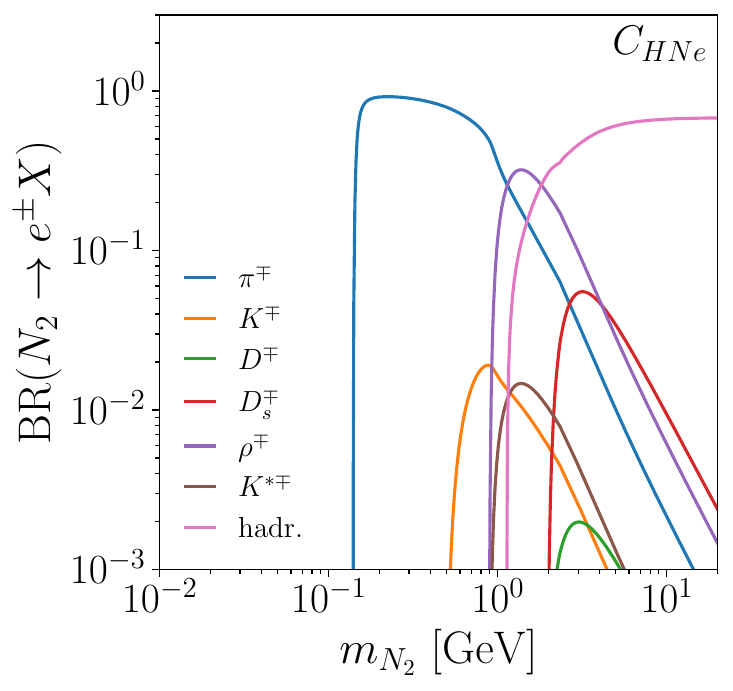}
    \caption{As in Fig.~\ref{fig:branching_1}, for $C_{NNB}$ (upper left), $C_{HN}$ (upper centre and right), $C_{NB}^\rho$ and $C_{NW}^\rho$ (lower left) and $C_{HNe}$ (lower centre and right).}
\label{fig:branching_2}
\end{figure}

If the minimum momentum transfer $q_{\text{min}}^2$ is instead below $Q = 1$~GeV, we compute the exclusive decay rates to the lightest hadronic final states.
Firstly, off-shell $Z$ and $W^\pm$ via the operators $Q_{HN}$ and $Q_{HNe}$ trigger the two-body decays to light pseudoscalar mesons, i.e. $N_j \to N_i P^0$ and $N_j \to \ell_\rho^\mp P^\pm$, respectively.
The generic rates for these processes are
\begin{align}
\Gamma_{N_i P^0} &= \frac{m_{N_j}^3\sqrt{\lambda_{iP}}}{64\pi}\bigg\{\Big[(1 - y_i^2)^2 - (1 + y_i^2)y_P^2\Big]\bigg|\sum_\alpha f_P^\alpha \big(L_{\underset{ij\alpha\alpha}{Nq}}^{V,RL} - L_{\underset{ij\alpha\alpha}{Nq}}^{V,RR}\big)\bigg|^2\nonumber \\
&\hspace{6em} - 2y_i y_P^2\text{Re}\Big[\Big(\sum_\alpha f_P^\alpha \big(L_{\underset{ij\alpha\alpha}{Nq}}^{V,RL} - L_{\underset{ij\alpha\alpha}{Nq}}^{V,RR})\Big)^2\Big]\bigg\} \,, \label{eq:pseudoscalar_NC}\\
\Gamma_{\ell_\rho^\mp P^\pm} &= \frac{m_{N_j}^3\sqrt{\lambda_{\rho P}}}{128\pi} \Big[(1 - y_\rho^2)^2 - (1 + y_\rho^2)y_P^2\Big]  \bigg|\sum_{\alpha,\beta} f_P^{\alpha\beta} V_{\alpha\beta}L_{\underset{\rho j\alpha\beta}{eNud}}^{V,RL}\bigg|^2 \,, \label{eq:pseudoscalar_CC}
\end{align}
where $\lambda_{ab} = \lambda(1, y_a^2, y_b^2)$.
We consider the lightest pseudoscalar mesons $\pi^0$, $\eta$, $\eta'$, $\pi^\pm$, $K^\pm$, $D^\pm$ and $D_s^\pm$.
In Eqs.~\eqref{eq:pseudoscalar_NC} and \eqref{eq:pseudoscalar_CC}, we make use of the decay constants of these states, defined via
\begin{align}
\bra{0}\bar{q}_\alpha\gamma_\mu \gamma_5 q_\alpha\ket{P} &= i f_P^\alpha p_\mu \,, \quad \bra{0}\bar{u}_\alpha\gamma_\mu \gamma_5 d_\beta\ket{P} = i f_P^{\alpha\beta} p_\mu \,,
\end{align}
with
\begin{align}
f_\pi^u &= -f_\pi^d = \frac{f_{\pi}^{ud}}{\sqrt{2}} = \frac{f_\pi}{\sqrt{2}}\,, \quad f_{K}^{us} = f_K \,, \quad f_{D}^{cd} = f_D \,,\quad f_{D_s}^{cs} = f_{D_s} \,, \nonumber \\
f_\eta^u &= f_\eta^d = \frac{c_8 f_8}{\sqrt{6}} - \frac{s_0 f_0}{\sqrt{3}}\,, \quad f_\eta^s = -\sqrt{\frac{2}{3}}c_8 f_8 - \frac{s_0 f_0}{\sqrt{3}} \,, \nonumber \\
f_{\eta'}^u &= f_{\eta'}^d = \frac{s_8 f_8}{\sqrt{6}} + \frac{c_0 f_0}{\sqrt{3}}\,, \quad f_{\eta'}^s = -\sqrt{\frac{2}{3}}s_8 f_8 + \frac{c_0 f_0}{\sqrt{3}} \,,
\end{align}
and $f_P^\alpha = f_P^{\alpha\beta}= 0$ otherwise.
In addition to off-shell $Z$ and $W^\pm$ from the operators $Q_{HN}$ and $Q_{HNe}$, vector meson final states are also induced by the dipole operators $Q_{NNB}$, $Q_{NB}$ and $Q_{NW}$.
The decay rates for the two-body processes $N_{j} \to N_i V^0$, $N_{j} \to \nu_\rho V^0$ and $N_{j} \to \ell_\rho^\mp V^\pm$ are given, respectively, by
\begin{align}
\Gamma_{N_i V^0} &= \frac{m_{N_j}^3\sqrt{\lambda_{iV}}}{64\pi}\bigg\{16\bigg[2(1 - y_i^2)^2 - (1 + y_i^2)y_V^2 - y_V^4\Big] \bigg|\sum_\alpha e^2 Q_q \frac{f_V^\alpha}{m_V} d_{\underset{ij}{NN\gamma}}\bigg|^2 \nonumber \\
&\hspace{6em} + \Big[(1 - y_i^2)^2 + (1 + y_i^2)y_V^2 - 2y_V^4\Big]\bigg|\sum_\alpha f_V^\alpha \big(L_{\underset{ij\alpha\alpha}{Nq}}^{V,RL} + L_{\underset{ij\alpha\alpha}{Nq}}^{V,RR}\big)\bigg|^2\nonumber \\
&\hspace{6em} + 6 y_i y_V^2 \text{Re}\bigg[16\bigg(\sum_\alpha e^2 Q_q \frac{f_V^\alpha}{m_V} d_{\underset{ij}{NN\gamma}}\bigg)^2 + \Big(\sum_\alpha f_V^\alpha \big(L_{\underset{ij\alpha\alpha}{Nq}}^{V,RL} + L_{\underset{ij\alpha\alpha}{Nq}}^{V,RR}\big)\Big)^2\bigg]\bigg\} \,, \nonumber \\
\Gamma_{\nu_\rho V^0} &= \frac{m_{N_j}^3(1 - y_V^2)}{4\pi}(2 - y_V^2 - y_V^4)\bigg|\sum_\alpha e^2 Q_q \frac{f_V^\alpha}{m_V} d_{\underset{\rho j}{\nu N\gamma}}\bigg|^2 \,, \nonumber \\
\Gamma_{\ell_\rho^{\mp} V^\pm} &= \frac{m_{N_j}^3 \sqrt{\lambda_{\rho V}}}{128\pi} \Big[(1 - y_\rho^2)^2 + (1 + y_\rho^2)y_V^2 - 2y_V^4\Big] \bigg|\sum_{\alpha,\beta} f_V^{\alpha\beta} V_{\alpha\beta} L_{\underset{\rho j\alpha\beta}{eNud}}^{V,RL}\bigg|^2 \,.
\end{align}
As for the pseudoscalar final states, we utilise the decay constants defined via
\begin{align}
\bra{0}\bar{q}_\alpha\gamma_\mu q_\alpha\ket{V} &=  f_V^\alpha m_V\epsilon_\mu\,, \quad \bra{0}\bar{u}_\alpha\gamma_\mu d_\beta\ket{V} = f_V^{\alpha\beta} m_V\epsilon_\mu \,,
\end{align}
For the considered states $\rho^0$, $\omega$, $\phi$, $\rho^\pm$ and $K^{*\pm}$, we have
\begin{align}
f_\rho^u &= -f_\rho^d = \frac{f_\rho^{ud}}{\sqrt{2}}= \frac{f_\rho}{\sqrt{2}} \,, \quad f_\omega^u = f_\omega^d = \frac{f_\omega}{\sqrt{2}}\,, \quad f_{\phi}^s = f_\phi \,, \quad f_{K^*}^{us} = f_{K^*} \,,
\end{align}
and $f_V^\alpha = f_V^{\alpha\beta}= 0$ otherwise.

Finally, we construct the total decay width of the heavier HNL $N_j$ as
\begin{align}
\Gamma_{N_j} &= \Gamma_{N_i \gamma} + \Gamma_{\nu_\rho\gamma} + \Gamma_{N_i \ell_\alpha^- \ell_\beta^+} + \Gamma_{\nu_\rho \ell_\alpha^- \ell_\beta^+} \nonumber \\
&\hspace{1.3em} + \Theta_{N_i N_j}(\Gamma_{N_i P^0} + \Gamma_{N_i V^0}) + (1 -  \Theta_{N_i N_j})\Gamma_{N_i + \text{hadr.}} \nonumber \\
&\hspace{1.3em} + \Theta_{\nu_\rho N_j}\Gamma_{\nu_\rho V^0} + (1 - \Theta_{\nu_\rho N_j})\Gamma_{\nu_\rho + \text{hadr.}} \nonumber \\
&\hspace{1.3em} + 2\Theta_{\ell_\rho^- N_j}(\Gamma_{\ell_\rho^- P^+} + \Gamma_{\ell_\rho^- V^+}) + 2(1 - \Theta_{l_\rho^- N_j})\Gamma_{\ell_\rho^- + \text{hadr.}} \,,
\label{eq:total_decay_width}
\end{align}
with implicit summation over all final state flavour indices.
Here, we make use of a function $\Theta_{ab}$ which smoothly interpolates between $1$ for $q^2_{\text{min}} = (m_a - m_b)^2 < Q^2$ and $0$ for $q_{\text{min}}^2 > Q^2$.

Using the expressions above, the expected proper decay length and branching fractions of the heavier HNL $N_2$ can be computed.
Of particular interest are the branching fractions to leptonic $\ell^+\ell^-$ final states, which are considered in Sec.~\ref{sec:analysis} for DV searches at Belle~II.
In Figs.~\ref{fig:branching_1} and \ref{fig:branching_2}, we plot the branching fractions of $N_2$ as a function of $m_{N_2}$ assuming one nonzero Wilson coefficient at a time.
For the operators involving two HNLs (inducing the decays $N_2 \to N_1 X$), the branching fractions are shown for $\delta = 1$ (solid) and $\delta = 0.1$ (dashed).
In some cases, the leptonic branching fractions are trivial, such as $\text{BR}(N_2 \to N_1 e^+ e^-) = 1/2$ for $C_{lN}$ and $\text{BR}(N_2 \to N_1 e^+ e^-/\nu_e e^+e^-) = 1$ for $C_{eN}$ and $C_{lNle}^{ee}$, as long as $m_{N_2} > m_{N_1/\nu} + 2m_e$.
The branching fractions for $N_2 \to \nu_\rho e^+e^-$ and $N_2 \to \nu_e e^\pm \ell_\rho^\mp$ are identical for the coefficients $C_{lNle}^{e\rho}$ and $C_{lNle}^{\rho e}$.
For the operators in Fig.~\ref{fig:branching_2}, the leptonic branching fractions are reduced above the $\pi^0/\pi^\pm$ and $\rho^0/\rho^\pm$ thresholds, and slowly decrease as a function of $m_{N_2}$ as inclusive hadronic decays increase in importance.

%%%%%%%%%%%%%%%%%%%%%%%%%%%%%%%%%%%%%%%%%%%%%%%%%%%%%%%%%
\subsection{HNL kinematics}
\label{app:kinematics}
%%%%%%%%%%%%%%%%%%%%%%%%%%%%%%%%%%%%%%%%%%%%%%%%%%%%%%%%%

Consider the HNL production and decay,
\begin{align}
e^+e^- \to N_1 N_2/\nu N_2\,,\quad N_2 \to N_1 \ell^+\ell^-/\nu\ell^+\ell^- \,,
\end{align}
where the mass splitting between $N_2$ and $\nu/N_1$ is $\delta = (m_{N_2} - m_{N_1/\nu})/m_{N_2}$.
In principle, there are enough constraints to solve all of the kinematics up to an ambiguity in $\delta$.
There are 12 unknowns in the four-momenta of $N_2$ and $N_1/\nu$, but eight constraints arise from four-momentum conservation at the $N_2$ production and decay vertices, i.e.
\begin{align}
\label{eq:production_constraint}
m_{N_1/\nu}^2 &= s + m_{N_2}^2 - 2E_i E_{N_2} + 2|\vec{p}_i||\vec{p}_{N_2}|\cos\theta \,,  \\
m_{N_1/\nu}^2 &= m_{N_2}^2 + m_{\ell\ell}^2 - 2E_{N_2}E_{\ell\ell} + 2|\vec{p}_{N_2}||\vec{p}_{\ell\ell}|\cos\alpha_{\ell\ell} \,.
\label{eq:decay_constraint}
\end{align}
Here, $s = E_i^2 - |\vec{p}_i|^2$ is the squared CM energy, with the laboratory frame energy $E_i = E_{e^+} + E_{e^-}$ and momentum $|\vec{p}_i| = |\vec{p}_{e^+} + \vec{p}_{e^-}|$, and $m_{\ell\ell}^2 = E_{\ell\ell}^2 - |\vec{p}_{\ell\ell}|^2$ is the invariant mass squared of the $\ell^+\ell^-$ pair, with the laboratory frame energy $E_{\ell\ell}$ and momentum $|\vec{p}_{\ell\ell}|$.
The angle $\theta$ is between the momenta of the outgoing $N_2$ and the incoming $e^-$, and $\alpha_{\ell\ell}$ is the pointing angle defined in Eq.~\eqref{eq:pointing_angle}.
Additionally, there are two constraints from the HNL flight direction $\hat p_{N_2}$ (inferred from the $N_2$ decay vertex position) and a constraint due to $N_2$ being on-shell, i.e.
\begin{align}
\label{eq:N2_onshell}
m_{N_2} = \sqrt{E_{N_2}^2 - |\vec{p}_{N_2}|^2} \,.
\end{align}
Combining these constraints, the $N_2$ energy can be expressed as a linear function of the $N_2$ three-momentum by subtracting Eq.~\eqref{eq:decay_constraint} from Eq.~\eqref{eq:production_constraint} and rearranging to find
\begin{align}
\label{eq:EN2_pN2}
E_{N_2} = A |\vec{p}_{N_2}| + B \,,
\end{align}
where,
\begin{align}
A &= \frac{c_\theta |\vec{p}_i| - c_\alpha |\vec{p}_{\ell\ell}|}{E_i - E_{\ell\ell}} \,, \quad B = \frac{s - m_{\ell\ell}^2}{2(E_i - E_{\ell\ell})} \,,
\end{align}
with $c_\theta = \cos\theta$ and $c_\alpha = \cos\alpha_{\ell\ell}$. 
Note that Eq.~\eqref{eq:EN2_pN2} is independent of the unknown mass splitting ratio $\delta$.
Adding Eqs.~\eqref{eq:production_constraint} and \eqref{eq:decay_constraint}, the $N_2$ three-momentum then satisfies
\begin{align}
\label{eq:pN2_quadratic}
a |\vec{p}_{N_2}|^2 + b |\vec{p}_{N_2}| + c = 0 \,,
\end{align}
with,
\begin{align}
a &= 2\delta(2 - \delta)(A^2 - 1) \,, \nonumber \\
b &= 2\left(A \left(2\delta(2-\delta)B - E_i - E_{\ell\ell}) + c_\theta |\vec{p}_i| + c_\alpha |\vec{p}_{\ell\ell}|\right)\right) \,, \nonumber \\
c& = s + m_{\ell\ell}^2 + 2B\left(\delta(2 - \delta)B - E_i - E_{\ell\ell}\right) \,.
\end{align}
From Eq.~\eqref{eq:pN2_quadratic} can be found two $|\vec{p}_{N_2}|$ solutions, $p_{N_2}^\pm = (- b \pm \sqrt{S})/2a$, with the discriminant 
\begin{equation}
S = b^2 - 4 a c \,, 
\label{eq:S}
\end{equation}
%%%
and thus two $E_{N_2}$ solutions ($E_{N_2}^\pm$) from Eq.~\eqref{eq:EN2_pN2}.
Using Eqs.~\eqref{eq:N2_onshell}, this allows us to obtain two $m_{N_2}$ solutions ($m_{N_2}^\pm$), up to an ambiguity in the value of $\delta$.

Solutions for the mass $m_{N_2}$ can also be obtained for the backgrounds considered in this work, i.e. from photon conversions or displaced $K_L^0$ decays.
The are are two scenarios where the equations above do not yield real and positive solutions for $m_{N_2}$.
The first is if the discriminant is negative, $S < 0$, and the second is if the energy solution is smaller than the momentum, i.e. $E_{N_2}^\pm < p_{N_2}^\pm$.
We introduce the variable
\begin{align}
T = \frac{\text{Re}[(m_{N_2}^+)^2 + (m_{N_2}^-)^2]}{2|m_{N_2}^+m_{N_2}^-|}\,,
\label{eq:T}
\end{align}
which is negative for $E_{N_2}^\pm < p_{N_2}^\pm$. The variable has the limiting value $|T| \to 1$ for $S \to 0$.

%%%%%%%%%%%%%%%%%%%%%%%%%%%%%%%%%%%%%%%%%%%%%%%%%%%%%%%%%
\section{\texttt{FeynRules} and \texttt{UFO} model files}
\label{app:feynrules}
%%%%%%%%%%%%%%%%%%%%%%%%%%%%%%%%%%%%%%%%%%%%%%%%%%%%%%%%%

To simulate the HNL signal in Secs.~\ref{sec:constraint} and \ref{sec:DVsens}, we use a \texttt{FeynRules} implementation of the $d = 5$ and $d = 6$ $\nu$SMEFT operators in Table~\ref{tab:vSMEFT-operators}, to generate a \texttt{UFO} file for use in \madgraph. Our \texttt{FeynRules} model file \texttt{vSMEFT\_BelleII.fr} will be made available in the \texttt{FeynRules} model database~\cite{feynrules}.

We note that \madgraph~does not currently support multi-fermion operators containing Majorana fields~\cite{Alwall:2014hca}. However, there are two possible methods to circumvent this issue. Firstly, one may introduce heavy auxiliary and non-propagating scalar and vector fields with renormalisable couplings which reproduce the desired Wilson coefficients of the $\nu$SMEFT operators~\cite{Cottin:2021lzz}. Alternatively, as first used in Ref.~\cite{Bolton:2025tqw}, one may define the HNL fields as Dirac fermions and write the Lagrangian in a form that, for example, the cross section for $e^+e^- \to N_1 \bar{N}_2 + \bar{N}_1 N_2$ reproduces the cross section for $e^+e^- \to N_1 N_2$. In this work, we use the latter method in the \texttt{FeynRules} model file \texttt{vSMEFT\_BelleII.fr}, as outlined in this appendix. However, in the final stages of this project, the \texttt{FeynRules} implementation of Ref.~\cite{Titov:2025bjl} using the auxiliary field method provided an update making available all of the operators in Table~\ref{tab:vSMEFT-operators}, and is therefore a viable alternative.

In \texttt{vSMEFT\_BelleII.fr}, we write the physical HNL fields $N_1$ and $N_2$ as Dirac fermions, carrying lepton number $L = 1$, as:
\begin{lstlisting}
F[21] == {
 ClassName       -> N1,
 SelfConjugate   -> False,
 Mass            -> {MN1, 0.0},
 Width           -> {WN1, 0.0},
 QuantumNumbers  -> {Q -> 0, LeptonNumber -> 1},
 PropagatorType  -> Straight,
 PropagatorArrow -> Forward,
 PDG             -> {6000012},
 ParticleName    -> {"N1"},
 FullName        -> {"Lightest HNL"}
},

F[22] == {
 ClassName       -> N2,
 SelfConjugate   -> False,
 Mass            -> {MN2, 2.0},
 Width           -> {WN2, Automatic},
 QuantumNumbers  -> {Q -> 0, LeptonNumber -> 1},
 PropagatorType  -> Straight,
 PropagatorArrow -> Forward,
 PDG		 -> {6000014},
 ParticleName    -> {"N2"},
 FullName        -> {"Heavier HNL"}
}
\end{lstlisting}
To demonstrate the method, we explore one example of nonzero Wilson coefficients, which have a NP scale $\Lambda$ determined by the input parameter \texttt{LAM}:
\begin{lstlisting}
LAM == {
 ParameterType    -> External,
 ComplexParameter -> False,
 BlockName        -> NPSCALE,
 Value            -> 1000.0,
 TeX              -> \[CapitalLambda],
 Description      -> "Scale of NP"
}
\end{lstlisting}
Firstly, for the operators $Q_{lN}$ and $Q_{HN}$ coupling the HNLs $N_1$ and $N_2$, we define as inputs the real parts (imaginary parts are also inputs) of the corresponding Wilson coefficients:
\begin{lstlisting}
(********** vSMEFT BLOCK PARAMETERS **********)

(********** ClN **********)

ClNee12R == {
 ParameterType    -> External,
 ComplexParameter -> False,
 BlockName	  -> FourFermion,
 OrderBlock	  -> 3,
 Value            -> 0.0,
 InteractionOrder -> {NP, 1},
 TeX              -> Subscript[C, lN, e, e, 1, 2, R],
 Description      -> "Vector LR e-e-N1-N2 coupling real"
},

(********** CHN **********)

CHN12R == {
 ParameterType    -> External,
 ComplexParameter -> False,
 BlockName        -> Higgs,
 OrderBlock       -> 3,
 Value            -> 0.0,
 InteractionOrder -> {NP, 1},
 TeX              -> Subscript[C, HN, 1, 2, R],
 Description      -> "Higgs N1-N2-H-H coupling real"
}
\end{lstlisting}
With these external parameters, in the \texttt{FeynRules} model file we then perform internally the matching in Table~\ref{tab:Matching} to the $\nu$LEFT coefficients in Table~\ref{tab:vLEFT-operators}. Thus, the model file is only valid for processes with energies much below the electroweak scale, as is the case at Belle~II. First, the operator $Q_{HN}$ induces an effective $Z$ coupling to $N_1$ and $N_2$:
\begin{lstlisting}
ZNR12R == {
 ParameterType    -> Internal, 
 Value            -> - vev^2/2 * CHN12R, 
 InteractionOrder -> {QED,1}, 
 TeX              -> Subscript[Z, N, 1, 2, R, R],
 Description      -> "RH N1-N2-Z coupling real"
}
\end{lstlisting}
Then, for example, the induced $\nu$LEFT coefficient of the vector operator $\mathcal{O}_{Ne}^{V,RL}$ is:
\begin{lstlisting}
(********** vLEFT BLOCK PARAMETERS **********)

(************** N1-N2 Couplings **************)

LNeVRL12eeR == {
 ParameterType    -> Internal,
 Value            -> ClNee12R - gZ^2/MZ^2 * ZNR12R * ZeL,
 InteractionOrder -> {NP, 1},
 TeX              -> Subscript[L, Ne, VRL, 1, 2, e, e, R],
 Description      -> "Vector RL N1-N2-e-e coupling real"
}
\end{lstlisting}
There are, of course, other $\nu$LEFT operators induced by $Q_{lN}$ and $Q_{HN}$; both generate the operator $\mathcal{O}_{N\nu}^{V,RL}$ and the latter also the operators $\mathcal{O}_{Ne}^{V,RR}$, $\mathcal{O}_{Nq}^{V,RL}$ and $\mathcal{O}_{Nq}^{V,RR}$. All necessary matching is performed in the model file.

For the operator $\mathcal{O}_{Ne}^{V,RL}$, we demonstrate how the correct behaviour of Majorana HNLs is provided by the model file. The Lagrangian for Majorana HNLs can be written as
\begin{align}
\mathcal{L} &\supset L_{\underset{12ee}{Ne}}^{V,RL}(\bar{N}_1 \gamma_\mu P_R N_2)(\bar{e}\gamma^\mu P_L e) + \text{h.c.} \nonumber \\
& = \frac{1}{2}\big(\bar{N}_1 \gamma_\mu (L_{\underset{12ee}{Ne}}^{V,RL} P_R - L_{\underset{12ee}{Ne}}^{V,RL*}P_L) N_2\big)(\bar{e}\gamma^\mu P_L e) + \text{h.c.} \,,
\label{eq:LNeVRL_Lagrangian}
\end{align}
where we note that
\begin{align}
L_{\underset{21ee}{Ne}}^{V,RL} = L_{\underset{12ee}{Ne}}^{V,RL*} \,.
\end{align}
The first and second lines are equivalent forms, using the relation $\bar{N}_2 \gamma_\mu P_R N_1 = - \bar{N}_1 \gamma_\mu P_L N_2$ for Majorana fields. The amplitude for the process $e^+e^-\to N_1 N_2$ is then
\begin{align}
i\mathcal{M}\big|_{\text{Maj}} &= \frac{i}{2}\big[\bar{u}_{N_1} \gamma_\mu (L_{\underset{12ee}{Ne}}^{V,RL} P_R - L_{\underset{12ee}{Ne}}^{V,RL*}P_L) v_{N_2}\big] \big[\bar{v}_{e} \gamma^\mu P_L u_{e}\big] \nonumber \\
&\hspace{1.3em} + \frac{i}{2}\big[\bar{u}_{N_2} \gamma_\mu (L_{\underset{12ee}{Ne}}^{V,RL*} P_R - L_{\underset{12ee}{Ne}}^{V,RL}P_L) v_{N_1}\big] \big[\bar{v}_{e} \gamma^\mu P_L u_{e}\big] = \frac{i}{2}\big(\mathcal{M}_{12} + \mathcal{M}_{21}\big) \,,
\end{align}
where we note that $\mathcal{M}_{12} = \mathcal{M}_{21}$. The cross section for $e^+e^-\to N_1 N_2$ is proportional to
\begin{align}
d\sigma(e^+e^-\to N_1 N_2)\big|_{\text{Maj}} \propto \frac{1}{4}|\mathcal{M}_{12} + \mathcal{M}_{21}|^2 = |\mathcal{M}_{12}|^2 \,.
\end{align}
This cross section can be reproduced by writing the Lagrangian ($\texttt{L12}$) of the Dirac HNLs in the \texttt{FeynRules} model file like the second line of Eq.~\eqref{eq:LNeVRL_Lagrangian}:
\begin{lstlisting}
L12 := 1/Sqrt[2]/LAM^2 * (
(LNeVRL12eeR+I*LNeVRL12eeI)*({N1bar.Ga[mu].ProjP.N2}.{ebar.Ga[mu].ProjM.e})
-
(LNeVRL12eeR-I*LNeVRL12eeI)*({N1bar.Ga[mu].ProjM.N2}.{ebar.Ga[mu].ProjM.e})
+
(LNeVRL12eeR-I*LNeVRL12eeI)*({N2bar.Ga[mu].ProjP.N1}.{ebar.Ga[mu].ProjM.e})
-
(LNeVRL12eeR+I*LNeVRL12eeI)*({N2bar.Ga[mu].ProjM.N1}.{ebar.Ga[mu].ProjM.e})
)
\end{lstlisting}
The prefactor of $1/\sqrt{2}$ instead of $1/2$ in the Majorana case is the normalisation such that
\begin{align}
d\sigma(e^+e^-\to N_1 \bar{N}_2) + d\sigma(e^+e^-\to \bar{N}_1 N_2)\big|_{\text{Dirac}} &\propto \frac{1}{2}|\mathcal{M}_{12}|^2 + \frac{1}{2}|\mathcal{M}_{21}|^2 = |\mathcal{M}_{12}|^2 \,,
\end{align}
and thus the cross section for $e^+e^- \to N_1 N_2$ is reproduced by adding the cross sections for $e^+e^- \to N_1 \bar{N}_2$ and $e^+e^- \to \bar{N}_1 N_2$. The cross sections for $e^+e^- \to N_1 \bar{N}_2$ and $e^+e^- \to \bar{N}_1 N_2$ are identical, and the non-trivial dependence on the masses ($m_{N_2}$ and $\delta$) in the Majorana scenario is present in both, as the interference between the LH and RH chiral projection operators takes place in $|\mathcal{M}_{12}|^2 = |\mathcal{M}_{21}|^2$.

Using the \texttt{UFO} file \texttt{vSMEFT\_BelleII\_UFO} generated by \texttt{FeynRules}, we can now simulate $e^+e^-\to N_1 N_2$ by generating $e^+e^-\to N_1 \bar{N}_2 + \bar{N}_1 N_2$ in \madgraph. To do so, it is convenient to define the multiparticles \texttt{N1m} and \texttt{N2m} as $N_1+\bar{N}_1$ and $N_2 + \bar{N}_2$, respectively:
\begin{lstlisting}
import model vSMEFT_BelleII_UFO-WCs
define n1m n1 n1~
define n2m n2 n2~
\end{lstlisting}
As we implement all of the $\nu$SMEFT operators in Table~\ref{tab:vSMEFT-operators} in the model file, we include a restriction card \texttt{restrict\_WCs.dat} in the \texttt{UFO} file, allowing to fix the Wilson coefficients not considered to zero and reduce the number of diagrams generated by \madgraph. The process $e^+e^-\to N_1 N_2$ for Majorana HNLs is correctly simulated by running:
\begin{lstlisting}
generate e- e+ > n1m n2m
\end{lstlisting}
We have validated the simulated $e^+e^-\to N_1 N_2$ events and total cross sections for the operators in Table~\ref{tab:vSMEFT-operators} against the analytical cross section formulae in App.~\ref{subsec:production}.

The Lagrangian $\texttt{L12}$ in the model file for Dirac HNLs also reproduces the behaviour of the Majorana HNL decay $N_2 \to N_1 e^+e^-$. The amplitude for the process $N_2 \to N_1 e^+e^-$ from the Lagrangian in Eq.~\eqref{eq:LNeVRL_Lagrangian} is
\begin{align}
i\tilde{\mathcal{M}}\big|_{\text{Maj}} &= \frac{i}{2}\big[\bar{u}_{N_1} \gamma_\mu (L_{\underset{12ee}{Ne}}^{V,RL} P_R - L_{\underset{12ee}{Ne}}^{V,RL*}P_L) u_{N_2}\big] \big[\bar{u}_{e} \gamma^\mu P_L v_{e}\big] \nonumber \\
&\hspace{1.3em} + \frac{i}{2}\big[\bar{v}_{N_2} \gamma_\mu (L_{\underset{12ee}{Ne}}^{V,RL*} P_R - L_{\underset{12ee}{Ne}}^{V,RL}P_L) v_{N_1}\big] \big[\bar{u}_{e} \gamma^\mu P_L v_{e}\big] = \frac{i}{2}\big(\tilde{\mathcal{M}}_{12} + \tilde{\mathcal{M}}_{21}\big) \,,
\end{align}
where, as before, $\tilde{\mathcal{M}}_{12} = \tilde{\mathcal{M}}_{21}$. The decay rate for $N_2 \to N_1 e^+e^-$ is thus proportional to
\begin{align}
d\Gamma(N_2 \to N_1 e^+e^-)\big|_{\text{Maj}} \propto \frac{1}{4}|\tilde{\mathcal{M}}_{12} + \tilde{\mathcal{M}}_{21}|^2 = |\tilde{\mathcal{M}}_{12}|^2 \,.
\end{align}
From the Lagrangian \texttt{L12}, the decay rates for $N_2 \to N_1 e^+ e^-$ and $\bar{N}_2 \to \bar{N}_1 e^+ e^-$ are 
\begin{align}
d\Gamma(N_2 \to N_1 e^+e^-)\big|_{\text{Dirac}} &\propto \frac{1}{2}|\tilde{\mathcal{M}}_{12}|^2 \,, \nonumber \\
d\Gamma(\bar{N}_2 \to \bar{N}_1 e^+e^-)\big|_{\text{Dirac}} &\propto \frac{1}{2}|\tilde{\mathcal{M}}_{21}|^2 \,.
\end{align}
Thus, the correct decay rate for $N_2 \to N_1 e^+ e^-$ in the Majorana HNL scenario is found by adding the decay rates of $N_2 \to N_1 e^+e^-$ and $\bar{N}_2 \to \bar{N}_1 e^+e^-$. In \madgraph:
\begin{lstlisting}
generate n2m > n1m e+ e- 
\end{lstlisting}

Finally, it is possible to simulate both the production \textit{and} decay of $N_2$ as:
\begin{lstlisting}
generate e- e+ > n1m n2m, n2m > n1m e+ e-
\end{lstlisting}
which we use in Secs.~\ref{sec:constraint} and \ref{sec:DVsens} to generate the kinematics of the final state $e^+e^-$. One caveat with the above command is that $e^+e^- \to N_1 \bar{N}_2$ is followed by $\bar{N}_2 \to \bar{N}_1 e^+e^-$ and $e^+e^- \to \bar{N}_1 N_2$ by $N_2 \to N_1 e^+e^-$. The consequence is that the effective decay rate is a factor of $2$ smaller than required. However, as \madgraph~yields the cross sections multiplied by the branching fraction of the indicated decay, the factor of $2$ cancels with the total $N_2$ width. Additionally, non-trivial dependencies on the masses in the Majorana scenario are still present, as the interference between the LH and RH chiral projection operators occurs in both the production ($|\mathcal{M}_{12}|^2 = |\mathcal{M}_{21}|^2$) and decay ($|\tilde{\mathcal{M}}_{12}|^2 = |\tilde{\mathcal{M}}_{21}|^2$). We finally note that the method above permits the generation of LNV processes arising from $N_2$ production and decay (or involving virtual $N_2$), e.g. $\bar{u}d \to e^- N_2$ followed by $N_2 \to e^- \bar{u}d$ via the $\nu$LEFT operator $\mathcal{O}_{eNud}^{V,RL}$ induced by $Q_{HNe}$. However, we do not consider LNV processes in this work.

%%%%%%%%%%%%%%%%%%%%%%%%%%%%%%%%%%%%%%%%%%%%%%%%%%%%%%%%%
\section{Background estimate details}
\label{app:bkg_estimates}
%%%%%%%%%%%%%%%%%%%%%%%%%%%%%%%%%%%%%%%%%%%%%%%%%%%%%%%%%

%%%%%%%%%%%%%%%%%%%%%%%%%%%%%%%%%%%%%%%%%%%%%%%%%%%%%%%%%
\subsection{Photon conversion background}
\label{app:conversion}
%%%%%%%%%%%%%%%%%%%%%%%%%%%%%%%%%%%%%%%%%%%%%%%%%%%%%%%%%

In this appendix, we estimate the photon conversion background in the Belle~II detector, following closely the background estimation for dark photon searches at Belle~II in Ref.~\cite{Jaeckel:2023huy}.
We devise cuts on the kinematics of the outgoing $\ell^+\ell^-$ pair that strongly reduce this background while maintaining reasonably high signal efficiency in the HNL parameter space of interest.

The dilepton photoproduction process on a nuclear target $A$, i.e.
\begin{align}
\gamma(p_\gamma) + A(p_A) \to \ell^+(p_{\ell^+}) +\ell^-(p_{\ell^-}) + A(p_A') \,,
\end{align}
receives contributions from the Bethe-Heitler (BH) and timelike Compton scattering (TCS) processes, which depend on the invariant mass $m_{\ell\ell}$, with $m_{\ell\ell}^2 = (p_{\ell^+} + p_{\ell^-})^2$, and momentum transfer $q^2 = (p_A' - p_A)^2 = 2m_A(m_A - E_A') < 0$.
The BH cross section can be written as
\begin{align}
\label{eq:Bethe-Heitler}
\frac{d\sigma_{\text{BH}}}{dm_{\ell\ell} dq^2} = \frac{8\alpha^3 m_{\ell\ell}\beta_\ell}{(\mathfrak{s} - m_A^2)^2 q^4 (m_{\ell\ell}^2 - q^2)^4}\big|F_1(q^2)\big|^2\bigg[C_1 + C_2 \frac{1}{\beta_\ell}\log\frac{1-\beta_\ell}{1+\beta_\ell}\bigg] \,,
\end{align}
where $\mathfrak{s} = (p_\gamma + p_A)^2 =m_A (m_A + 2E_\gamma)$ with $E_\gamma$ the photon energy in the laboratory frame; $\beta_\ell = \sqrt{1 - 4m_\ell^2/m_{\ell\ell}^2}$; 
$F_{1}(q^2)$ is the charge form factor of the nucleus $A$, assuming a spin of $1/2$, i.e.
\begin{align}
\langle{p_A'|J_\mu(0)|p_A\rangle} = F_1(q^2) \bar{u}(p_A') \gamma_\mu u(p_A) \,;
\end{align}
and we have defined the factors
\begin{align}
C_1 &= q^2(\mathfrak{s} - m_A^2)(\mathfrak{s} - m_A^2 - m_{\ell\ell}^2 +q^2)(m_{\ell\ell}^4 + 6m_{\ell\ell}^2 q^2 + q^4 + 4m_\ell^2 m_{\ell\ell}^2)\,, \nonumber \\
& \hspace{1.3em} + \frac{1}{2}(m_{\ell\ell}^2 - q^2)^2(q^2 + 2m_A^2)\big((m_{\ell\ell}^2 + q^2)^2 + 4m_\ell^2 m_{\ell\ell}^2\big) \,,\nonumber \\
C_2 &= q^2 (\mathfrak{s} - m_A^2)(\mathfrak{s} - m_A^2 - m_{\ell\ell}^2 + q^2) (m_{\ell\ell}^4 + q^4 + 4 m_\ell^2 (m_{\ell\ell}^2 + 2 q^2 - 2m_\ell^2)) \nonumber \\
& \hspace{1.3em} + \frac{1}{2}(m_{\ell\ell}^2 - q^2)^2 (q^2 + 2m_A^2)\big(m_{\ell\ell}^4 + q^4 + 4m_\ell^2 (m_{\ell\ell}^2 - 2m_\ell^2)\big) \,.
\end{align}
We note that, for the photon energies relevant for the estimation of the photon conversion background, the spin of the nucleus has a negligible impact on the scattering cross section.
The charge form factors for the nuclei in the Belle~II detector, detailed shortly, are taken from Ref.~\cite{DeVries:1987atn}.

The TCS cross section is instead given, for near-real and -forward kinematics (satisfying $m_{\ell\ell}^2, |q^2| \ll \mathfrak{s}$) by
\begin{align}
\label{eq:TCS}
\frac{d\sigma_{\text{TCS}}}{dm_{\ell\ell} dq^2} = \frac{4m_A^2\alpha^3 \beta_\ell}{(\mathfrak{s} - m_A^2)^2 m_{\ell\ell}} \bigg(1 - \frac{\beta_\ell^2}{3}\bigg)\bigg|\frac{f(\nu)}{\alpha}\bigg|^2 \,,
\end{align}
with $\nu \equiv p_\gamma \cdot (p_A + p_A')/(2m_A) = E_\gamma - (m_{\ell\ell}^2 - q^2)/(4m_A)$.
Here, $f(\nu)$ is the spin-averaged forward Compton amplitude, with a real part given by
\begin{align}
\text{Re} f(\nu) = -\frac{Z^2 \alpha}{m_A} + \frac{\nu^2}{2\pi^2}\fint_0^\infty d\nu'\frac{\sigma_{\gamma A}(\nu')}{\nu^{\prime 2} - \nu^2} \,,
\end{align}
where $\sigma_{\gamma A}$ is the unpolarised cross section for the total photoabsorption on the nucleus $A$ with atomic number $Z$ and mass $m_A$, and the Cauchy principal value of the integral is taken.
The optical theorem instead yields the imaginary part
\begin{align}
\text{Im} f(\nu) = \frac{\nu}{4\pi} \sigma_{\gamma A}(\nu) \,.
\end{align}
As in Ref.~\cite{Jaeckel:2023huy}, we assume that the total photoabsorption cross section per nucleon $\sigma_{\gamma A}/A$ is the same for all nuclei above the pion threshold, $\nu \geq m_\pi$~\cite{Ahrens:1985hxw,Hutt:1999pz}.
Thus, we take $\sigma_{\gamma A} = A \sigma_{\gamma p}$, with the cross section $\sigma_{\gamma p}$ for the proton taken from Ref.~\cite{Gryniuk:2015eza}.
For low energy photons, $E_\gamma \lesssim 0.1$~GeV, the amplitude is dominated by the elastic part $f(\nu) = - Z^2\alpha/m_A$, corresponding to the contribution of protons in the nucleus as point-like charges.
For intermediate energy photons, $0.1 \lesssim E_\gamma/\text{GeV} \lesssim 2$, the amplitude is dominated by resonances in $\sigma_{\gamma A}$.
Finally, for photons with $E_\gamma \gtrsim 2~\text{GeV}$, the amplitude tends towards behaviour well described by Regge theory~\cite{Donnachie:1992ny}.
We assume that the TCS cross section is given by Eq.~\eqref{eq:TCS} for momentum transfers up to $|q^2| = 1/R_A^2$, where $R_A = 1.2 A^{1/3}~\text{fm}$ is the nuclear radius, and neglect the process for larger values.

We now investigate the possible sources of single photons for the photon conversion process, so that the final-state signature of a displaced $\ell^+\ell^-$ and missing energy is replicated.
We first consider the conversion photons produced by the processes $e^+e^- \to \gamma\gamma$ and radiative Bhabha scattering $e^+e^-\to e^+e^-\gamma$.
To be consistent with our signal signature, the second photon and the $e^+e^-$ pair in these two final states, respectively, must not be detected.
To interact in the Belle~II pixel vertex detector (PXD), silicon vertex detector (SVD) or central drift chamber (CDC) and produce a detectable 2-track vertex, the conversion photon must have a polar angle in the laboratory frame in the range $17^\circ < \theta_\gamma < 150^\circ$.

For $e^+e^-\to\gamma\gamma$, the other outgoing photon will always be within the coverage of the electromagnetic calorimeter (ECL), $12.4^\circ < \theta_\gamma < 155.1^\circ$, but may nonetheless escape detection.
To estimate the probability for this, we rely on Ref.~\cite{Belle-II-photon-eff}, which reports that the efficiency to detect a photon with $E_\gamma > 0.5$~GeV in the barrel region angular range $50^\circ < \theta_\gamma < 110^\circ$ is at least 99\%.
By requiring that the conversion photon is within $43.6^\circ < \theta_\gamma < 101.5^\circ$, we ensure that the other photon will always satisfy $50^\circ < \theta_\gamma < 110^\circ$, and we can assume a 1\% probability for it to escape undetected.
We note that the endcap calorimeter regions can also be used, increasing the signal acceptance, but we conservatively ignore them due to lack of public information regarding their efficiency for photon detection.
Simulating the process $e^+e^-\to\gamma\gamma$ in \madgraph, we obtain an effective cross section in this region of phase space of $\sigma_{\gamma\gamma} = 0.71$~nb.
As either of the two final-state photons can induce the conversion process, we multiply this cross section by a factor of two.
From the $2\to 2$ kinematics, the photon energies are given by $\gamma(1 - \beta \cos\theta_\gamma)E_\gamma = \sqrt{s}/2$, with the boost factors $\gamma = 1/\sqrt{1-\beta^2}$ and $\beta = (E_{e^-} - E_{e^+})/(E_{e^-} + E_{e^+}) = 0.273$.
The conversion photons therefore have energies in the range $4.8~\text{GeV} < E_\gamma < 6.3~\text{GeV}$.

To estimate the background from radiative Bhabha scattering $e^+e^-\to e^+e^-\gamma$, we simulate the process in \madgraph~with a nonzero electron mass to regulate the $t$-channel divergence and a generator-level cut on the photon energy, $E_\gamma > 1$~MeV, to evade the infrared divergence from soft photon emission.
In addition, less energetic photons can not induce the $\gamma A \to e^+ e^- A$ process.
For events in which the outgoing photon satisfies $43.6^\circ < \theta_\gamma < 101.5^\circ$ and the outgoing $e^+e^-$ do not enter the SVD or CDC, i.e., $\theta_{e^\pm} < 17^\circ$ or $\theta_{e^\pm} > 150^\circ$, we find an effective cross section of $\sigma_{e^+e^-\gamma} = 4.0$~nb.
Unlike $e^+e^-\to\gamma\gamma$, the distribution of photon energies from $e^+e^-\to e^+e^-\gamma$ is peaked at small $E_\gamma$ values and falls quickly as $E_\gamma$ increases.

Additional sources of conversion photons are also considered.
From the background studies for the monophoton signature at Belle~II~\cite{Belle-II:2018jsg} (Sec.~16.2) and at BABAR~\cite{BaBar:2014zli}, we conclude that other photon sources, e.g. $e^+e^- \to f\bar{f}\gamma$ ($f = \mu, \tau, \nu, q$), $e^+e^- \to \text{hadr.}+\gamma$, $e^+e^- \to \gamma\gamma\gamma$ and $e^+e^- \to e^+e^-\gamma\gamma$, as well as photons from machine backgrounds, are negligible with respect to $e^+e^-\to \gamma\gamma$ and $e^+e^- \to e^+e^-\gamma$, particularly after the cuts considered in the following.
Furthermore, we have verified that the number of photons from neutral pion production and decay is relatively small.
For example, we find through simulation in \madgraph~that the process $e^+e^-\to \gamma\pi^0 \to \gamma\gamma\gamma$ has a total cross section of $\sigma_{\gamma\gamma\gamma} = 2.8\times 10^{-6}$~nb, as it is suppressed due to the effective pion photon form factor scaling as $F(q^2) = \sqrt{2}f_\pi/q^2$, where $f_\pi = 0.13$~GeV is the pion decay constant.
Thus, this process can be safely neglected.
The process $e^+e^- \to e^+e^- \pi^0 \to e^+e^-\gamma\gamma$ is less suppressed, with a total cross section of $\sigma_{e^+e^-\gamma\gamma} = 1.5$~nb, but the effective cross section for one photon being in the range $43.6^\circ < \theta_\gamma < 101.5^\circ$ and the other being outside the acceptance of the ECL is considerably smaller, at $\sigma_{e^+e^-\gamma\gamma} = 0.45$~nb.
Photons from radiative Bhabha scattering always dominate the total photon conversion rate with respect to this process.

We now estimate the number of background events at a given transverse displacement $R$ in the Belle~II detector.
Firstly, we model the Belle~II detector in the plane transverse to the beam axis as concentric layers of materials as in Table~1 of Ref.~\cite{Jaeckel:2023huy}.
These consist of the beam pipe section (Au, Be and C layers), the SVD and PXD (N and Si), CDC inner wall (C) and the CDC main volume (W, Al, C, He).
For a photon conversion in a given layer, corresponding to a given displacement, we require that the photon is not absorbed by the preceding layers.
For layer $i$, the survival probability is
\begin{align}
\label{eq:survival_prob}
P_\gamma^i = \exp\left(- n_i \sigma_\gamma^i \frac{z_i}{\sin\theta_\gamma}\right) \,,
\end{align}
where $n_i$ and $z_i$ are the number density and radial width of the layer, respectively, both taken from Table~1 of Ref.~\cite{Jaeckel:2023huy}, and $\sigma_\gamma^i$ is the total photoabsorption cross section, taken from Ref.~\cite{NIST}.
The probability of the photon conversion $\gamma A \to \ell^+\ell^- A$ in the layer $i$, differential in the invariant mass $m_{\ell\ell}$ and momentum transfer $q^2$, is given by
\begin{align}
\label{eq:conv_prob}
\frac{dp_i}{dm_{\ell\ell}dq^2}(E_\gamma,\cos\theta_\gamma) = \left(\prod_j^{i - 1} P_\gamma^j\right)\left(1 - P_\gamma^i\right)\frac{1}{\sigma_\gamma^i} \frac{d\sigma_i}{dm_{\ell\ell} dq^2} \,,
\end{align}
where the differential cross section is the sum of the BH and TCS cross sections in Eqs.~\eqref{eq:Bethe-Heitler} and \eqref{eq:TCS}, respectively.
We make explicit that the conversion probability is a function of the photon energy $E_\gamma$ (through $\sigma_\gamma^i$ and $d\sigma_i/dm_{\ell\ell}dq^2$) and angle $\theta_\gamma$ (through $P_\gamma^i$).
Finally, we obtain the differential number of events in only the invariant mass $m_{\ell\ell}$ by integrating Eq.~\eqref{eq:conv_prob} over the kinematically-allowed $q^2$ range, i.e., 
\begin{align}
\label{eq:conv_prob_2}
\frac{dp_i}{dm_{\ell\ell}}(E_\gamma,\cos\theta_\gamma) = \int_{q^2_-(m_{\ell\ell})}^{q^2_+(m_{\ell\ell})}dq^2\frac{dp_i}{dm_{\ell\ell}dq^2}\Theta(\text{cuts}) \,,
\end{align}
with the maximum and minimum values
\begin{align}
q_\pm^2 (m_{\ell\ell}) = \frac{1}{2}\left[2m_A^2 + m_{\ell\ell}^2 - \mathfrak{s} - \frac{m_A^2(m_A^2 - m_{\ell\ell}^2)}{\mathfrak{s}} \pm \frac{(\mathfrak{s} - m_A^2)\lambda^{1/2}(\mathfrak{s}, m_A^2, m_{\ell\ell}^2)}{\mathfrak{s}}\right]\,.
\end{align}
The allowed range of the invariant mass is $m_{\ell\ell} \in [2m_\ell, \sqrt{\mathfrak{s}}-m_A]$.
We include in Eq.~\eqref{eq:conv_prob_2} a Heaviside theta function to account for the cuts considered in the following.

We now estimate the differential number of photon conversion events in layer $i$ as
\begin{align}
\label{eq:Nconv_differential}
\frac{dN_i}{dm_{\ell\ell}} = \mathcal{L} \sum_{s = \gamma\gamma, \, e^+e^-\gamma} n_s \sigma_s \frac{1}{N_{\text{ev}}}\sum_a \frac{dp_i}{dm_{\ell\ell}}(E_\gamma^a,\cos\theta_\gamma^a) \,,
\end{align}
where $\mathcal{L}$ is the integrated luminosity, $N_{\text{ev}}$ is the total number of simulated events included, and $\sigma_s$ is the cross section for the photon production process $s$ after applying the cuts described above.
We sum over the two photon sources considered and multiply by a factor $n_s$, with $n_{\gamma\gamma} = 2\times 0.01$ accounting for the two photons produced from $e^+e^-\to\gamma\gamma$ and the second-photon veto efficiency, and $n_{e^+e^-\gamma} = 1$.
We also average the differential photon conversion probability over the energies and angles of the simulated photon events after the cuts described above.

\begin{figure}
    \centering
    \includegraphics[width=0.49\linewidth]{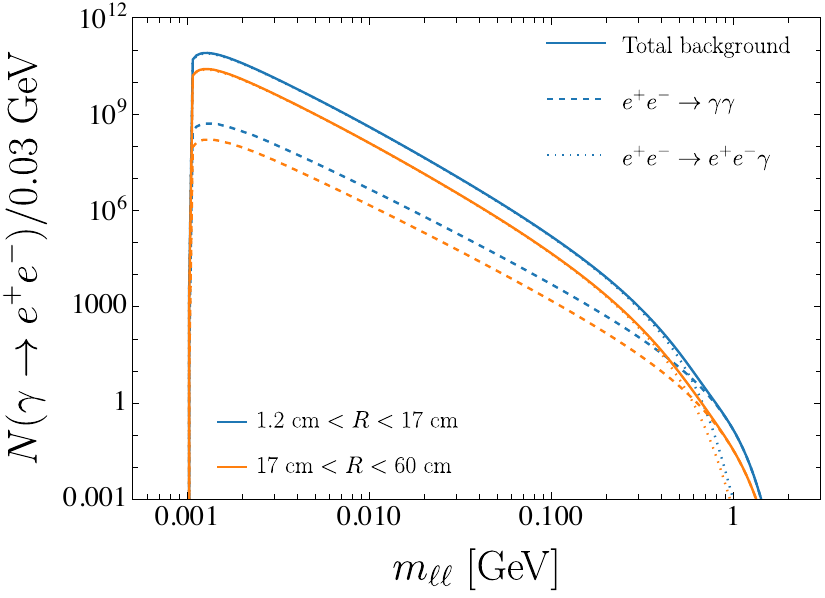}
\includegraphics[width=0.49\linewidth]{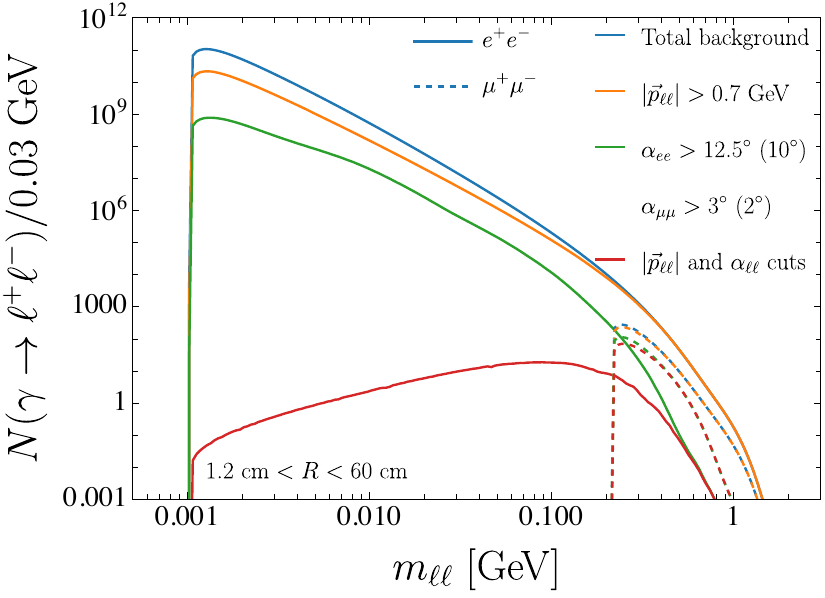}
    \caption{(Left) Number of photon conversion to $e^+e^-$ events per $0.03$~GeV dilepton-mass bin in the full Belle~II dataset of $50~\textrm{ab}^{-1}$ as a function of the invariant mass $m_{\ell\ell}$. The blue and orange lines indicate the expected background for $43.6^\circ < \theta_{\text{DV}} < 101.5^\circ$ in the SVD and PXD to the CDC inner wall (transverse displacement $1.2~\text{cm}< R < 60~\text{cm}$) and CDC only ($17~\text{cm}< R < 60~\text{cm}$), respectively. The dashed and dotted lines show the contributions from $e^+e^-\to\gamma\gamma$ and $e^+e^- \to e^+e^-\gamma$, respectively. (Right) The number of photon conversion events to $e^+e^-$ (solid) and $\mu^+\mu^-$ (dashed) for $50~\text{ab}^{-1}$, $43.6^\circ < \theta_{\text{DV}} < 101.5^\circ$ and $1.2~\text{cm}< R < 60~\text{cm}$ (blue) with the cut $|\vec{p}_{\ell\ell}| > 0.7$~GeV (orange), $\alpha_{\ell\ell}$ cuts in Eq.~\eqref{eq:cuts-appendix} (green) and all cuts simultaneously (red).}
    \label{fig:conversion_bkg}
\end{figure}

In Fig.~\ref{fig:conversion_bkg} (left), we plot the photon conversion $e^+e^-$  yield per $0.3$~GeV dilepton-mass bin for an integrated luminosity of $\mathcal{L} = 50~\text{ab}^{-1}$ in two ranges of the transverse displacement $R$: the first range is $1.2~\text{cm} < R < 17~\text{cm}$ (blue), which covers the SVD and PXD to the CDC inner wall.
The second is $17~\text{cm} < R < 60~\text{cm}$ (orange), corresponding to the CDC main volume only.
The former contains a factor $\sim 4$ more background due to the larger material density.
We do not consider the range $R > 60~\text{cm}$, where the efficiency of track-reconstruction efficiency is low and its quality is degraded, nor $R < 1.2~\text{cm}$, where background from promptly produced tracks becomes high.
The $e^+e^-\to\gamma\gamma$ (dashed) and radiative Bhabha scattering (dotted) processes can be seen to dominate the photon conversion background above and below $m_{\ell\ell}\sim 0.6$~GeV, respectively.

We now explore possible cuts to reduce the photon conversion background while retaining high efficiency for the HNL signal.
A straightforward approach to reduce the radiative Bhabha scattering contribution is to place a cut on the momentum $|\vec{p}_{\ell\ell}|$ of the outgoing $\ell^+\ell^-$ pair.
The pair energy $E_{\ell\ell}$ and momentum are given by 
\begin{align}
E_{\ell\ell} = E_\gamma + \frac{q^2}{2m_A}\,, \quad |\vec{p}_{\ell\ell}| = \sqrt{E_{\ell\ell}^2 - m_{\ell\ell}^2} \,.
\end{align}
Since $q^2 < 0$, the pair momentum must satisfy $|\vec{p}_{\ell\ell}| < E_\gamma$.
Both the BH and TCS scattering cross sections are peaked at $q^2$ values close to the kinematic threshold $|q^2_+| \ll 2m_A E_\gamma$, so most outgoing $\ell^+\ell^-$ are expected to have $E_{\ell\ell}\approx E_\gamma$.
The cut $|\vec{p}_{\ell\ell}| > |\vec{p}_{\ell\ell}|_{\text{cut}}$ eliminates the contribution of photons with energies $E_\gamma < \sqrt{|\vec{p}_{\ell\ell}|_{\text{cut}}^2 + m_{\ell\ell}^2}$.
In Fig.~\ref{fig:conversion_bkg} (right), the blue curves show the total $e^+e^-$ (solid) and $\mu^+\mu^-$ (dashed) conversion backgrounds per 0.03~GeV for $43.6^\circ < \theta_{\text{DV}} < 101.5^\circ$ and $1.2~\text{cm} < R < 60~\text{cm}$ without any further cuts, while the orange curves show the backgrounds with $|\vec{p}_{\ell\ell}| > 0.7$~GeV applied.
The reduction in the backgrounds is modest, because the photon conversion rate increases with $E_\gamma$ and the contributions of the energetic photons survive the cut. The $e^+e^- \to \gamma\gamma$ induced background is essentially unaffected by the $|\vec{p}_{\ell\ell}|$ cut.

Another variable of interest is the pointing angle $\alpha_{\ell\ell}$, defined as the angle between the $\ell^+\ell^-$ pair momentum and the direction from the interaction point to the DV, as in Eq.~\eqref{eq:pointing_angle}.
For the photon conversion process it is given by,
\begin{align}
\label{eq:pointing_angle_photon_conversion}
\cos\alpha_{\ell\ell} &= \hat{p}_\gamma \cdot \hat{p}_{\ell\ell} = \frac{E_\gamma E_{\ell\ell} - \frac{1}{2} (m_{\ell\ell}^2 - q^2)}{E_\gamma|\vec{p}_{\ell\ell}|} \,.
\end{align}
As the BH and TCS cross sections peak at small $|q^2|$,
% it is evident that
the $\ell^+\ell^-$ pair are produced with $\cos\alpha_{\ell\ell} \approx 1$ for photons with $E_\gamma \gg m_{\ell\ell}$, which is especially the case for the background induced by $e^+e^- \to \gamma\gamma$.
We therefore consider the impact of the cut $\alpha_{ee} > 12.5^\circ~(10^\circ)$ in the transverse displacement range $1.2~\text{cm}< R < 17~\text{cm}$ ($17~\text{cm}< R < 60~\text{cm}$), shown as green lines in Fig.~\ref{fig:conversion_bkg} (right).
The $\alpha_{ee}$ cut reduces the $e^+e^-$ background by around two orders of magnitude for $m_{\ell\ell} > 0.1$~GeV, where more energetic photons dominate the conversion rate.
For $m_{\ell\ell} < 0.1$~GeV, less energetic photons (which generally induce larger pointing angles) from radiative Bhabha scattering become more important and the reduction becomes less prominent. For the $\mu^+\mu^-$ channel, we consider the looser cut of $\alpha_{\mu\mu} > 3^\circ~(2^\circ)$, because the di-muon threshold already results in far fewer background events compared to the $e^+e^-$ channel for $m_{\ell\ell} < 0.5$~GeV.

Finally, we find that a combination of cuts on $|\vec{p}_{\ell\ell}|$ and $\alpha_{\ell\ell}$ lead to an increased reduction in the background by many orders of magnitude, as shown by the red lines in Fig.~\ref{fig:conversion_bkg} (right).
This can be understood as follows: events which survive the minimum $|\vec{p}_{\ell\ell}|$ cut are induced by more energetic photons, which lead to smaller values of $\alpha_{\ell\ell}$.
Thus, the cut on $\alpha_{\ell\ell}$ is more effective at reducing the remaining background after a cut on $|\vec{p}_{\ell\ell}|$.
To summarise, we find that the following cuts:
\begin{gather}
43.6^\circ < \theta_{\text{DV}} < 101.5^\circ  \,, \nonumber \\ 
|\vec{p}_{\ell\ell}| > 0.7~\text{GeV}  \,, \nonumber \\
\alpha_{\ell\ell} > \begin{cases}
12.5^\circ~(10^\circ) &~\ell\ell = ee \\
3^\circ~(2^\circ) &~\ell\ell = \mu\mu
\end{cases} \,,
\label{eq:cuts-appendix}
\end{gather}
can be placed on the outgoing $\ell^+\ell^-$ pair to reduce the photon conversion background in the transverse displacement range $1.2~\text{cm} < R < 60~\text{cm}$ and invariant mass region $m_{\ell\ell} < 0.5$~GeV to $\mathcal{O}(100)$ events for the $e^+e^-$ and $\mu^+\mu^-$ channels, as summarised in Tables~\ref{tab:background_reductions} and~\ref{tab:backgrounds} in Sec.~\ref{sec:DVbackground}. We refer to these as \textit{tight} cuts, with \textit{loose} cuts also considered.

%%%%%%%%%%%%%%%%%%%%%%%%%%%%%%%%%%%%%%%%%%%%%%%%%%%%%%
\subsection{$K_L^0$ background}
\label{app:KL-background}
%%%%%%%%%%%%%%%%%%%%%%%%%%%%%%%%%%%%%%%%%%%%%%%%%%%%%%

This appendix describes in detail our estimates for the background yield from semileptonic $K_L^0$ decays. We start with the $K_L^0$ production process $e^+ e^-\to K^0_L K^0_L$, for which we take the cross section to be the same as for $e^+e^- \to e^+e^- K_S^0 K_S^0$, studied in Ref.~\cite{Belle:2013eck}.
Thus, from Fig.~4 of Ref.~\cite{Belle:2013eck}, we expect the production of approximately $10^5$ events per $\textrm{ab}^{-1}$ within the tracking volume.

To estimate the $K_L^0$ momenta, and hence their decay lengths, we assume that the process is dominated by the $f_2'(1525)$ resonance, consistent with the results of Ref.~\cite{Belle:2013eck}.
Using \texttt{FeynRules}~\cite{Alloul:2013bka}, we write a \texttt{UFO}~\cite{Degrande:2011ua} file introducing the spin-2 meson $f_2'$ with  an effective coupling to photons, enabling $\gamma\gamma$-fusion to $f_2'$.
We then simulate the process $e^+e^- \to e^+e^-f_2'$ in \madgraph~\cite{Alwall:2014hca} to obtain the outgoing $f_2'$ kinematics.
In the $f_2'$ rest frame, the polar angle $\theta_K$ of the outgoing $K_L^0$ pair from $f_2' \to K_L^0K_L^0$ is d-wave dominated, following a $d\Gamma/d\cos\theta_K \propto \sin^4\theta_K$ distribution.
We sample the $K_L^0$ momenta in the $f_2'$ rest frame and boost to the laboratory frame according to the $f_2'$ momentum, yielding an average $K_L^0$ transverse momentum of 0.63~GeV.
The average $K_L^0$ transverse flight distance is then $\left<R\right>\approx 19$~m, 
and its probability to decay within $1.2~\text{cm} < R < 60~\text{cm}$ is 3.0\%.
For simplicity, we adopt an approach similar to that of Ref.~\cite{Dib:2019tuj} and take the probability for detecting the $K_L^0$ decay vertex within this volume to be 50\%.
Thus, the probability for detecting at least one of the two $K_L^0$ meson vertices is 3.0\%.
While detection of both vertices would lead to rejection of the event, the probability for this is small and hence neglected.
However, the event may be rejected if the other $K_L^0$ is detected via its interaction in the ECL or KLM, with a probability of 70\%~\cite{Stoetzer:2022}.
In summary, we expect that approximately $900$ vertices are detected per $\textrm{ab}^{-1}$.

The background in the $e^+e^-$ channel arises mainly from the decay $K_L^0\to \pi^\pm e^\mp\nu$, which has a branching fraction of 40\%~\cite{ParticleDataGroup:2024cfk}.
The probability of misidentifying the pion as an electron is approximately 0.25\%~\cite{Hanagaki:2001fz}.
Combining the above factors, we conclude that the background in the $e^+e^-$ channel can be reduced to approximately 0.90 events per $\textrm{ab}^{-1}$.
In the $\mu^+\mu^-$ signal channel, the background arises from $K_L^0\to \pi^\pm \mu^\mp\nu$, which has a branching fraction of 27\%~\cite{ParticleDataGroup:2024cfk}.
At the Belle experiment, the probability of misidentifying the pion as a muon was 3.5\% for track momenta above 0.7~GeV~\cite{Abashian:2002bd}.
This fake rate should be somewhat smaller at Belle~II, but we take it to be the same as for Belle.
Thus, the expected background in the $\mu^+\mu^-$ channel is 8.5 events per $\textrm{ab}^{-1}$.
The same branching fractions and pion fake rates lead to an expected background in the $e^\pm \mu^\mp$ channel of 13 events per $\textrm{ab}^{-1}$.

\begin{figure}[t!]
\centering
\includegraphics[width=0.475\linewidth]{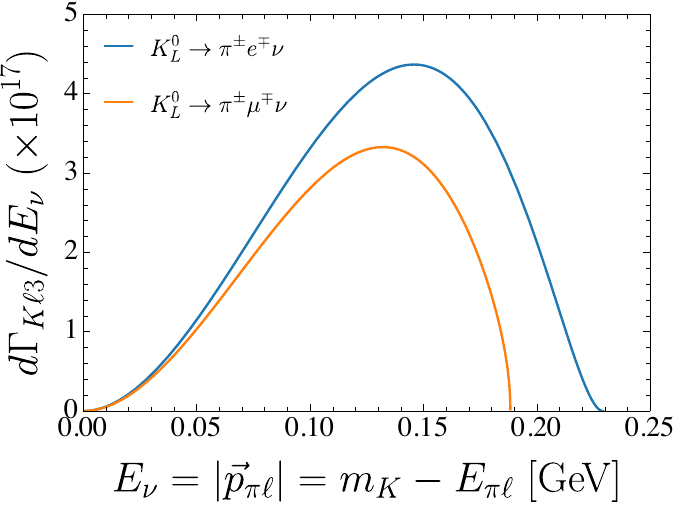}
\includegraphics[width=0.475\linewidth]{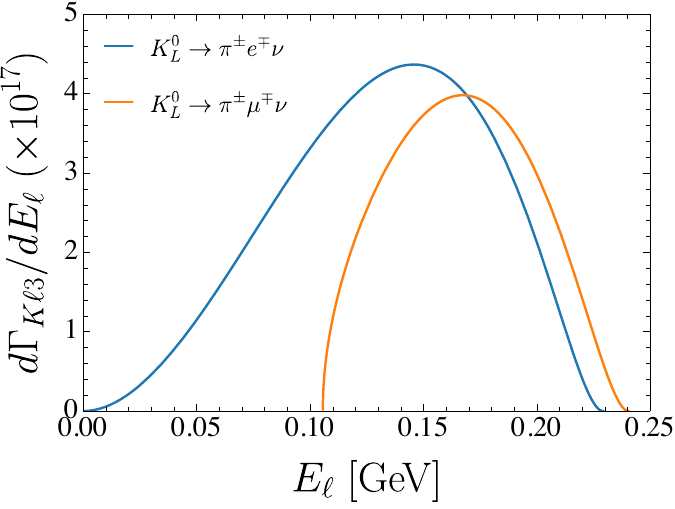}
\caption{For the decay $K_L^0 \to \pi^\pm \ell^\mp \nu$, the distributions of the neutrino energy $E_\nu$ (left), equal to the pair momentum $|\vec{p}_{\pi\ell}|$, and charged lepton energy $E_\ell$ (right) in the $K_L^0$ rest frame, for $\ell = e$ (blue) and $\ell = \mu$ (orange).}
\label{fig:KL_decays}
\end{figure}

We note that the background estimates above are before the cuts in Table~\ref{tab:DV_cuts} are applied.
To estimate the reduction in the background after these cuts, we make use of the distributions in the neutrino energy $E_\nu$ and charged lepton energy $E_\ell$ for the process $K_L^0\to \pi^\pm \ell^\mp\nu$ in the $K_L^0$ rest frame, given by~\cite{Bijnens:1994me}
\begin{align}
&\frac{d\Gamma_{K\ell3}}{dE_X} = \frac{G_F^2|V_{us}|^2}{64\pi^3 m_K^2} \nonumber\\
&\times \int dq^2\left\{A_X f_+^2(q^2) + \frac{2m_\ell^2(m_K^2 - m_\pi^2)}{q^2}B_X f_+(q^2)\left[f_0(q^2) - f_+(q^2)\right]
\right.\nonumber \\
&\left.\hspace{4.5em} + \frac{m_\ell^2(m_K^2 - m_\pi^2)^2}{q^2}\left(1 - \frac{m_\ell^2}{q^2}\right) \left[f_0(q^2) - f_+(q^2)\right]^2 \right\} \,,
\end{align}
where we define the factors
\begin{align}
A_\nu &= 8 m_K E_\nu(q^2 + m_K^2 - m_\pi^2 - m_\ell^2) - 16 m_K^2 E_\nu^2 - (q^2 - m_\ell^2)(4m_K^2 - m_\ell^2) \,, \nonumber \\
B_\nu &= 4m_K E_\nu + m_\ell^2 - q^2 \,,
\end{align}
for $X = \nu$ and
\begin{align}
A_\ell &= 8 m_K E_\ell(q^2 + m_K^2 - m_\pi^2 + m_\ell^2) - 16 m_K^2 E_\ell^2 - q^2(4m_K^2 + 3m_\ell^2) + m_\ell^2(4m_\pi^2 - m_\ell^2)\,, \nonumber \\
B_\ell &= q^2 - 4m_K E_\ell + 2m_K^2 - 2m_\pi^2 + m_\ell^2 \,,
\end{align}
for $X = \ell$.
Here, $q^2 = (p_K - p_\pi)^2$ is the squared momentum transfer and $f_+(q^2)$ and $f_0(q^2)$ are the vector and scalar $K\to\pi$ transition form factors, respectively, entering the hadronic matrix element as
\begin{align}
\langle \pi^+| \bar{u}\gamma_\mu s|K_L^0\rangle = (p_{K} + p_{\pi})_\mu f_+(q^2)+ \frac{m_K^2 - m_\pi^2}{q^2}(p_{K} - p_{\pi})_\mu \left[f_0(q^2) - f_+(q^2)\right]\,.
\end{align}
For the form factors, we take the parametrisations $f_+(q^2) = f_+(0)(1 + \lambda_+' q^2/m_\pi^2 + \lambda_+'' q^4/m_\pi^4)$ and $f_0(q^2) = f_0(0)(1 + \lambda_0 q^2/m_\pi^2)$ and use the numerical values of $\lambda_+'$, $\lambda_+''$ and $\lambda_0$ measured in Ref.~\cite{KTeV:2004ozu}.
In Fig.~\ref{fig:KL_decays}, the $K_L^0$ rest frame energies $E_\nu$ (left) and $E_\ell$ (right) are shown for $\ell = e$ (blue curves) and $\ell = \mu$ (orange).
We sample from these distributions, taking an isotropic distribution for the three-body decay plane, and boost to the laboratory frame according to the $K_L^0$ momenta found from the simulation of $e^+e^-\to e^+e^-f_2' \to e^+e^-K_L^0 K_L^0$.
For each event, it is possible to compute the pair momentum $|\vec{p}_{\pi\ell}|$, muon momentum $|\vec{p}_\mu|$, pion momentum $|\vec{p}_\pi|$, and pointing angle $\cos\alpha_{\ell\ell} = \hat{r}_{\text{DV}}\cdot\hat{p}_{\text{DV}} = \hat{p}_{K_L^0}\cdot\hat{p}_{\pi\ell}$.
After the $|\vec{p}_{\ell\ell}|$ and $|\vec{p}_\mu|$ cuts (the latter also applying to the pion if it is misidentified as a muon), shown in the centre column of Table~\ref{tab:background_reductions}, a slight reduction in the backgrounds is observed.
However, the addition of the pointing angle cuts in Eq.~\eqref{eq:conversion-cuts} provides a significantly larger suppression, especially for the $\mu^+\mu^-$ channel. This is because the heavy decay products of $K_L^0\to \pi^\pm \mu^\mp \nu$ are more collimated in the $K_L^0$ momentum direction. 
% For the $e^\pm\mu^-$ channel, we consider the same cut as for the $e^+e^-$ channel. 
After all the cuts, we find that an appreciable number of background events, a few tens, remains only for the $e^\pm\mu^\mp$ channel. 

The second dominant source of $K_L^0$ mesons is the process $e^+e^- \to  \gamma K_S^0 K_L^0$, which we expect to be dominated by $\phi \to K_S^0 K_L^0$.
The differential cross section in the polar angle  $\theta_\gamma^*$ of the outgoing photon in the CM frame can be approximated by using radiative return~(c.f. Ref.~\cite{Czyz:2010hj} for a detailed treatment) to the $\phi$ peak as
\begin{align}
\label{eq:epemtogammaKLKS}
\frac{d\sigma}{d\cos\theta_\gamma^*} = \frac{\alpha m_\phi \Gamma_\phi \sigma_{\phi}}{s^2(s - m_\phi^2)} \left[\frac{s^2 + m_\phi^4}{\sin^2\theta_\gamma^*} - \frac{(s - m_\phi^2)^2}{2}\right] \,,
\end{align}
where $\sigma_\phi = 12\pi \text{BR}(\phi \to e^+e^-)\text{BR}(\phi \to K_S^0 K_L^0)/m_\phi^2 = 1409$~nb is the peak cross section for $e^+ e^- \to \phi \to K_S^0 K_L^0$~\cite{BaBar:2014uwz} and the hard photon has the CM frame energy $E_\gamma^* = (s - m_\phi^2)/2\sqrt{s}$.
Due to the large boost of the $\phi$ meson in the laboratory frame,
% and the low $Q$ of the decay
we can assume that the outgoing $K_S^0 K_L^0$ are highly collimated in the direction of the outgoing $\phi$.
Thus, for $K_L^0$ decays in the angular range $43.6^\circ < \theta_{\text{DV}} < 101.5^\circ$, the outgoing photon is approximately confined to $50^\circ < \theta_{\gamma} < 110^\circ$ where it can be detected with a 99\% detection efficiency, resulting in the event being vetoed.
As we did for the photon conversion background from $e^+e^-\to \gamma\gamma$, we can then take a 1\% probability for the photon not being detected.
Integrating Eq.~\eqref{eq:epemtogammaKLKS} over $50^\circ < \theta_{\gamma} < 110^\circ$ after boosting to the laboratory frame yields an effective cross section of $\sigma = 0.25~\text{pb}$, and thus $2.5 \times 10^5$ events per $\text{ab}^{-1}$ within the tracking volume.

Equivalently to the $e^+e^-\to e^+e^-K_L^0K_L^0$ background, we use \texttt{FeynRules} to write a \texttt{UFO} file introducing an effective vector coupling of the $\phi$ meson to $e^+e^-$ and simulate the process $e^+e^-\to\gamma\phi$ in \madgraph.
The $\phi$ meson is transversely polarised and thus the distribution of the $K_L^0$ polar angle in the $\phi \to K_S^0 K_L^0$ rest frame (defined with respect to the direction of the $\phi$ momentum) is given by $d\Gamma/d\cos\theta_K \propto \sin^2\theta_K$.
Boosting the $K_L^0$ to the laboratory frame according to the simulated $\phi$ momentum, we obtain an average transverse momentum of 2.5~GeV, and thus the probability for an average $K_L^0$ to decay within $1.2~\text{cm} < R < 60~\text{cm}$ is 0.8\%.
To reproduce the signal, the outgoing $\gamma$ and $K_S^0$ should not be detected and the $\pi^\pm$ from the $K_L^0$ decay must be misidentified as a charged lepton.
The $K_S^0$ can decay via $K_S^0 \to \pi^+\pi^-$ with a branching fraction of 69.2\%.
We take the probability for both its tracks to escape detection to be 1\%, which appears realistic despite the highly boosted $K_S^0$, if one uses new neural-network-based track-finding algorithms~\cite{Reuter:2026pdv}.
We assume that the combination of a smaller branching fraction and the presence of four detectable photons with a total energy averaging 2.5~GeV makes the decay $K_S^0 \to \pi^0 \pi^0$ subdominant.
Combining these factors, we find that the $e^+e^- \to \gamma K_S^0 K_L^0$ process yields a negligible background for the $e^+e^-$, $\mu^+\mu^-$ and $e^\pm\mu^\mp$ channels before the momentum and angular cuts are applied. Thus, we only consider $K_L^0$ background events from the process $e^+e^- \to e^+e^- K_L^0K_L^0$.

%%%%%%%%%%%%%%%%%%%%%%%%%%%%%%%%%%%%%%%%%%%%%%%%%%%%%%%%%
\subsection{Pointing angle resolution}
\label{app:alpha-resolution}
%%%%%%%%%%%%%%%%%%%%%%%%%%%%%%%%%%%%%%%%%%%%%%%%%%%%%%%%%

Since in Eq.~(\ref{eq:cuts-appendix}) we require a large minimum value of the pointing angle $\alpha_{\ell\ell}$, the impact of the experimental resolution on $\alpha_{\ell\ell}$ should be negligible.
Nevertheless, we estimate the resolution, $\sigma_\alpha$ in the following.
With $\cos\alpha_{\ell\ell} = \hat r_{\textrm{DV}}\cdot \hat p_{\ell\ell}$, contributions to $\sigma_\alpha$ can arise from mismeasurement of either of the unit vectors $\hat{r}_{\textrm{DV}}$ and $\hat{p}_{\ell\ell}$.
For simplicity, we take the different contributions to be uncorrelated.
When our resolution estimates are performed for a single $\ell^\pm$ track, they are then multiplied by $\sqrt{2}$ to approximately account for the impact of both tracks in the DV.

Mismeasurement of $\hat p_{\ell\ell}$ is dominated by mismeasurement of the azimuthal ($\phi$) and polar ($\theta$) angles of the $\ell^+\ell^-$ tracks when their opening angle is small, i.e., for $m_{\ell\ell} / |\vec p_{\ell\ell}| \ll 1$.
For track momenta $|\vec{p}_{\ell^\pm}|$ greater than 2~GeV, Ref.~\cite{Brown:2008ccb} reports the track angular resolutions  $\sigma_\phi \approx \sigma_{\cot\theta} \approx 0.5\times 10^{-3}$~rad.
We note that $\sigma_{\cot\theta}=\sigma_\theta / \sin^2\theta$, and that for $\theta > 50^\circ$, $\sin^2\theta > 0.59$.
For simplicity, we take the worst-case scenario, $\sigma_\alpha=\sqrt{2}\times 0.75\times 10^{-3}$, for all tracks regardless of their polar angle.
We note that these values were obtained at the BABAR experiment for cosmic muons that pass near the IP and thus traverse all tracking layers.
For lack of corresponding information on Belle~II, we assume that the resolutions also approximately hold at Belle~II, ignoring its improvements over BABAR.
For tracks that originate far from the IP, particularly in the CDC, the measurement resolution is worsened by the fewer and less precise hits used to reconstruct the tracks, yet improved due to reduced multiple scattering in detector material.
On the whole, we take these values to  approximately hold for all cases of interest here.
For lower values of the track momentum $p = |\vec{p}_{\ell}|$ and speed $\beta = |\vec{p}_{\ell}|/E_{\ell}$, the resolution is expected to worsen as $1/\beta p$ due to the increase in multiple scattering~\cite{ParticleDataGroup:2024cfk}.

For large opening angles, corresponding to large $m_{\ell\ell} / |\vec p_{\ell\ell}|$, the $\alpha_{\ell\ell}$ resolution is more strongly affected by mismeasurement of the absolute track momenta $p = |\vec{p}_{\ell}|$ and additional momentum fluctuations that arise from final-state radiation and bremsstrahlung.
The relative momentum resolution for pions at Belle~II is shown in Fig.~17 of Ref.~\cite{BelleIITrackingGroup:2020hpx} to be $\sigma_p/p \approx 2.5\times 10^{-3}$ for transverse momenta $p_T\gtrsim 1$~GeV.
The resulting impact on the resolution of $\alpha_{\ell\ell}$ is $\sigma_\alpha \approx m_{\ell\ell} \sigma_p/p^2$, where we have simplified the treatment by neglecting the track masses and taking both tracks to have the same momentum, and have also added their resolutions in quadrature.
For the case of photons conversions from $e^+e^-\to \gamma\gamma$, we take a benchmark track momentum of $p = 2$~GeV.
The measured track momentum suffers additional fluctuations, mostly downward, due to final-state radiation and bremsstrahlung, particularly for electrons.
Based on  the invariant-mass distributions for $J/\psi$ decays to two leptons seen in Figs.~4.3 and~4.4 of Ref~\cite{Fuji:2021}, we take $\sigma_\alpha$ to be larger by a factor of two for $e^+e^-$ pairs produced in photon conversions.

Mismeasurement of $\alpha_{\ell\ell}$ is also impacted by mismeasurement of $\hat r_{\textrm{DV}}$, for which one contribution is the size of the SuperKEKB collision beamspot, which is 6, 0.04, and 150~$\mu$m in the $x$, $y$, and $z$ directions, respectively~\cite{Belle-II:2018jsg}.
For simplicity, we take the benchmark value of $75~\mu$m, approximately accounting for the fact that the plane of the conversion tracks is random.
The other contribution is the resolution of the DV position.
For DVs that are near the IP and tracks with $p_T\approx 1$~GeV, we take this resolution to be the track impact-parameter resolution,  approximately $30~\mu$m, from Fig.~17 of Ref.~\cite{BelleIITrackingGroup:2020hpx}.
Taking both contributions in quadrature, we obtain $\sigma_\alpha\approx 80 / r_{\textrm{DV}}$, with $r_{\text{DV}} = |\vec{r}_{\text{DV}}|$.
As $r_{\textrm{DV}}$ grows, the decrease in this ratio is partly balanced by the resolution degradation due to loss of detector hits.
In particular, for DVs that take place in the CDC, Table~1.3 of Ref.~\cite{Belle-II:2010dht} reports the spatial resolution to be $100~\mu$m in the $r\phi$ direction and $2~$mm in the $z$ direction.
Accounting for the random di-track plane orientation, we take the resolution in this region to be $\sigma_\alpha=1~\textrm{mm}/r_{\textrm{DV}}$.

Summarising these effects and making simplifying assumptions in the conservative direction, we estimate the pointing-angle resolution for photon conversions to be
\begin{align}
\sigma_\alpha \approx \left(1 \oplus 2.5 \frac{m_{\ell\ell}}{p} \cdot b \oplus \frac{\sigma_d}{r_{\textrm{DV}}} \right)\, \textrm{mrad}.
\end{align}
Here the first term corresponds to the track-angle resolution.
The second term arises from the track-momentum resolution, where $p$ is the track momentum, which we take to be approximately 2~GeV for the $e^+e^-\to\gamma\gamma$ background, and with $b=2$ for electrons and 1 for muons.
The third term is due to the vertex resolution, with $\sigma_d=30~\mu$m for $r_{\textrm{DV}}<140$~mm, so that the tracks have precise silicon hits, and $\sigma_d=1~$mm otherwise.
For further simplicity, we take only the dominant third term in the worst-case scenario, $\sigma_d / r_{\textrm{DV}} = 1~\textrm{mm} / 140~\textrm{mm} = 7$~mrad.
Clearly, this resolution is much smaller than the selected cuts on $\alpha_{\ell\ell}$ and thus can be ignored in the subsequent discussion.
The same is expected to be true for the $e^+e^-\to e^+e^-\gamma$ induced background.

%%%%%%%%%%%%%%%%%%%%%%%%%%%%%%%%%%%%%%%%%%%%%%%%%%%%%%%%%
\section{Constraints on $C_{HNe}$}
\label{app:CHNe}
%%%%%%%%%%%%%%%%%%%%%%%%%%%%%%%%%%%%%%%%%%%%%%%%%%%%%%%%%

%
\begin{figure}[t!]
\centering
\includegraphics[width=0.6\linewidth]{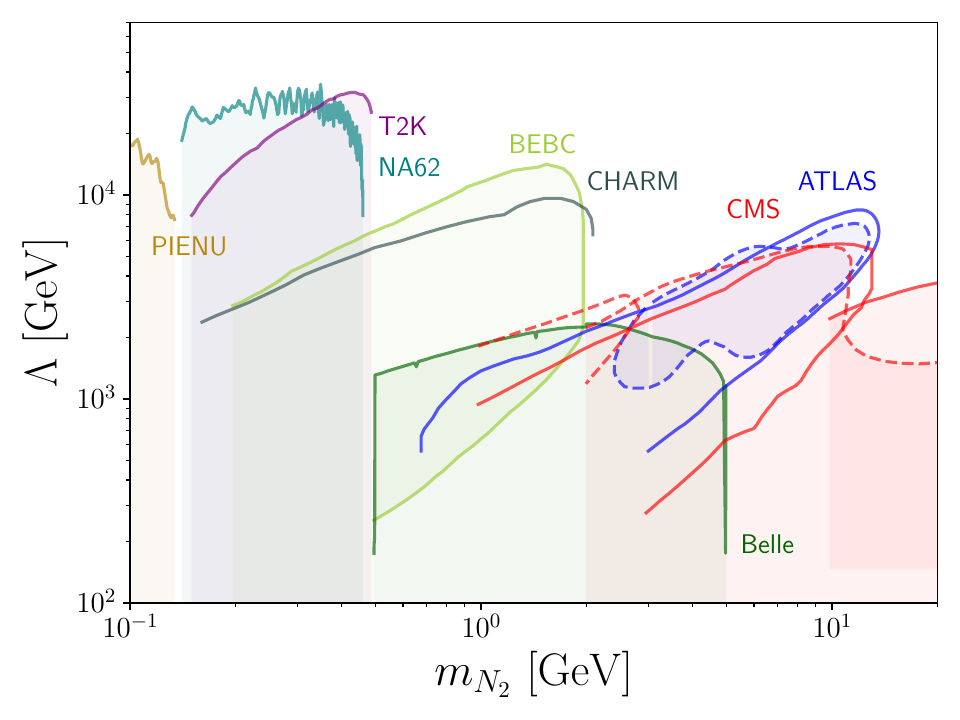}
\caption{Current constraints in the $\Lambda$ vs. $m_{N_2}$ plane for the Wilson coefficient $C_{HNe}$, shown as a combined shaded grey region in Fig.~\ref{fig:DV_MG_nuN2} (top left).}
\label{fig:CHNe_constraints}
\end{figure}

In contrast to the other $\nu$SMEFT operators, the parameter space for $C_{HNe}$ is constrained by a plethora of experiments.
Showing all of these individually as for the other operators in Figs.~\ref{fig:DV_MG_N1N2} and~\ref{fig:DV_MG_nuN2} would completely overshadow the Belle~II sensitivity contours.
Hence, the upper left plot in Fig.~\ref{fig:DV_MG_nuN2} shows only a combined excluded region of parameter space.

In Fig.~\ref{fig:CHNe_constraints}, all constraints are shown individually. These bounds are found by rescaling the upper bounds on the electron-flavour active-sterile mixing, as performed in Ref.~\cite{Fernandez-Martinez:2023phj}. Bounds are shown from PIENU~\cite{PIENU:2017wbj}, NA62~\cite{NA62:2020mcv}, T2K~\cite{T2K:2019jwa}, BEBC~\cite{Barouki:2022bkt}, CHARM~\cite{CHARM:1985nku,CHARMII:1994jjr}, Belle~\cite{Belle:2013ytx}, ATLAS~\cite{ATLAS:2022atq,ATLAS:2025uah}, CMS~\cite{CMS:2022fut,CMS:2023jqi,CMS:2024xdq,CMS:2024ake} and PMNS unitarity~\cite{Fernandez-Martinez:2016lgt}. There are additional strong constraints from $0\nu\beta\beta$ decay experiments, which have been explored in detail in Refs.~\cite{Deppisch:2014zta,Deppisch:2020ztt,Dekens:2020ttz,Dekens:2021qch}.

%%%%%%%%%%%%%%%%%%%%%%%%%%%%%%%%%%%%%%%%%%%%%%%%%%%%%%%%%
\section{Sensitivity dependence on pointing angle cuts}
\label{app:loose_cuts}
%%%%%%%%%%%%%%%%%%%%%%%%%%%%%%%%%%%%%%%%%%%%%%%%%%%%%%%%%

%
\begin{figure}[t!]
\centering
\includegraphics[width=0.45\linewidth]{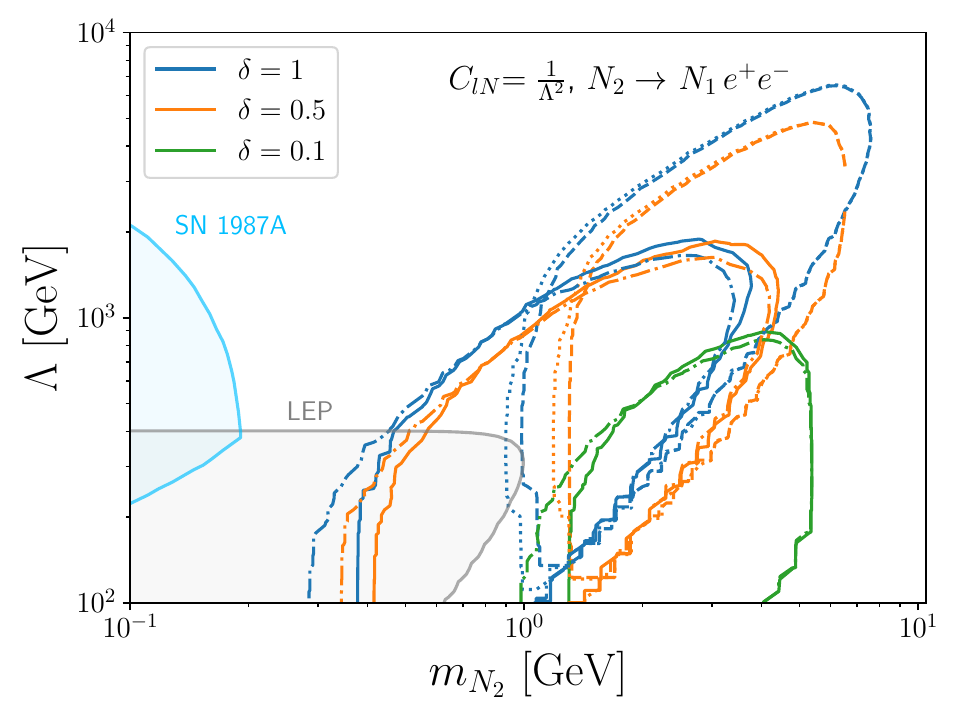}
\includegraphics[width=0.45\linewidth]{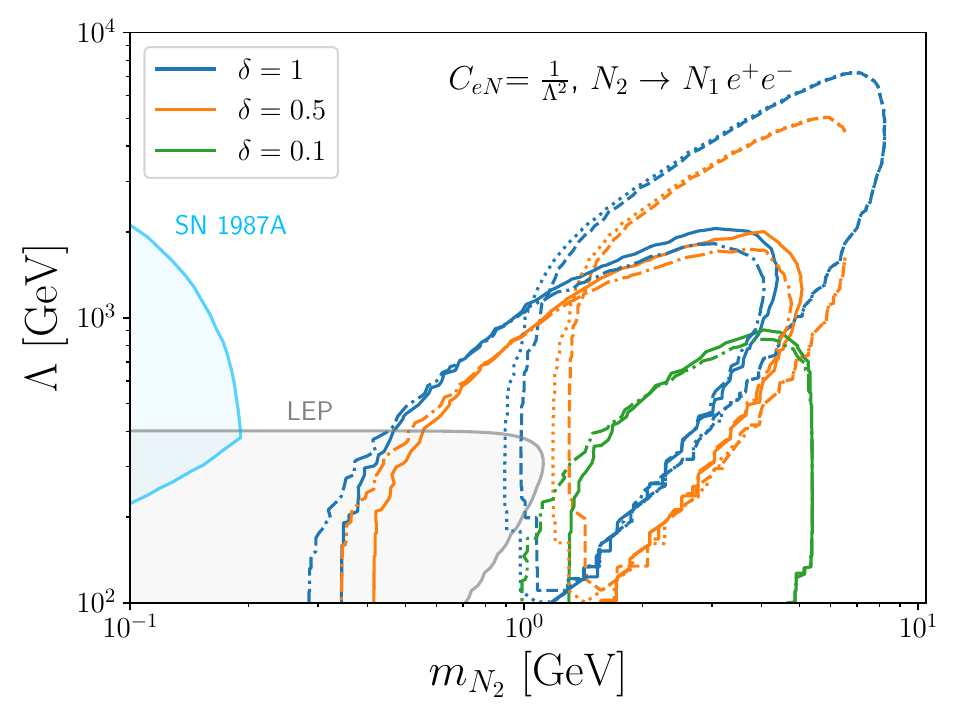}
\includegraphics[width=0.45\linewidth]{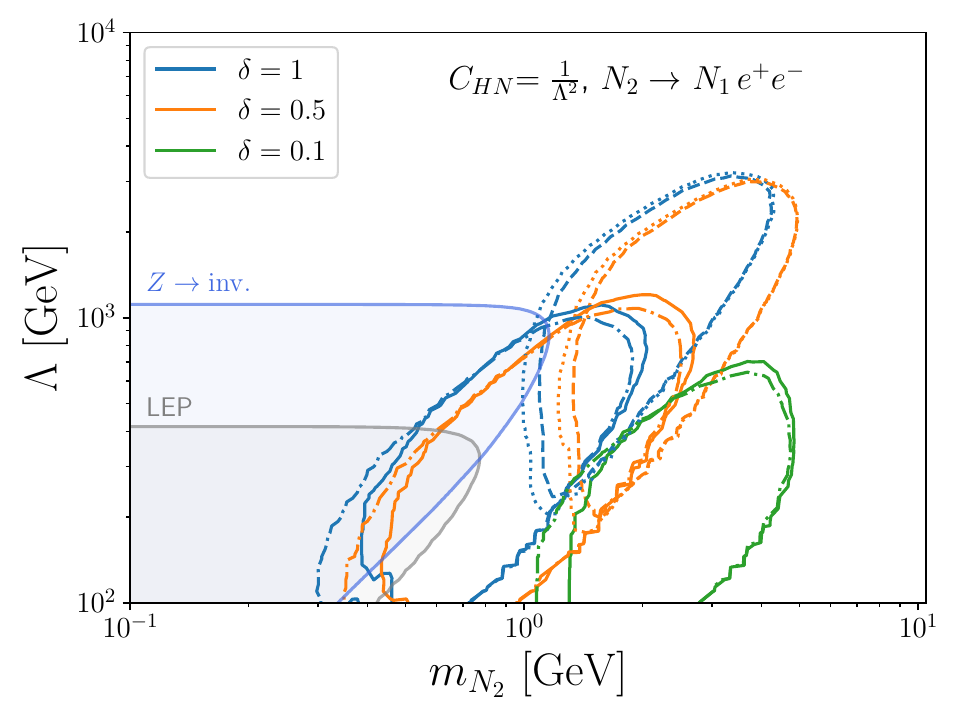}
\includegraphics[width=0.45\linewidth]{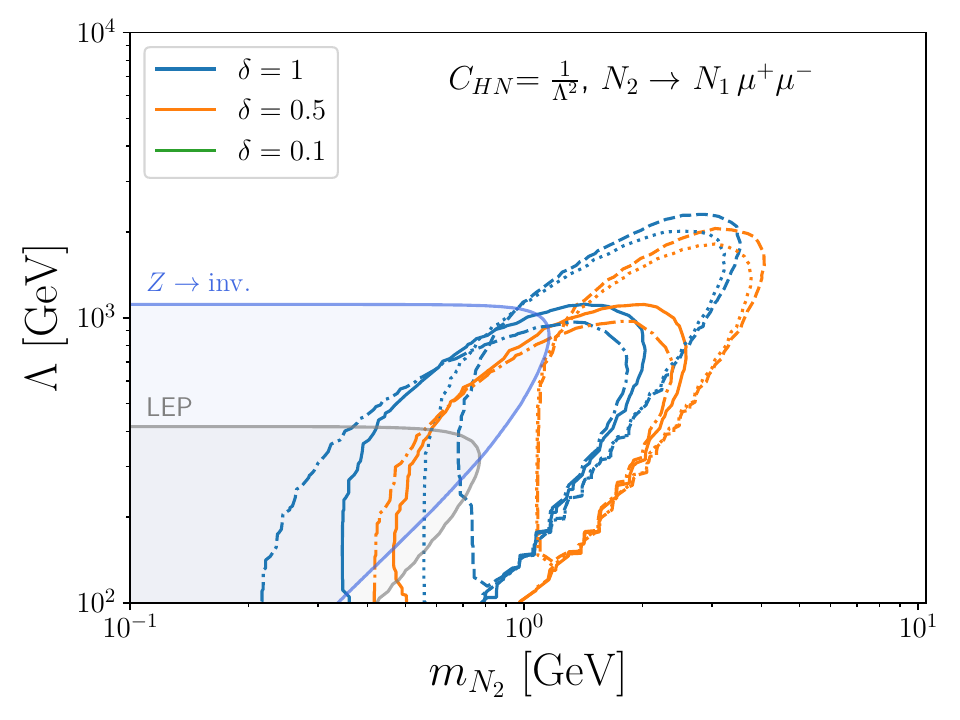}
\caption{Realistic sensitivity contours from a DV signal in the $\Lambda$ vs. $m_{N_2}$ plane for different mass splittings $\delta$ and for different single $\nu$SMEFT operators switched on at a time: $Q_{lN}$ (top left), $Q_{eN}$ (top right) and $Q_{HN}$ in a $e^+e^-$ (bottom left) and $\mu^+ \mu^-$ (bottom right) final state. All three operators allow for decays $N_2 \rightarrow N_1 e^+ e^-$ with variable mass splitting $\delta$, and $C_{HN}$ additionally for $N_2 \rightarrow N_1 \mu^+ \mu^-$ and in principle also $N_2 \rightarrow N_1 \tau^+ \tau^-$, but for the latter the mass threshold is too high to observe a significant amount of signal events. The solid (dash-dotted) curves correspond to the low invariant mass region $m_{\ell\ell} < 0.5$~GeV and the dashed (dotted) lines to the high invariant mass region $m_{\ell\ell} > 0.5$~GeV, assuming tight (loose) $\alpha_{\ell\ell}$ cuts.}
\label{fig:DV_MG_N1N2_tight+loose}
\end{figure}
\begin{figure}[t!]
\centering
\includegraphics[width=0.45\linewidth]{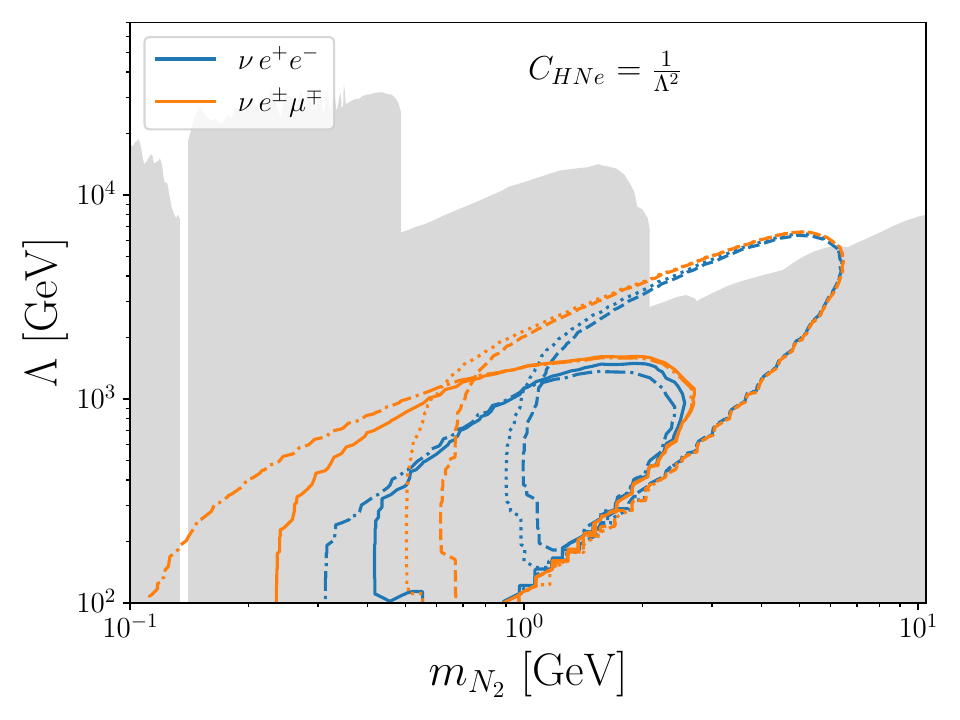}
\includegraphics[width=0.45\linewidth]{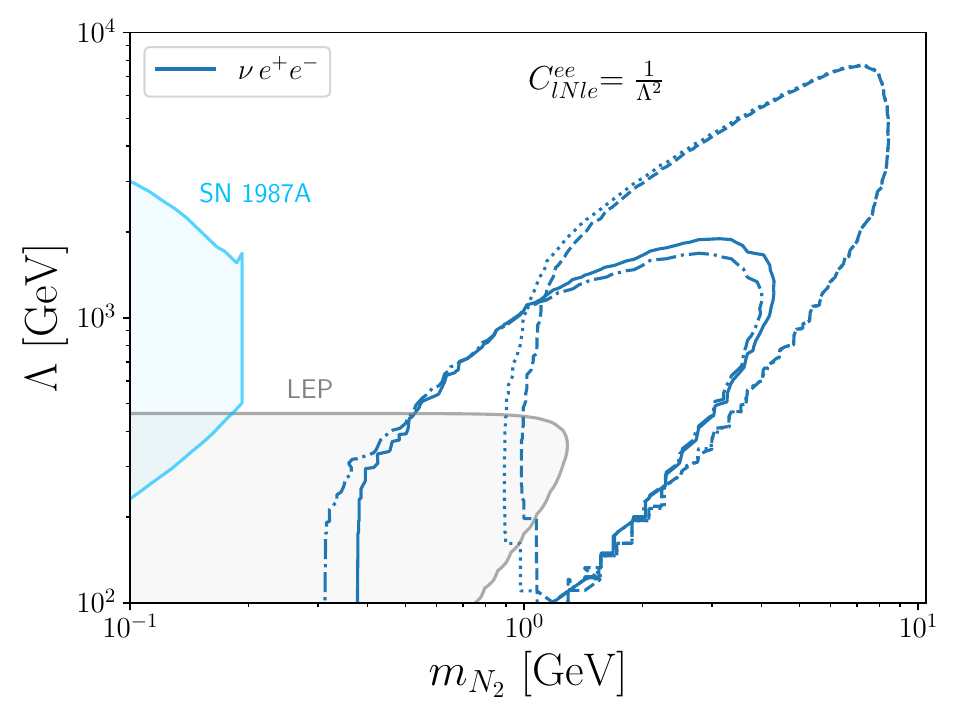}
\includegraphics[width=0.45\linewidth]{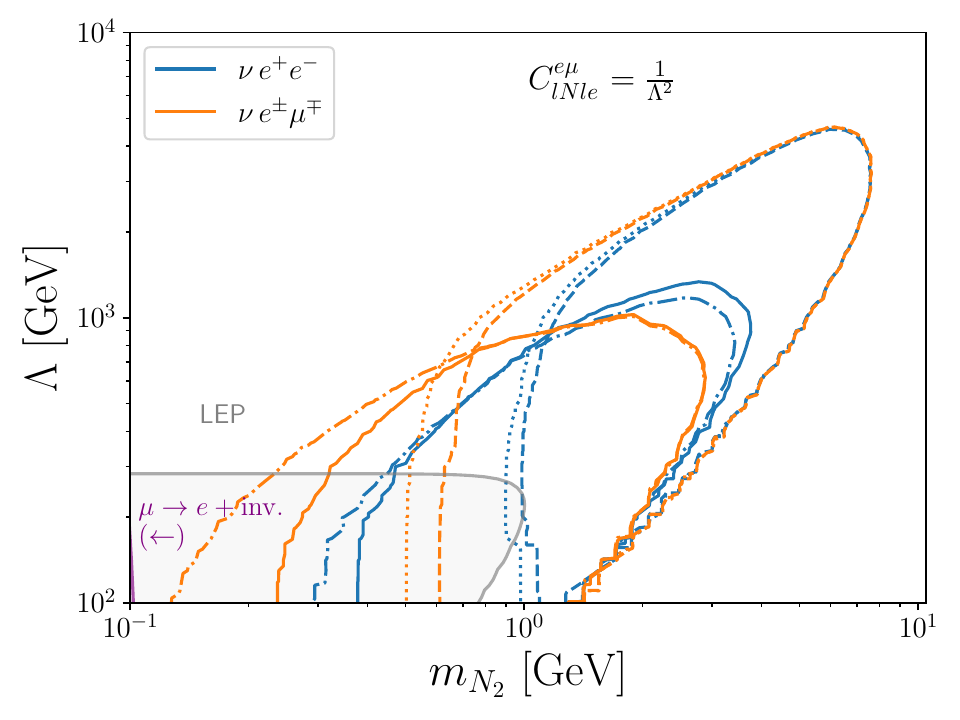}
\includegraphics[width=0.45\linewidth]{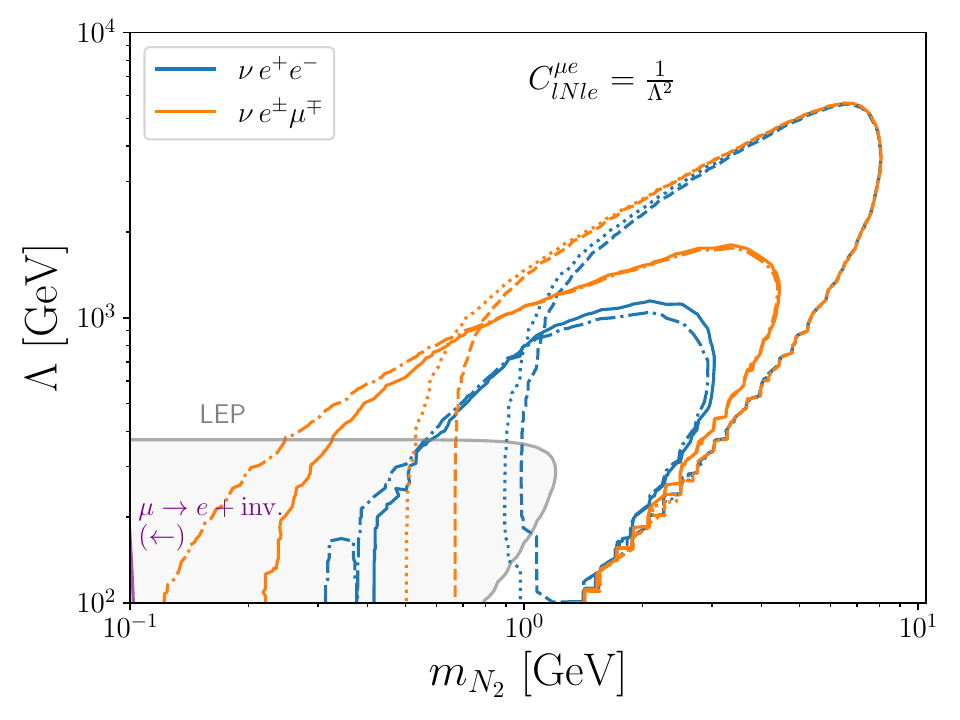}
\includegraphics[width=0.45\linewidth]{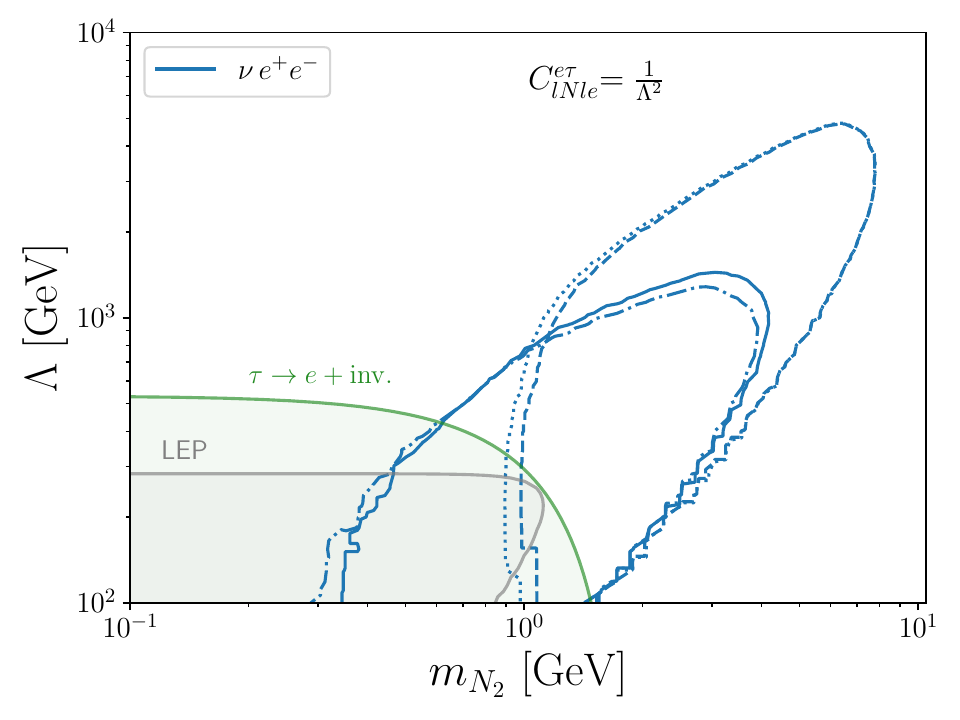}
\includegraphics[width=0.45\linewidth]{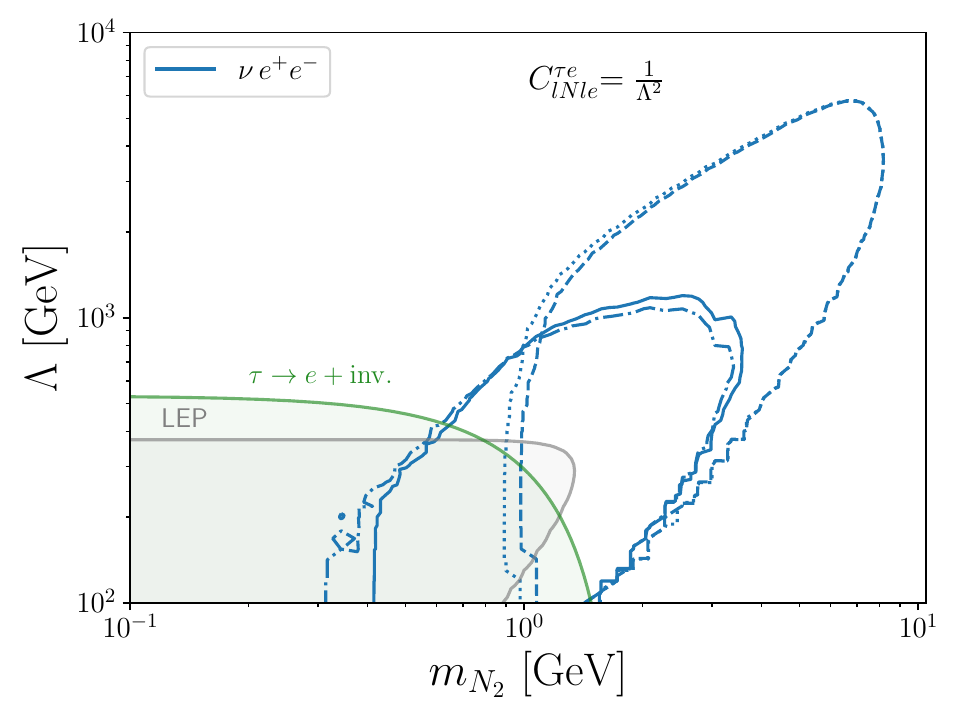}
\caption{Realistic sensitivity contours from a DV signal in the $\Lambda$ vs. $m_{N_2}$ plane for different single operators switched on at a time: $Q_{HNe}$ (top left), $Q_{lNle}^{ee}$ (top right), $Q_{lNle}^{e \mu}$ (centre left), $Q_{lNle}^{\mu e}$ (centre right), $Q_{lNle}^{e \tau}$ (bottom left) and $Q_{lNle}^{\tau e}$ (bottom right). All operators allow for decays $N_2 \rightarrow \nu e^{\pm} \ell^{\mp}$. The solid (dash-dotted) curves correspond to the low invariant mass region $m_{\ell\ell} \leq 0.5$~GeV and the dashed (dotted) lines to the high invariant mass region $m_{\ell\ell} > 0.5$~GeV, assuming tight (loose) $\alpha_{\ell\ell}$ cuts. Different colours denote contours from different final states.}
\label{fig:DV_MG_nuN2_tight+loose}
\end{figure}

Figs.~\ref{fig:DV_MG_N1N2_tight+loose} and \ref{fig:DV_MG_nuN2_tight+loose} show the dependence of the sensitivity contours on the cuts on the pointing angle $\alpha_{\ell\ell}$, where the solid and dashed lines correspond to the low- and high-mass regions, respectively, using the tight cuts on $\alpha_{\ell\ell}$, and the dash-dotted and dotted curves to the same regions for loose cuts on $\alpha_{\ell\ell}$.
It is striking that the difference is mostly visible in the low-mass region, in which the cuts drastically affect the number of background events. In general, one sees that for tighter cuts, higher scales $\Lambda$ can be probed at the mass at which the curves peak, while for looser cuts the sensitivity extends towards smaller mass $m_{N_2}$. This can be understood as the peak scale $\Lambda$ being more strongly affected by the number of background events, while events for small masses $m_{N_2}$ are more likely to have small $\alpha_{\ell\ell}$.

\bibliography{bibliography}
\end{document}